\documentclass[usenatbib,a4paper]{mn2e}

\usepackage{psfig}
\usepackage{url}
\usepackage{units}
\usepackage{graphicx}
\usepackage{amssymb}

\title[First IPTA Data Release]{The International Pulsar Timing Array:
  First Data Release}

\author[J.~P.~W.~Verbiest et.~al.]
{J.~P.~W.~Verbiest,$^{1,2}$\thanks{E-mail:
    verbiest@physik.uni-bielefeld.de}
  L.~Lentati,$^3$                     
  G.~Hobbs,$^4$                       
  R.~van~Haasteren,$^5$               
  P.~B.~Demorest,$^6$                 
  \newauthor
  G.~H.~Janssen,$^7$                  
  J.-B.~Wang,$^8$                     
  G.~Desvignes,$^2$
  R.~N.~Caballero,$^2$
  M.~J.~Keith,$^{9}$                  
  \newauthor
  D.~J.~Champion,$^2$    
  Z.~Arzoumanian,$^{10}$              
  S.~Babak,$^{11}$                    
  C.~G.~Bassa,$^{7}$
  N.~D.~R.~Bhat,$^{12}$               
  \newauthor
  A.~Brazier,$^{13,14}$               
  P.~Brem,$^{11}$
  M.~Burgay,$^{15}$                   
  S.~Burke-Spolaor,$^{6}$
  S.~J.~Chamberlin,$^{16}$            
  \newauthor
  S.~Chatterjee,$^{14,17}$            
  B.~Christy,$^{18}$                  
  I.~Cognard,$^{19,20}$               
  J.~M.~Cordes,$^{17}$                
  S.~Dai,$^{4,21}$                    
  \newauthor
  T.~Dolch,$^{22,14,17}$              
  J.~A.~Ellis,$^{5}$
  R.~D.~Ferdman,$^{23}$               
  E.~Fonseca,$^{24}$                  
  J.~R.~Gair,$^{25}$                  
  \newauthor
  N.~E.~Garver-Daniels,$^{26}$        
  P.~Gentile,$^{26}$
  M.~E.~Gonzalez,$^{27}$              
  E.~Graikou,$^{2}$
  L.~Guillemot,$^{19,20}$
  \newauthor
  J.~W.~T.~Hessels,$^{7,28}$          
  G.~Jones,$^{29}$                    
  R.~Karuppusamy,$^{2}$
  M.~Kerr,$^{4}$
  M.~Kramer,$^{2,9}$
  \newauthor
  M.~T.~Lam,$^{17}$
  P.~D.~Lasky,$^{30}$                 
  A.~Lassus,$^{2}$
  P.~Lazarus,$^{2}$
  T.~J.~W.~Lazio,$^{5}$
  K.~J.~Lee,$^{31}$                   
  \newauthor
  L.~Levin,$^{26,9}$
  K.~Liu,$^{2}$
  R.~S.~Lynch,$^{32}$                 
  A.~G.~Lyne,$^{9}$                   
  J.~Mckee,$^{9}$
  M.~A.~McLaughlin,$^{26}$
  \newauthor
  S.~T.~McWilliams,$^{26}$            
  D.~R.~Madison,$^{33}$               
  R.~N.~Manchester,$^4$
  C.~M.~F.~Mingarelli,$^{34,2}$       
  \newauthor
  D.~J.~Nice,$^{35}$                  
  S.~Os{\l}owski,$^{1,2}$
  N.~T.~Palliyaguru,$^{36}$           
  T.~T.~Pennucci,$^{37}$              
  B.~B.~P.~Perera,$^{9}$
  \newauthor
  D.~Perrodin,$^{15}$
  A.~Possenti,$^{15}$
  A.~Petiteau,$^{38}$                 
  S.~M.~Ransom,$^{33}$
  D.~Reardon,$^{30,4}$
  \newauthor
  P.~A.~Rosado,$^{39}$                
  S.~A.~Sanidas,$^{28}$
  A.~Sesana,$^{40}$                   
  G.~Shaifullah,$^{2,1}$
  R.~M.~Shannon,$^{4,12}$             
  \newauthor
  X.~Siemens,$^{41}$                  
  J.~Simon,$^{41}$
  R.~Smits,$^{7}$
  R.~Spiewak,$^{41}$
  I.~H.~Stairs,$^{24}$
  B.~W.~Stappers,$^{9}$
  \newauthor
  D.~R.~Stinebring,$^{42}$            
  K.~Stovall,$^{43}$                  
  J.~K.~Swiggum,$^{26}$
  S.~R.~Taylor,$^5$
  G.~Theureau,$^{19,20,44}$           
  \newauthor
  C.~Tiburzi,$^{2,1}$
  L.~Toomey,$^{4}$
  M.~Vallisneri,$^{5}$
  W.~van~Straten,$^{39}$
  A.~Vecchio,$^{40}$
  Y.~Wang,$^{45}$                     
  \newauthor
  L.~Wen,$^{46}$                      
  X.~P.~You,$^{47}$                   
  W.~W.~Zhu$^{2}$
  and X.-J.~Zhu$^{46}$
  \\
}
\begin{document}

\date{Accepted. Received ; in original form }

\pagerange{\pageref{firstpage}--\pageref{lastpage}} \pubyear{2015}

\maketitle

\label{firstpage}

\begin{abstract}
  The highly stable spin of neutron stars can be exploited for a
  variety of (astro-)physical investigations. In particular arrays of
  pulsars with rotational periods of the order of milliseconds can be
  used to detect correlated signals such as those caused by
  gravitational waves. Three such ``Pulsar Timing Arrays'' (PTAs) have
  been set up around the world over the past decades and collectively
  form the ``International'' PTA (IPTA). In this paper, we describe
  the first joint analysis of the data from the three regional PTAs,
  i.e.\ of the first IPTA data set. We describe the available PTA
  data, the approach presently followed for its combination and
  suggest improvements for future PTA research. Particular attention
  is paid to subtle details (such as underestimation of measurement
  uncertainty and long-period noise) that have often been ignored but
  which become important in this unprecedentedly large and
  inhomogeneous data set. We identify and describe in detail several
  factors that complicate IPTA research and provide recommendations
  for future pulsar timing efforts. The first IPTA data release
  presented here (and available online) is used to demonstrate the
  IPTA's potential of improving upon gravitational-wave limits placed
  by individual PTAs by a factor of $\sim 2$ and provides a $2-\sigma$
  limit on the dimensionless amplitude of a stochastic GWB of
  $1.7\times 10^{-15}$ at a frequency of $1\,{\rm yr}^{-1}$. This is
  1.7 times less constraining than the limit placed by
  \citep{srl+15ltd}, due mostly to the more recent, high-quality data
  they used.
\end{abstract}


\section{Introduction}\label{sec:intro}
The stable and regular rotation of pulsars, combined with their
lighthouse-like radiation beams enable a wide variety of pulsar timing
experiments of (astro-)physical interest \citep[see][for an
overview]{lk05}. Of particular interest is the use of pulsar timing
arrays to detect correlated signals, such as those caused by
gravitational waves. In the following, the technique of pulsar timing
is explained in some detail (Section~\ref{ssec:PT}), followed by the
potential sources of gravitational waves that our experiment might be
expected to be sensitive to (Section~\ref{ssec:GWPT}). The sensitivity
scaling laws for such GW-detection efforts are described in
Section~\ref{ssec:PTASens} and this provides a clear case for
combining data from as many telescopes as possible, which is the
subject of this paper, introduced in Section~\ref{ssec:DC}.

\subsection{Pulsar Timing}\label{ssec:PT}
The process of pulsar timing is fundamentally dependent on an accurate
description of everything that affects the times of arrival (ToAs) of
the pulsed radiation at the telescope. In addition to a time standard
and the Solar-System ephemerides (which predict the positions and
masses of the Solar-System bodies at any given point in time, to the
degree this information is available), pulsar timing
requires knowledge of the pulsar's spin and spin-down, its position
and proper motion, its distance, the number of dispersing electrons in
the interstellar medium along the propagation path of the radio waves
and (unless the pulsar is solitary) multiple orbital parameters. All
of these parameters are included in a so-called ``timing model'',
which can be used to predict the phase of the pulsar's periodic signal
at any point in time. For a full description of the technique of
pulsar timing, we refer the interested reader to \citet{lk05} and for
a complete derivation of the formulae included in pulsar timing
models, \citet{ehm06} is recommended. In the following, we will
merely highlight the aspects that are directly relevant to the further
analysis presented in this paper.

To determine the arrival times from the observations, a
high-signal-to-noise ratio (S/N) ``template'' profile (i.e.\
phase-resolved pulse shape) is constructed
through coherent addition of the highest-quality data. This template (or an
analytic version derived from it) is then used as a phase reference
against which all other observations are timed through
cross-correlation \citep{tay92}.  The differences between the measured
ToAs and those predicted by the timing model are the ``timing
residuals'', which are the unmodelled difference between the
observations and the theory.
It is the investigation of these timing residuals that allows
additional science (i.e.\ all the science that is not yet included in
the timing model) to be derived.

The amount of information that can be derived from the timing
residuals of any given pulsar varies strongly. In particular, some
binary pulsars are more interesting as they may yield information on
the binary system, such as the pulsar and companion masses, whereas
solitary pulsars can typically at best provide their spin period,
spindown, parallax and proper motion. Non-pulsar-specific correlated
signals, however, should be encoded in the timing residuals of all
pulsars. Three such signals are of particular interest to pulsar
timing array (PTA) experiments \citep{fb90,thk+15}:
\begin{itemize}
\item A monopolar signal, which affects all pulsars equally, would be
  caused by an error in the Earth-based time
  standards
  \footnote{See \citet{kr14} and \citet{pp14} however for a
    potentially different origin.}. Recently, \citet{hcm+12Ltd} used
  PTA data to constrain this signal.
\item A dipolar signal, which would be caused by an imperfection in
  our models of the Solar System. Since the ToAs are necessarily
  corrected for the Earth's motion around the Solar-System
  barycentre,
  incomplete information on the masses and positions of Solar-System
  bodies would cause errors in the timing
  residuals. \citet{chm+10Ltd} made a first attempt at measuring
  such a signal in PTA data. 
\item A quadrupolar signal, as would be caused by GWs, which distort
  space-time in a quadrupolar fashion and therefore affect the ToAs of
  pulsar signals in a quadrupolar way
  \citep{hd83}\footnote{\citet{hd83} also showed that the effect of
    the GWs on the timing is fully characterised by their effect at
    the time the pulsar signal is emitted (the so-called ``pulsar
    term'') and at the time the signal is received (the so-called
    ``Earth term''). In the absence of highly precise information on
    the distances of the pulsars in the array, only the Earth term is
    correlated, in which case the GW effect is not a purely
    quadrupolar signal, but a quadrupolar signal with an equally
    strong white noise component.}. An overview of recent analyses on
  such signals is given in Section~\ref{ssec:GWPT}.
\end{itemize}

In order to detect the extremely weak effects listed above in the timing
residuals, it is important to have very high precision and accuracy in
the measured ToAs. Two sources of white noise in the pulse
observations determine this precision and accuracy. The first of these
is radiometer noise, which affects ToA precision and can be quantified
(in the case of a simple Gaussian or rectangular pulse shape) with the
radiometer equation for pulsar timing
\citep[after][]{lk05}: 
\begin{equation}\label{eq:radiom}
  \sigma_{\rm Radiom} = k \frac{S_{\rm sys}P\delta^{3/2}}{S_{\rm
      mean}\sqrt{t_{\rm int} n_{\rm pol}\Delta f}},
\end{equation}
with $k$ a correction factor accounting for digitisation losses
($k \approx 1$ for modern systems, but for some of the older, one- or
two-bit systems $k \approx 1.2$); 
$S_{\rm sys}=T_{\rm sys}/G = 2k_{\rm B}T_{\rm sys}/A_{\rm eff}$ the
system equivalent flux density which depends on the system
temperature $T_{\rm sys}$, the telescope's effective collecting area
$A_{\rm eff}$ and Boltzmann's constant, $k_{\rm B}$. $P$ is the pulse
period of the pulsar, $\delta = W/P$ is the pulsar's duty cycle (pulse
width $W$ divided by pulse period), $S_{\rm mean}$ is the flux density of
the pulsar averaged over its pulse period, $n_{\rm pol}$ is the number
of polarisations observed and $t_{\rm int}$ and $\Delta f$ are
respectively the duration and bandwidth of the observation. The second
white-noise contribution is pulse-phase jitter, also known as SWIMS
\citep{ovh+11,ovdb13} and affecting both the ToA accuracy and
precision. SWIMS are relevant in any system that has sufficient
sensitivity to detect individual pulses from pulsars, as it quantifies
the stability of pulsar pulse shapes on short timescales, given by:
\begin{equation}\label{eq:jitter}
\sigma_{\rm Jitter} \propto \frac{f_{\rm J} W_{\rm eff} \left( 1 +
    m_{\rm I}^2\right)}{\sqrt{N_{\rm p}}},
\end{equation}
with $f_{\rm J}$ the jitter parameter, which needs to be determined
experimentally \citep{lkl+12,sod+14ltd}; $W_{\rm eff}$ the
pulse width; $m_{\rm I} = \sigma_{\rm E}/\mu_{\rm E}$ the modulation
index, defined by the mean ($\mu_{\rm E}$) and standard deviation
($\sigma_{\rm E}$) of the pulse-energy distribution; and
$N_{\rm p} = t_{\rm int}/P$ the number of pulses in the observation, which
equals the total observing time divided by the pulse period.

Consequently, the highest-precision timing efforts ideally require
rapidly rotating pulsars ($P \lesssim 0.03$\,s) with high relatively flux
densities ($S_{\rm 1.4\,GHz}\gtrsim0.5$\,mJy) and narrow pulses
($\delta \lesssim 20$\%) are observed at sensitive
($A_{\rm eff}/T_{\rm sys}$) telescopes with wide-bandwidth receivers
($\Delta f$) and for long integration times
($t_{\rm int} \gtrsim 30$\,min).

\subsection{Gravitational-Wave Detection with Pulsar
  Timing}\label{ssec:GWPT} 
In order to detect the correlated signals in pulsar timing data, an
array of millisecond pulsars (MSPs) must be observed with large,
sensitive telescopes. Such a ``pulsar timing array''
\citep[PTA,\footnote{Where originally the acronym ``PTA'' was purely
  defined as the set of pulsars that comprise the experiment, more
  recently the same acronym has been used to refer to the
  collaborations that carry out these experiments. We continue this
  convention of having one acronym to refer to both the set of pulsars
  and the scientific collaboration.}][]{rom89,fb90} has currently been
set up in three different places. Specifically, the Australian Parkes
PTA \citep[PPTA,][]{mhb+13ltd} is centred on the Parkes radio
telescope (PKS); the European PTA \citep[EPTA,][]{dcl+16ltd} uses the
five\footnote{The Sardinia Radio Telescope in Italy is also part of
  the EPTA collaboration, but had not yet commenced routine scientific
  observations during the timespan covered by the data presented in
  this work.} major European centimetre-wavelength telescopes (see
Table~\ref{tab:PTAs} for details); and the North-American Nanohertz
Observatory for GWs \citep[NANOGrav,][]{abb+15ltd} uses the Robert
C.~Byrd Green Bank Telescope (GBT) and the 305-m William E.\ Gordon
Telescope of the Arecibo Observatory  (AO). Combined, these three PTAs
form the International PTA \citep[or IPTA, as previously described
by][]{haa+10ltd,man13} and presently observe 49 pulsars (see
Table~\ref{tab:PSRs} for details) in the quest for the aforementioned
correlated signals and for GWs in particular.

\begin{table*}
  \begin{center}
    \caption{Sources of IPTA data. For each regional PTA the telescopes used in the current data set are listed, along with the typical time between observations, the
      number of pulsars observed at each telescope and the observing frequencies used (rounded to the nearest \unit[100]{MHz} and limited to a single band per GHz
      interval). The final two columns give the MJD range over which observations are included in the current combination. The \protect\citet{ktr94} data set is not part of a PTA
      as such, but refers to the publicly available data sets on PSRs~J1857+0943 and J1939+2134, which have also been included in the combined IPTA data. Note that all
      three PTAs are ongoing efforts that continue to extend their data sets. Because the present IPTA combination did not run in parallel to these individual efforts and
      took significantly more time than a data combination at the level of an individual PTA, these constituent data sets are often significantly outdated. A follow-up
      effort to create a second IPTA data combination containing all the most recent data available to the three individual PTAs is ongoing.}
    \label{tab:PTAs}
    \begin{tabular}{l|lcccll}
      \hline
      PTA & Telescope (code) & \multicolumn{1}{c}{Typical}          & Number  & Observing   & \multicolumn{1}{c}{Earliest}         & \multicolumn{1}{c}{Latest}      \\
          &           & \multicolumn{1}{c}{cadence}          & of      & Frequencies & \multicolumn{1}{c}{Date}             & \multicolumn{1}{c}{Date}        \\
          &           & \multicolumn{1}{c}{($\unit{weeks}$)} & Pulsars & (GHz)       & \multicolumn{1}{c}{(MJD, Gregorian)} & \multicolumn{1}{c}{(MJD, Gregorian)} \\
      \hline
      \hline
      EPTA & Effelsberg (EFF) & 4        &   18      & 1.4, 2.6      & 50360 (04 Oct 1996) & 55908 (13 Dec 2011)\\
          & Lovell (JBO)      & 3        &   35      & 1.4           & 54844 (13 Jan 2009) & 56331 (08 Feb 2013)\\
          & Nan\c cay Radio Telescope (NRT) & 2 & 42 & 1.4, 2.1      & 47958 (08 Mar 1990) & 55948 (22 Jan 2012)\\
          & Westerbork (WSRT) & 4        &   19      & 0.3, 1.4, 2.2 & 51386 (27 Jul 1999) & 55375 (28 Jun 2010)\\
      \hline
      NANOGrav & Green Bank Telescope (GBT) & 4 & 10 & 0.8, 1.4      & 53216 (30 Jul 2004) & 55122 (18 Oct 2009)\\
          & Arecibo (AO)      & 4     & \phantom{0}8 & 0.3, 0.4, 1.4, 2.3 & 53343 (04 Dec 2004) & 55108 (04 Oct 2009)\\
      \protect\citet{zsd+15ltd} & GBT \& AO & 2  &\phantom{0}1      & 0.8, 1.4, 2.3 & 48850 (16 Aug 1992) & 56598 (02 Nov 2013)\\      
      \hline
      PPTA & Parkes (PKS)     & 2        &   20      & 0.6, 1.4, 3.1 & 49373 (21 Jan 1994) & 56592 (27 Oct 2013)\\
      \hline
      \protect\citet{ktr94} & Arecibo (AO)    & 2  & \phantom{0}2 & 1.4, 2.3 & 46436 (06 Jan 1986) & 48973 (17 Dec 1992)\\
      \hline
    \end{tabular}
  \end{center}
\end{table*}

\begin{table*}
  \begin{center}
    \caption{The pulsars in the first IPTA data release with their basic properties. Given are the name in the J2000 system (note the following pulsars also have B1950
      names: PSRs J1824$-$2452A, J1857+0943, J1939+2134 and J1955+2908 are respectively known as PSRs B1821$-$24A, B1855+09, B1937+21 and B1953+29), the pulse period in
      milliseconds, the orbital period in days, the dispersion measure in $\unit{cm^{-3}\,pc}$, the flux at $\unit[1.4]{GHz}$ in mJy (if available) and the most likely
      distance in kpc, either from distance measurements compiled by \protect\citet{vwc+12} or (indicated by $^{\dag}$) from the Galactic electron density model of
      \protect\citet{cl02}, assuming a 20\% uncertainty. The next three columns indicate by ``X'' the PTAs in which the pulsar is observed and the final column gives the relevant
      references for the given
      data, typically including the discovery paper, the most recent timing analysis and where relevant a paper with the VLBI astrometry or flux density measurement. The
      references are as follows:
      (1) \protect\cite{lzb+00},     
      (2) \protect\cite{aaa+09eltd}, 
      (3) \protect\cite{lkn+06},     
      (4) \protect\cite{bhl+94},     
      (5) \protect\cite{aaa+10bltd}, 
      (6) \protect\cite{hlk+04},     
      (7) \protect\cite{tbms98},     
      (8) \protect\cite{nbf+95},     
      (9) \protect\cite{kxl+98},     
      (10) \protect\cite{dyc+14},    
      (11) \protect\cite{vl14},      
      (12) \protect\cite{jlh+93},    
      (13) \protect\cite{vbv+08},    
      (14) \protect\cite{mhb+13ltd}, 
      (15) \protect\cite{dvtb08},    
      (16) \protect\cite{bjd+06},    %
      (17) \protect\cite{lnl+95},    %
      (18) \protect\cite{vbc+09},    
      (19) \protect\cite{cnst96},    %
      (20) \protect\cite{sna+02},    %
      (21) \protect\cite{bjb+97},    %
      (22) \protect\cite{lzc95}, 
      (23) \protect\cite{nss+05},
      (24) \protect\cite{nll+95},    %
      (25) \protect\cite{lwj+09},    
      (26) \protect\cite{hbo06},     
      (27) \protect\cite{jbo+07},    
      (28) \protect\cite{llb+96},    %
      (29) \protect\cite{llww05},    %
      (30) \protect\cite{fwc93},     %
      (31) \protect\cite{eb01b},     
      (32) \protect\cite{jsb+10},    %
      (33) \protect\cite{jac04a},    
      (34) \protect\cite{fwe+12},    
      (35) \protect\cite{sfl+05},    %
      (36) \protect\cite{fsk+04},    %
      (37) \protect\cite{lfl+06},    %
      (38) \protect\cite{lbm+87},    %
      (39) \protect\cite{hfs+04},    %
      (40) \protect\cite{fsk+10},    %
      (41) \protect\cite{gsf+11},    
      (42) \protect\cite{srs+86},    %
      (43) \protect\cite{fw90},      
      (44) \protect\cite{jbv+03},    
      (45) \protect\cite{tsb+99},    
      (46) \protect\cite{bkh+82},    
      (47) \protect\cite{bbf83},     %
      (48) \protect\cite{ntf93},     %
      (49) \protect\cite{nss01},     %
      (50) \protect\cite{rtj+96},
      (51) \protect\cite{spl04},
      (52) \protect\cite{cam95a},    
      (53) \protect\cite{wdk+00},
      (54) \protect\cite{cnt93},
      (55) \protect\cite{cnt96},
      (56) \protect\cite{nt95},
      (57) \protect\cite{van13}.
    }
    \label{tab:PSRs}
    \vspace{-2em}
    \begin{tabular}{lrrrrlcccl}
      \hline 
      J2000 & Pulse  & Orbital & Dispersion           & Flux Density & Distance & EPTA & NANOGrav & PPTA & Reference(s) \\
      Name  & Period & Period  & Measure              & at $\unit[1.4]{GHz}$ & (kpc)        &   &   &   &              \\
            & (ms)   & (days)  & ($\unit{cm^{-3}\,pc}$) & (mJy)        &                 &   &   &   &              \\
      \hline
      J0030+0451   &  4.865 &    -- &             4.33 &           0.6 & $0.28^{+0.10}_{-0.06}$ & X & X &   & (1, 2, 3) \\
      J0034$-$0534 &  1.877 &   1.6 &            13.77 &           0.6 & $0.5\pm0.1^{\dag}$     & X &   &   & (4, 5, 6, 7)      \\
      J0218+4232   &  2.323 &   2.0 &            61.25 &           0.9 & $3.2^{+0.9}_{-0.6}$    & X &   &   & (8, 6, 9, 10, 11) \\
      J0437$-$4715 &  5.757 &   5.7 &             2.64 &         149.0 & $0.156\pm0.001$        &   &   & X & (12, 13, 14, 15)  \\
      J0610$-$2100 &  3.861 &   0.3 &            60.67 &           0.4 & $3.5\pm0.7^{\dag}$     & X &   &   & (16)         \\
                                                              
      J0613$-$0200 &  3.062 &   1.2 &            38.78 &           2.3 & $0.9^{+0.4}_{-0.2}$    & X & X & X & (17, 18, 14) \\
      J0621+1002   & 28.854 &   8.3 &            36.60 &           1.9 & $1.4\pm0.3^{\dag}$     & X &   &   & (19, 20)     \\
      J0711$-$6830 &  5.491 &    -- &            18.41 &           1.4 & $0.9\pm0.2^{\dag}$ &   &   & X & (21, 18, 14)          \\
      J0751+1807   &  3.479 &   0.3 &            30.25 &           3.2 & $0.4^{+0.2}_{-0.1}$    & X &   &   & (22, 23, 9)  \\
      J0900$-$3144 & 11.110 &  18.7 &            75.70 &           3.8 & $0.5\pm0.1^{\dag}$ & X &   &   & (16)           \\
                                                              
      J1012+5307   &  5.256 &   0.6 &             9.02 &           3.0 & $0.7^{+0.2}_{-0.1}$    & X & X &   & (24, 25, 9) \\
      J1022+1001   & 16.453 &   7.8 &            10.25 &           1.5 & $0.52^{+0.09}_{-0.07}$ & X &   & X & (19, 18, 57) \\
      J1024$-$0719 &  5.162 &    -- &             6.49 &           1.5 & $0.49^{+0.12}_{-0.08}$ & X &   & X & (21, 18, 14, 26) \\
      J1045$-$4509 &  7.474 &   4.1 &            58.17 &           2.2 & $0.23^{+0.17}_{-0.07}$ &   &   & X & (4, 18, 14) \\ 
      J1455$-$3330 &  7.987 &  76.2 &            13.57 &           1.2 & $0.5\pm0.1^{\dag}$ & X & X &   & (17, 6, 7)    \\
                                                              
      J1600$-$3053 &  3.598 &  14.3 &            52.33 &           2.4 & $2.4^{+0.9}_{-0.6}$   & X & X & X & (27, 18, 14)  \\
      J1603$-$7202 & 14.842 &   6.3 &            38.05 &           4.2 & $1.2\pm0.2^{\dag}$ &   &   & X & (28, 18, 14)  \\
      J1640+2224   &  3.163 & 175.5 &            18.43 &           2.0 & $1.2\pm0.2^{\dag}$ & X & X &   & (29, 9)         \\
      J1643$-$1224 &  4.622 & 147.0 &            62.41 &           5.0 & $0.42^{+0.09}_{-0.06}$ & X & X & X & (17, 18, 14)  \\
      J1713+0747   &  4.570 &  67.8 &            15.99 &           7.4 & $1.05^{+0.06}_{-0.05}$ & X & X & X & (30, 18, 14)  \\
                                                              
      J1721$-$2457 &  3.497 &    -- &            47.76 &           0.6 & $1.3\pm0.3^{\dag}$ & X &   &   & (31, 32)        \\
      J1730$-$2304 &  8.123 &    -- &             9.62 &           3.9 & $0.5\pm0.1^{\dag}$ & X &   & X & (17, 18, 14)  \\
      J1732$-$5049 &  5.313 &   5.3 &            56.82 &           1.3 & $1.4\pm0.3^{\dag}$ &   &   & X & (31, 18, 14)  \\
      J1738+0333   &  5.850 &   0.4 &            33.77 &            -- & $1.5\pm 0.1$       & X &   &   & (33, 34)        \\
      J1744$-$1134 &  4.075 &    -- &             3.14 &           3.3 & $0.42\pm0.02$      & X & X & X & (21, 18, 14)  \\
                                                              
      J1751$-$2857 &  3.915 & 110.7 &            42.81 &           0.1 & $1.1\pm0.2^{\dag}$  & X &   &   & (35)              \\
      J1801$-$1417 &  3.625 &    -- &            57.21 &           0.2 & $1.5\pm0.3^{\dag}$ & X &   &   & (36, 37)        \\
      J1802$-$2124 & 12.648 &   0.7 &           149.63 &           0.8 & $2.9\pm0.6^{\dag}$ & X &   &   & (36, 40)        \\
      J1804$-$2717 &  9.343 &  11.1 &            24.67 &           0.4 & $0.8\pm0.2^{\dag}$ & X &   &   & (28, 6, 9)    \\
      J1824$-$2452A&  3.054 &    -- &           120.50 &           1.6 & $5\pm1^{\dag}$ &   &   & X & (38, 18, 14)  \\
                                                              
      J1843$-$1113 &  1.846 &    -- &            59.96 &           0.1 & $1.7\pm0.3^{\dag}$  & X &   &   & (39)       \\
      J1853+1303   &  4.092 & 115.7 &            30.57 &           0.4 & $2.09\pm0.4^{\dag}$ & X & X &   & (36, 41, 35)     \\
      J1857+0943   &  5.362 &  12.3 &            13.30 &           5.9 & $0.9\pm0.2$         & X & X & X & (42, 18, 14, 43) \\
      J1909$-$3744 &  2.947 &   1.5 &            10.39 &           2.6 & $1.26\pm0.03$       & X & X & X & (44, 18, 14)     \\
      J1910+1256   &  4.984 &  58.5 &            38.06 &           0.5 & $2.3\pm0.5^{\dag}$ & X & X &   & (36, 41, 35)          \\
                                                              
      J1911+1347   &  4.626 &    -- &            30.99 &           0.1 & $1.2\pm0.2^{\dag}$ & X &   &   & (28, 45, 9)           \\
      J1911$-$1114 &  3.626 &   2.7 &            30.98 &           0.5 & $2.1\pm0.4^{\dag}$ & X &   &   & (36, 37)      \\
      J1918$-$0642 &  7.646 &  10.9 &            26.55 &           0.6 & $1.2\pm0.2^{\dag}$ & X & X &   & (31, 32)       \\
      J1939+2134   &  1.558 &    -- &            71.04 &          13.8 & $5^{+2}_{-1}$      & X &   & X & (46, 18, 14, 43)    \\
      J1955+2908   &  6.133 & 117.3 &           104.58 &           1.1 & $4.6\pm0.9^{\dag}$ & X & X &   & (47, 41, 9)           \\
                                                              
      J2010$-$1323 &  5.223 &    -- &            22.16 &           1.6 & $1.0\pm0.2^{\dag}$ & X &   &   & (27)         \\
      J2019+2425   &  3.935 &  76.5 &            17.20 &            -- & $1.5\pm0.3^{\dag}$ & X &   &   & (48, 49)        \\
      J2033+1734   &  5.949 &  56.3 &            25.08 &            -- & $2.0\pm0.4^{\dag}$  & X &   &   & (50, 51)        \\
      J2124$-$3358 &  4.931 &    -- &             4.60 &           2.4 & $0.30^{+0.07}_{-0.05}$ & X &   & X & (21, 18, 14)  \\
      J2129$-$5721 &  3.726 &   6.6 &            31.85 &           1.6 & $0.4^{+0.2}_{-0.1}$ &   &   & X & (28, 18, 14)  \\
                                                              
      J2145$-$0750 & 16.052 &   6.8 &             9.00 &           9.3 & $0.57^{+0.11}_{-0.08}$ & X & X & X & (4, 18, 14)   \\
      J2229+2643   &  2.978 &  93.0 &            23.02 &           0.9 & $1.5\pm0.3^{\dag}$ & X &   &   & (52, 53, 9)   \\
      J2317+1439   &  3.445 &   2.5 &            21.91 &           4.0 & $0.8\pm0.2^{\dag}$ & X & X &   & (54, 55, 9)   \\
      J2322+2057   &  4.808 &    -- &            13.37 &            -- & $0.8\pm0.2^{\dag}$  & X &   &   & (48, 56)        \\
      \hline
    \end{tabular}
  \end{center}
\end{table*}

The search for GW signals in pulsar timing data is pursued along
several lines, according to the types of predicted GW sources. In the
past \citep[see, e.g.][]{hd83,fb90,ktr94,jhlm05}, isotropic and
incoherent GW backgrounds were considered in a pulsar timing
context. Such a gravitational-wave background (GWB) could arise in
three different ways. Firstly, it could be the gravitational
equivalent to the cosmic microwave background: a GW background arising
from the era of graviton decoupling in the early Universe
\citep{gri05,bb08}. Secondly, various processes involving cosmic
strings could cause a GW background at frequencies detectable by PTAs
\citep[and reference therein]{sbs12}. Finally, hierarchical
galaxy-formation models predict a large number of supermassive
black-hole (SMBH) binaries in the Universe's history. This population
would produce a GW background of particular astrophysical interest and
its predicted amplitude and frequency range may well lie within reach
of current PTA sensitivity \citep{rr95a,ses13}.

In addition to stochastic sources of GWs, several types of single
sources could be detectable by PTA efforts as well. Clearly nearby
SMBH binaries would be detectable if they stand out above the
aforementioned background \citep{svv09}, but in addition to those,
bursts of GWs might be detected as well, arising from a periastron
passage in a highly eccentric SMBH binary \citep{fl10}, cusps in
cosmic strings \citep{dv00} or single SMBH merger events
\citep{set09,vl10,pbp10}. Interestingly, in the case of a single SMBH
merger, the merger event itself is likely undetectable to PTAs, but
its gravitational memory effect \citep{fav09} might be detectable.

At present, the most constraining limit from pulsar timing on the
stochastic GW background, is a 95\%-confidence upper limit of
$1.0\times 10^{-15}$ on the dimensionless strain
amplitude\footnote{Note that all limits quoted here are at a
  reference frequency of $\unit[1]{yr^{-1}}$ or \unit[32]{nHz} and
  where needed assume a spectral index for the characteristic strain
  spectrum of $-2/3$, as expected from an incoherent superposition of
  SMBH binary signals.},
that was obtained by \citet{srl+15ltd} and based on data from the
PPTA. Competitive limits of 
$1.5\times 10^{-15}$ 
and
$3\times 10^{-15}$ 
have been placed by
\citet{abb+15bltd} 
and
\citet{ltm+15ltd}, 
respectively, based on the NANOGrav and EPTA data. Single-source
limits have recently been derived by \citet{bps+15ltd} from the EPTA
data, by \citet{abb+14ltd} from the NANOGrav data and by
\citet{zhw+14} from the PPTA data, in all cases showing that all
proposed binary SMBH systems are still well below current sensitivity
levels. Similar conclusions were reached for GW burst events
\citep{whc+15}. Most recently, \citet{tmg+15ltd} used the quadrupolar
correlation signal to probe the anisotropy and granularity of the
background and placed the first constraints on this.

\subsection{PTA Sensitivity}\label{ssec:PTASens}
Because the GW background from SMBH binaries is better-founded and
predicted to be stronger than the other backgrounds; and because the
burst events are predicted to be extremely rare
\citep{set09,vl10,pbp10}, PTA research has so far focussed on
detecting single SMBH binaries or a stochastic background composed of
these. In the low-S/N regime where the gravitational wave background
contributes less power to the data than the other noise sources
outlined in Section~\ref{ssec:PT}, \citet{sejr13} derived that the S/N
of a PTA's detection sensitivity scales as
\begin{equation}\label{eq:PTASens}
S/N \propto NCA^2T^{13/3}/\sigma^2,
\end{equation}
where $N$ is the number of pulsars in the array, $C$ the cadence
(i.e.\ the inverse of the typical observing periodicity), $A$ the
expected amplitude of the GW background, $T$ the length of the pulsar
timing data set and $\sigma$ the root-mean-square (RMS) of the timing
residuals. Clearly the length of the data set is of great importance,
as is the timing precision (hence further strengthening the
requirement for large, sensitive radio telescopes). In the
intermediate regime, where GWs start to stand out beyond the noise,
this scaling law changes and the number of pulsars becomes far more
relevant:
\begin{equation}\label{eq:PTASens2}
S/N \propto N C^{3/26} A^{3/13} T^{1/2} / \sigma^{3/13}.
\end{equation}
For single SMBH binary sources, the sensitivity would scale as
$A\sqrt{NTC}/\sigma$ \citep{lwk+11}, also strongly dependent on the
timing precision. Either single sources or a background of
gravitational waves could realistically be expected for the first
detection, as demonstrated by \citet{rsg15}.

The above scaling laws indicate several clear ways of improving the
sensitivity of PTAs to GWs in the near future. Specifically, the
sensitivity can be improved by: adding more pulsars to the array
(i.e.\ increasing $N$), as can be achieved particularly through
pulsar surveys which discover previously-missed MSPs with good
potential for high-precision timing (see Figure~\ref{fig:Disc} and
Section~\ref{ssec:survey});
increasing the observing cadence, $C$, which can be accomplished
through pooling of observing resources, i.e.\ by combining data from
multiple telescopes; increasing the time-span of the observations,
$T$, which can be done through the addition of archival data or
continued observing; and improving the timing precision (i.e.\
lowering $\sigma$), which can be done through hardware improvements,
increased integration times and bandwidths; and generally by using the
most sensitive telescopes available. (More advanced improvements to
the analysis method will also strongly impact timing precision. A list
of some advances currently under investigation will be presented in
Section~\ref{ssec:Improvement}.)

A substantial gain in sensitivity could be expected from combining the
data sets from the three existing PTAs. This should improve our
sensitivity through all factors mentioned above (except the amplitude
of the GWs, which is independent of the observing strategy), given
existing complementarity between the three PTAs. 
Such a combination
is, however, a technical challenge for a number of reasons 
that are explained in detail throughout this paper. 
%
%
%

\subsection{Data Combination}\label{ssec:DC}
In this paper, we provide a detailed analysis of the steps that
are involved in an IPTA data-combination project. When combining
data from different telescopes and collaborations, in principle the
steps should be well-defined and straightforward, namely:
\begin{itemize}
\item concatenate ToAs and merge timing models, or select the best
  timing model as a starting point;
\item insert phase offsets between ToAs of different instruments that
  have not otherwise been aligned;
\item correct ToA uncertainties (which are often underestimated);
\item correct time-variable interstellar dispersion;
\item estimate the covariances between arrival time estimates owing to
  low-frequency timing noise;
\item re-fit the timing model.
\end{itemize}
However, in practice many of these steps have to be iterated or
performed simultaneously, which is often complicated by
inconsistencies in the data and lack of (meta-)data.

To correctly and straightforwardly perform the steps listed above in
future IPTA efforts, we therefore discuss the complications
and shortcomings of current PTA data sets and provide recommendations
that will facilitate IPTA research in the future. Specifically, we
briefly describe the current state of the IPTA and its technical
set-up in Section~\ref{sec:IPTA}; list specifics of the data sets
currently available and discuss the practical difficulties inherent to
this present data set in Section~\ref{sec:Data}. Since the current
state of IPTA data combination leaves much to be desired (a situation
we attempt to remedy in this work), the data set presented here is
relatively outdated and therefore not optimally sensitive to
GWs. Nevertheless, to illustrate the difference the IPTA can provide,
we present limits on the strength of a GWB, both for the individual
PTA data sets and the combined data set, in
Section~\ref{sec:GWlimits}. As the goal of our work is to ease PTA
research in the future, we present a summary of challenges and
expected progress beyond the present work, on both technical and
analytic fronts, in Section~\ref{sec:Future}; and
Section~\ref{sec:Conc} concludes the paper with a list of projects
based on combined IPTA data sets.  A detailed list of recommendations
for pulsar timing projects is presented in Appendix~\ref{sec:Format},
where we propose a ``best practice'' for pulsar timing formats and
methods.

\section{The IPTA}\label{sec:IPTA}
The IPTA consists of three regional PTAs: the EPTA, NANOGrav and the
PPTA, as listed in Table~\ref{tab:PTAs}. These three arrays are
complementary in their capabilities, most specifically in their sky
coverage 
and in their observing frequencies, which are crucial for correction
of time-variable interstellar effects, as described in more detail in
Section~\ref{ssec:Comb}. Furthermore, the combined data from these
three PTAs can increase the average observing cadence by a factor of
up to six, further improving the sensitivity to GWs.

\subsection{The IPTA Source List}\label{ssec:Sources}
\begin{figure*}
  \includegraphics[angle=-90.0,width=17cm]{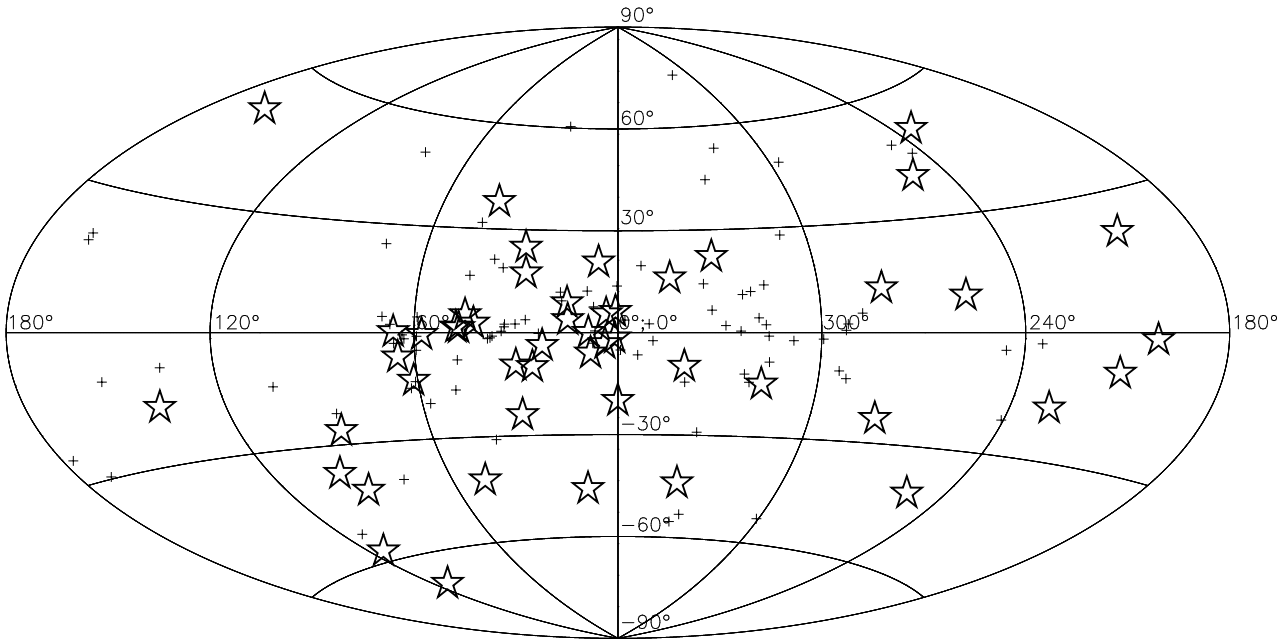}
  \caption{Plot of the Galactic distribution of all presently known
    Galactic-disk MSPs \protect\citep[as found in the ATNF pulsar catalogue
    version 1.51,][]{mhth05}. MSPs currently part of the IPTA are
    indicated by star symbols, all other Galactic MSPs that are
    detected in radio and do not inhabit globular clusters, are
    indicated by crosses. Galactic latitude is on the vertical axis;
    Galactic longitude on the horizontal axis, increasing leftward,
    with the Galactic centre at the origin. Several newly discovered
    pulsars that are presently being evaluated in terms of potential
    timing precision, fill holes in the current PTA distribution,
    particularly at high Northern latitudes and in the Galactic
    anti-centre.}
  \label{fig:Sky}
\end{figure*}

\begin{figure}
  \includegraphics[width=6cm,angle=-90.0]{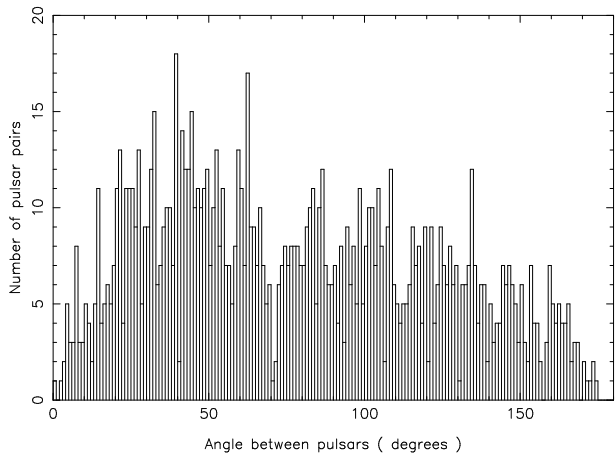}
  \caption{Histogram of the angular separation between the IPTA
    pulsars. Even though the pulsars are not spread out evenly on the
    sky (see Figure~\ref{fig:Sky}), the angular separation between
    pulsars has a relatively uniform coverage. In this histogram,
    every bin corresponds to one degree.}
  \label{fig:HDCoverage}
\end{figure}

The combined source list of the current IPTA data release contains 49
MSPs, of which 14 are solitary and 35 are in binary orbits. The binary
MSPs are mostly orbited by helium white dwarfs (28 systems), with six
CO white-dwarf binaries and one black-widow system
(PSR~J1610$-$2100). The global placement of our telescopes allows IPTA
pulsars to be spread across the entire sky, as shown in
Figure~\ref{fig:Sky}. Because the known MSP population is concentrated
in the Galactic disk and in the inner Galaxy, the IPTA sources also
cluster in those regions. (Note this clustering is not necessarily
physical, but partly a consequence of the inhomogeneous surveying
performed so far.) In the search for isotropic stochastic correlated
signals, the sky position of pulsars is not in itself of importance,
but the distribution of angular separations between pulsar pairs does
impact the sensitivity \citep{hd83}\footnote{Note that for anisotropic
  searches, the absolute sky positions do
  matter.}. Figure~\ref{fig:HDCoverage} shows the histogram of the
angular separations in the IPTA sample and Table~\ref{tab:separations}
shows the pairs of pulsars with the largest and smallest angular
distances on the sky. Clearly small angles, up to $\sim 70^{\circ}$
are most densely sampled, but the angular sampling is overall quite
uniform, notwithstanding the apparent clustering of our pulsars
towards the inner Galaxy. An important point of note, however, is that
for many practical purposes only a subset of these 49 pulsars may be
used. Specifically, only a handful of these pulsars dominate
constraints on GWs, which is primarily a consequence of the wide range
in timing precision obtained on these sources, something that is not
taken into account in the theoretical analyses mentioned in
Section~\ref{ssec:PTASens} but which has been considered in the
context of observing schedule optimisation \citep{lbj+12}.

\begin{table}
  \begin{center}
    \caption{Pulsar pairs with the largest and smallest angular
      separations on the sky.}
    \label{tab:separations}
    \begin{tabular}{ccc}
      \hline
      Pulsar     & Pulsar     & Angular    \\
      Name       & Name       & Separation \\
      (J2000)    & (J2000)    & (degrees)  \\
      \hline
      J1910+1256 & J1911+1347 & 0.88 \\
      J1721$-$2457 & J1730$-$2304 & 2.79 \\
      J1751$-$2857 & J1804$-$2717 & 3.32 \\
      J1853+1303 & J1857+0943 & 3.47 \\
      J1853+1303 & J1910+1256 & 4.14 \\
      \hline
      J0621+1002 & J1843$-$1113 & 174.5\\
      J0621+1002 & J1801$-$1417 & 173.5\\
      J0751+1807 & J2010$-$1323 & 173.4\\
      J1012+5307 & J2129$-$5721 & 172.6\\
      J0613$-$0200 & J1738+0333 & 171.1\\
      \hline
    \end{tabular}
  \end{center}
\end{table}

\begin{figure}
  \includegraphics[width=6cm,angle=-90.0]{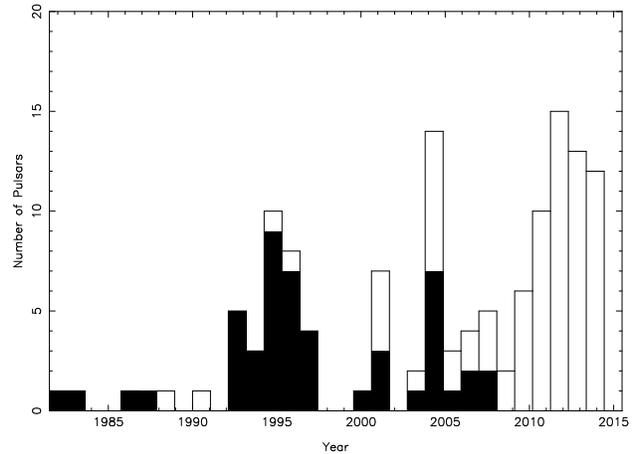}
  \caption{Histogram showing the discovery dates of all known MSPs
    belonging to the Galactic disk population \protect\citep[as featured in
    the ATNF pulsar catalogue][]{mhth05}. The MSPs contained in the
    current IPTA data set are indicated in black. Multiple new
    discoveries are not included in this first IPTA data combination
    as the constituent data sets are slightly aged and the timing
    baselines for most recently discovered pulsars were too short to
    significantly add to PTA work at the time, but more recently many
    of these sources have been included in PTAs, e.g. by
    \protect\citet{abb+15ltd} and these sources will be contained in future IPTA
    work.}
  \label{fig:Disc}
\end{figure}

As can be seen in Figure~\ref{fig:Disc}, recent surveys have resulted
in a very strong growth of the known MSP population. Before these new
MSPs can be usefully employed in PTA analyses, however, their timing
models must be adequately determined and their timing precision needs
to be evaluated.  For these reasons (and the strong dependence of GW
sensitivity on the timing baseline, as discussed in
Section~\ref{ssec:PTASens}), the current data set is dominated by MSPs
discovered in the mid-1990s and early 2000s.  Many more MSPs are
already being monitored by the various PTAs, but these are not
effective for GW detection efforts yet and are excluded from the
present work. Some preliminary results on those new discoveries were
recently presented by \citet{abb+15ltd} and included in the IPTA
source list of \citet{man13}. The complete list of MSPs contained in
the first IPTA data release, is given in Table~\ref{tab:PSRs}, along
with some basic characteristics.

\subsection{Constituent Data Sets}\label{ssec:sets}
As listed in Table~\ref{tab:PTAs}, the IPTA data set is a combination
of the data sets presented by the three PTAs independently: the
NANOGrav five-year data set \citep{dfg+13ltd}, spanning from 2005 to
2010; the extended PPTA Data Release 1 \citep{mhb+13ltd}, ranging from
1996 to February 2011; and the EPTA Data Release 1.0
\citep{dcl+16ltd}, covering 1996 to mid-2014; complemented by the
publicly available data from \citet{ktr94} on PSRs~J1857+0943 and
J1939+2134 (timed from their discoveries in 1982 and 1984
respectively, until the end of 1992) and the extended NANOGrav data on
PSR~J1713+0747 \citep[extended from its discovery in 1992 to the end
of 2013]{zsd+15ltd}
\footnote{The analysis of further archival data
  from the Arecibo telescope is ongoing and will likely further extend
  the baseline and increase the cadence for other pulsars too; but
  inclusion of these data is left for a future paper.}. These data
sets typically average observations in both frequency and time,
leading to a single ToA per pulsar, observation and telescope. There
are three exceptions to this: the \citet{dfg+13ltd} NANOGrav data are
timed without frequency averaging, so each frequency-channel provides
a single ToA; the \citet{zsd+15ltd} data were partially averaged in
time (up to 30 minutes) and frequency (final frequency resolution
dependent on the observing frequency and instrument used); and
observations made with the Parkes dual 10/50-cm receiver result in two
ToAs: one per observing band.

Two differences exist between the data presented here and those
published by the individual PTAs. 
The PPTA data differ for PSR~J1909$-$3744 as the initial version
published by \citet{mhb+13ltd} had instrumental offsets fixed at
values that were sub-optimal for this high-precision data set. The
updated PSR~J1909$-$3744 data used in our analysis have these offsets
determined from the data and have been extended with more recent
observations; this version of the PPTA data is described in more
detail by \citet{srl+15ltd}. The EPTA data differ as the data set
described by \citet{dcl+16ltd} contains additional 
digital-filterbank data for several pulsars.
This subset of the EPTA data does add some more ToAs, though their
precision is limited given the low sensitivity of the instrument. This
limits the contribution to the IPTA data set as a whole, justifying
its exclusion from our analysis.

Finally, to ensure consistency between pulsars and improve the
analysis, all timing models made use of the DE421 Solar-System
ephemeris model \citep{fwb09}, used a solar-wind density model with a
density of 4 electrons per cubic cm at \unit[1]{AU} \citep{yhc+07b}
and were referred to the TT(BIPM2013) time scale using barycentric
coordinate time \citep[TCB][]{hem06}.

\section{Creating The IPTA Data Set}\label{sec:Data}
In the analysis of long, high-precision pulsar timing data sets, four
fundamental challenges arise. 

Firstly, delays in or changes to observing hardware cause time offsets
between different telescopes and observing systems, which are derived
from the data through fitting of arbitrary offsets (so-called
``jumps'') between systems, which can lower the sensitivity to signals
of interest.

Secondly, imperfections in the data analysis and relevant algorithms
as well as possible environmental and elevation-dependent effects,
conspire with noise and noise-like artefacts in the observations to
corrupt the estimation of ToA uncertainties. This is particularly a
problem for pulsars that scintillate strongly. Scintillation is a
propagation effect caused by the ionised interstellar medium (IISM)
and to first order results in order-of-magnitude variations in the
observed flux density of a pulsar, making the ToA uncertainty highly
variable, too. The strength of scintillation depends strongly on the
observing frequency, distance to the pulsar and the nature of the IISM
between us and the pulsar in question. For a more complete overview of
scintillation (and some of its higher-order effects), the interested
reader is referred to \citet{ric90} and \citet{sti13}.

For sources that do not show significant scintillation this problem is
limited, since uncertainties could simply be ignored without much loss
of information; but since the IPTA MSP sample consists of mostly
nearby sources (see Table~\ref{tab:PSRs}), scintillation does
occur\footnote{Note that scintillation can combine with
  frequency-dependent variations in the pulse profile shape to cause
  systematic corruptions to ToAs. As discussed in
  Appendix~\ref{sec:Format}, approaches to prevent such corruptions
  have recently been developed.}, especially for the brightest and
most precisely timed MSPs. Ignoring the ToA uncertainties thereby
worsens timing precision (i.e.\ the RMS of the data set and its
sensitivity to timing parameters) dramatically, implying that a more
accurate estimate of the timing uncertainties is needed. This problem
is compounded by the large variation in the types of calibration that
have been applied to the data. Because pulsar emission is typically
highly polarised, imperfections in the receiver systems can cause
corruptions to the pulse shape if the systems are not properly
calibrated for polarisation \citep{sbm+97,van13}. These effects are
strongly receiver and telescope dependent and so are not equally
important for each data set. Furthermore, the different levels at
which the IPTA data have been calibrated imply that any
calibration-related imperfections will affect different subsets quite
differently, thereby adding importance to the underestimation of ToA
uncertainties. 

Thirdly, because pulsars are high-velocity objects, the lines of
sight to them move slightly through the Galaxy during our observing
campaign. This combines with small-scale structures in the IISM and
results in time-variable, frequency-dependent variations in
ToAs. These variations may be accounted for in the pulsar timing
model, provided multi-frequency data are available at all times;
alternatively a mathematical description needs to be used to
interpolate between (or extrapolate from) multi-frequency epochs.

The fourth and final challenge for long-term, high-precision pulsar
timing is low-frequency noise. This does not directly affect the
precision of the ToAs themselves, but can significantly distort the
timing model and complicates combination of data sets that are not
(fully) overlapping in time. Low-frequency noise could have
instrumental origins \citep[which can be correlated between
pulsars][]{van13} or might be intrinsic to the pulsar, as is the case
for slow pulsars \citep{hlk10}. This unexplained, long-term noise is
of particular concern for PTAs as PTA projects are long-term projects
by definition.

In this section, we first describe each of these issues in detail,
along with the approach taken to measure and correct these in the IPTA
data (Section~\ref{ssec:Comb}). Subsequently, in
Section~\ref{ssec:Data}, the results from our analysis are presented
and any shortcomings of the present data set in this regard are
identified. Many of these shortcomings could be avoided or limited in
future (large-scale) pulsar timing projects, provided some ``rules of
best practice'' are followed. A list of such recommendations
is presented in Appendix~\ref{sec:Format}. 

\subsection{Complications of IPTA Data Combination}\label{ssec:Comb}
Each of the IPTA's constituent data sets is highly inhomogeneous,
combining a large number of different telescopes and/or data recording
systems and observing frequencies. In addition to this, the observing
cadence is often highly irregular and occasionally observations at a
particular observatory or observing frequency are interrupted entirely
for instrumental upgrades (see Figure~\ref{fig:Coverage}). For the
longer data sets especially, observing set-ups (central observing
frequencies, bandwidths, integration times and cadences) changed in
time, making the statistical properties of these data sets highly
non-stationary. These aspects greatly complicate any analysis and make
the properties of the three PTA data sets very different. Consequently
each of the PTAs has developed its own tools and practices to correct
the four main challenges listed earlier, but by design these
approaches are often hard to extend to the data from the other
collaborations. To best account for all described effects
simultaneously, we chose to employ the recently developed
\textsc{temponest} software \citep{lah+14} in our
analysis. \textsc{temponest} is an extension to the \textsc{tempo2}
software package \citep{hem06} that performs the timing analysis
within a Bayesian framework. Further details are given below and by
\citet{lah+14}.

\subsubsection{Definition of Systemic Offsets}\label{sssec:jumps}
Time delays in the signal chain between the telescope's focus (where
the pulsar signal is first received) and the hardware that
applies a time stamp (which can be traced to a time standard) to the data,
are supposedly constant
in time, but can differ greatly between different observing systems
and telescopes. Methods to measure these time offsets between
different systems, at a level of precision beyond the presently
achieved pulsar timing precision, are being developed
\citep{mhb+13ltd,abb+15ltd}, but are as yet in their infancy and not
widely adopted, or only applicable to data from multiple systems on a
single telescope. 
Consequently, in combining heterogeneous data, all observing set-ups
that could have different instrumental delays must be aligned by
subtraction of a constant phase offset that is part of the timing
model\footnote{Assuming that offsets are within a pulse period; and
  that larger offsets have already been corrected.}.

To this end, homogeneous \emph{systems} were identified within the
data. A system in this context is defined as a unique combination of
observing telescope, recording system and receiver (or centre
frequency) used. For the EPTA telescopes the receiver information was
not always available, so if multiple receivers were used
interchangeably at the same centre frequency (as is the case in
particular for the Effelsberg 100-m radio telescope), this was ignored
and both receivers were considered the same. In the case of historic
PPTA data, a further complication arose since the earliest data were
analysed by \citet{vbc+09}, whereas more recent data (from the same
observing systems) were analysed through independent pipelines,
thereby introducing another arbitrary phase offset. In these cases
distinction was made between versions of the same system at different
times. In some cases fewer than five ToAs were identified as a single
system. Such systems (and their ToAs) were removed from the analysis
as they add very little information, particularly after
determining a systemic offset and uncertainty factors (see the next
sub-section).

Because \citet{mhb+13ltd} did determine some instrumental time delays
at high precision, these PPTA systems were bound together in
\emph{groups} and offsets within such groups were not determined,
except for PSRs~J0437$-$4715, J1713+0747 and J1909$-$3744, which
are more sensitive to these offsets than the independent measurements
made by \citet{mhb+13ltd}. 
Subsequently constant time offsets between all groups and un-grouped
systems were determined. 
A discussion of the measured offsets (and mostly of the limitations of
the available data sets in this regard) is given in
Section~\ref{ssec:Data} and suggested improvements for future work on
this topic are listed in Appendix~\ref{sec:Format}.

\subsubsection{Determining the Measurement Uncertainties}\label{sssec:Unc}
It has long been known that the uncertainties of ToAs do not
accurately describe their scatter \citep{lvk+11}. There are two known
reasons for this, though more unidentified reasons may exist. Firstly,
the standard approach to ToA determination proposed by \citet{tay92}
does not determine the formal uncertainty on the ToAs, but instead
calculates an approximate value that underestimates the true error in
the low-S/N regime. Secondly, in the high-S/N regime, pulse-phase
jitter (or SWIMS) will become relevant and add an extra noise
component to the ToAs (see Section~\ref{ssec:PT}).
The resulting underestimation of ToA uncertainties has a direct impact
on the uncertainties of the timing model parameters. More importantly
for the IPTA, if the underestimation is different for different
telescopes or receiving systems, then different sub-sets will
effectively be weighted more strongly than others, without actual
justification.

Two standard statistical approaches can be used to amend this
situation. Firstly, ToA uncertainties can simply be multiplied by a
system-dependent factor, the so-called ``error factor'' or EFAC. This
approach might be justified in case a S/N-dependent underestimation of
the ToA uncertainty is present. Alternatively, uncertainties can be
increased through quadrature addition of a constant noise level, the
so-called ``quadrature-added error'' or EQUAD. This approach is mostly
justified in the high-S/N regime, where pulse-phase jitter adds a
random variation to the ToAs, which is unquantified by the Gaussian
noise in the off-pulse region of the observation \citep{ovh+11}, or
in case a (possibly instrumental) noise floor exists \citep[as
e.g. shown in Figure~2 of][]{vbb+10ltd}. 

Historically, EQUADs have been applied before EFACs \citep{ehm06}:
\begin{equation}\label{eq:EQUADs}
\sigma_{\rm new} = F\sqrt{Q^2+\sigma_{\rm old}^2}
\end{equation}
where $F$ and $Q$ are the EFAC and EQUAD values, respectively. This
may seem counter-intuitive given the physical reasoning laid out
above. The \textsc{temponest} software implements the application in
reverse order, namely \citep{lah+14}:
\begin{equation}\label{eq:TNEQUAD}
\sigma_{\rm new} = \sqrt{Q^2+F^2\sigma_{\rm old}^2}.
\end{equation}
In practice, both of these approaches are too simplified to be optimal
\citep[as discussed in detail by][]{sod+14ltd}, since jitter noise
(and therefore some part of the EQUAD) should decrease with the square
root of the integration length and the mechanisms underlying the need
for an EFAC are still relatively poorly
quantified. 

A third correction factor for ToA uncertainties is the ``error
correction factor'' or ECORR, introduced by \citet{abb+14ltd} and
described in detail by \citet{vv14}. This factor accounts for
pulse-phase jitter in two ways: it functions as an EQUAD factor in the
determination of uncertainties; and it takes into consideration
correlations between simultaneous ToAs taken at different observing
frequencies. In particular for the NANOGrav data this factor is
important, as the highly sensitive NANOGrav data are split in
frequency bands that are narrower than the bandwidth of pulse-phase
jitter, implying the jitter component is fully correlated between
ToAs \citep{ovh+11}. 

For the IPTA data combination, EFAC and EQUAD values were derived for
all systems (as defined in Section~\ref{sssec:jumps}) and ECORR values
were determined for all NANOGrav systems. In doing so, we used the
\textsc{temponest} definition of EQUAD and EFAC and will do so
henceforth. Practically this makes no difference for the EFAC, but in
the case of the EQUADs, the values that we report must be divided by
the EFAC value in order to obtain the equivalent quantity according to
the \textsc{tempo2} definition.

\subsubsection{Modelling Interstellar Dispersion Variability}\label{sssec:DMvar}
Because of dispersion in the IISM, radio signals undergo a
frequency-dependent delay when traversing ionised clouds in our Galaxy
\citep{lk05}:
\begin{equation}\label{eq:DM}
t = D \times \frac{{\rm DM}}{f^2},
\end{equation}
with $D = \unit[4.148808\times 10^3]{MHz^2cm^3s/pc}$, $f$ the
observing frequency and the dispersion measure
${\rm DM} = \int_0^dn_{\rm e}(l)\,{\rm d}l$ the integrated electron density
between us and the pulsar along the line of sight. This effect in
itself has little impact on high-precision timing, but because of the
high spatial velocities of pulsars and because of the Earth's motion
around the Sun, the lines of sight to our pulsars sample changing
paths through density variations in the IISM, thereby making this
delay time-variable. Such a time-variable signal clearly does
affect pulsar timing efforts, especially on the longest time scales,
where both the IISM effects \citep{ars95} and the GW background
\citep{ses13} are strongest\footnote{Note that, depending on the GW
  source population, it has been shown that the GW background may peak
  at higher frequencies, too \citep{en07,shmv04}.}.

\begin{figure*}
  \includegraphics[width=23cm,angle=-90.0]{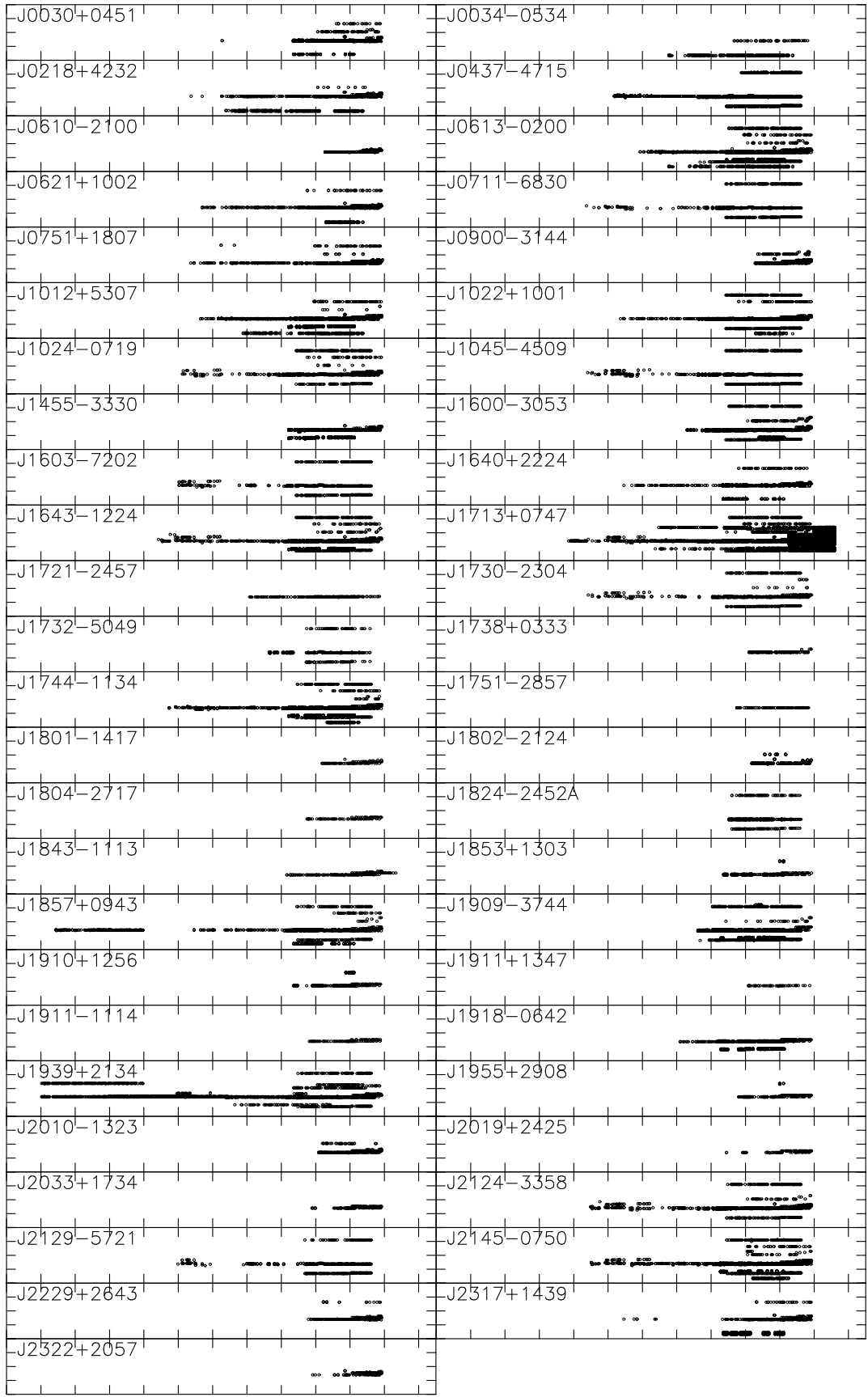}
  \caption{Plot of temporal and frequency coverage for all
    pulsars. The ranges of each sub-plot are identical and cover the
    MJD-range 45000 to 57500 (31 January 1982 to 22 April 2016) in X
    and 0 to \unit[4]{GHz} in Y. Tick marks are at 1000-day and 1-GHz
    intervals. Note that only the centre frequency of each ToA is
    plotted (i.e. the bandwidth is ignored) and that many pulsars have
    only a few years of data at a single frequency.}
  \label{fig:Coverage}
\end{figure*}

Correcting these interstellar delays (henceforth referred to as ``DM
variations'') is not necessarily problematic, provided adequate
multi-frequency data are available at all times. In reality, however,
multi-frequency data are often intermittent or lacking altogether (as
can be seen in Figure~\ref{fig:Coverage}), or are of insufficient
quality. This has made corrections for DM variations a significant
problem, which has been dealt with in a variety of ways in the past.

Traditionally, time-derivatives of DM were included in the timing
model \citep[e.g. by][]{cbl+95}, but in case of sufficiently dense
sampling, smoothed time series have also occasionally been applied
\citep{ktr94}. More recently such smoothing has been developed further
\citep{yhc+07,kcs+13ltd}, but this approach only really works well if
the multi-frequency sampling is relatively homogeneous throughout the
data set. Furthermore, the most recent of these developments
\citep{kcs+13ltd} does not take into consideration the uncertainties
of the individual DM measurements. The issue of DM correction becomes
even more complex in highly sensitive wideband systems, where the
frequency-dependence of the pulse profile shape introduces possible
correlations with the measured DM \citep{pdr+14,ldc+14}, causing
\citet{dfg+13ltd} to propose a correction method specifically aimed at
such data, but difficult to apply to less sensitive, more narrow-band
observations.

\textsc{temponest} does not indirectly correct for DM variability,
like most previous methods did, but directly implements a spectral
model of the DM variations and obtains the posterior probability
distribution for the model parameters that define its power spectrum,
taking into account the entire data set rather than individual
observing epochs one at a time.
Specifically, for the IPTA data combination discussed here,
two-parameter power law models\footnote{Note that the spectral shape
  is fundamentally free and that different spectral models can be
  evaluated. A complete comparison of the evidence for different DM
  spectral models will be presented in a paper by
  \citet{lea+15}; here we merely use the most likely model, which is a
  power law.} with $f^{-2}$ scaling were evaluated and included in the
final timing models in case significant evidence for such variations
existed. In addition to such a power-law model, an annual DM variation
was evaluated for PSR~J0613$-$0200, because \citet{kcs+13ltd}
identified such a trend; and individual DM ``events'' (i.e.\
short-term changes that do not follow the power-law model but do have
a $f^{-2}$ behaviour) were evaluated for PSRs~J1603$-$7202 and
J1713+0747, in agreement with \citet{kcs+13ltd} and \citet{dcl+16ltd}
respectively. Details of the DM event models are given by
\citet{lea+15}. Contrary to \citet{kcs+13ltd}, our analysis showed no
evidence for annual DM variations in excess of our power-law model,
for PSR~J0613$-$0200. This is primarily caused by the fact that our
power-law model already contains DM variations at the periodicity of a
year, while the analysis by \citet{kcs+13ltd} quantified the total
power of DM variations on a yearly timescale, rather than the excess
DM variations beyond a power-law model.

\subsubsection{Evaluation of Intrinsic Pulsar Timing Instabilities}\label{sssec:Instab}
A final difficulty in long-term, high-precision pulsar timing is the
presence of intrinsic pulsar timing noise. Such long-period noise has
long been documented in slow pulsars \citep[e.g.][]{bgh+72} and a few
exceptional MSPs also display this property \citep{ktr94}, though most
MSPs have to date shown surprising levels of stability
\citep{vbc+09,mhb+13ltd}. As time spans become longer and
instrumentation becomes more sensitive, however, instabilities and
their associated low-frequency noise become clearer and start to
affect subsequent pulsar analyses \citep{vbv+08,chc+11} and in
particular the search for long-period GWs. This is particularly so if
predictions of steep-spectrum timing noise in MSPs hold true
\citep{sc10}. In order to cope with this, as part of the data
combination, individual low-frequency noise models that do not depend
on the observing frequency were determined for each
pulsar. Furthermore, in order to accommodate the possibility of
instabilities in the observing hardware, the presence of low-frequency
noise in every observing system independently was
investigated.

As for the DM modelling, we only consider power-law models and refer
to \citet{lea+15} for a full comparison of spectral models. The
results of our analysis are summarised in Section~\ref{ssec:Data}.

\subsection{Determination of Pulsar Timing Parameters}
In addition to the group and system offsets, the EFACs and EQUADs, the
DM spectra and low-frequency noise, all traditional parameters of the
pulsar timing model such as pulse period and spindown, astrometric
position and proper motion, parallax (where detectable), dispersion
measure and any orbital parameters, are jointly evaluated by
\textsc{temponest}. Especially for binary pulsars, a wide variety of
orbital parameters (and relativistic time-derivatives thereof) could
be included in the timing model. Parameters that were not detected
with at least 90\% confidence, were not included in the timing
models. In some cases apparently relativistic terms can have geometric
causes \citep[e.g.\ a binary pulsar with high proper motion could be
observed to have an anomalous periastron advance, see][]{kop95,kop96}. We
have not undertaken the interpretation of such terms and translation
into geometric parameters (inclination, longitude of the ascending
node) if this was not already done by the authors of the respective
input data sets, as this does not affect our results and as this
astrophysical interpretation of the timing signatures may decrease the
stability of the pulsar timing fit (adding more timing parameters
without additional information). In these cases, we have used
whichever timing model parameters were used by the individual PTAs.

For all determined parameters, this analysis results in probability
distributions obtained through marginalisation over the entire
parameter space. For the 
deterministic parameters (these are all the parameters except those
quantifying the white noise, red noise and DM variations), this
marginalisation was done analytically, using the linearised timing
model as already implemented in \textsc{tempo2}. For
the 
stochastic parameters (i.e.\ the white noise, red noise and DM
variations), the marginalisation was done numerically. The results are
discussed in the following section.

\subsection{The Combined Data Set: Results}\label{ssec:Data}
The combined data set is available in the additional on-line material
and on the internet at \url{http://www.ipta4gw.org}. It is provided in
three different forms:
\begin{itemize}
\item Combination ``A'': a raw form that has jumps, but no EFACs,
  EQUADs, DM or red-noise models included;
\item Combination ``B'': a default ``\textsc{tempo2}'' form which
  includes jumps, EFACs, \textsc{tempo2}-format EQUADs (i.e.~following
  Equation~\ref{eq:EQUADs}), a DM model implemented through DM-offset
  flags (``-dmo'') added to the ToA lines, a red-noise model in the
  form of a spectral model compatible with the Cholesky
  \textsc{tempo2} code introduced by \citet{chc+11}, but no ECORRs;
\item Combination ``C'': a \textsc{temponest} combination with JUMPS,
  EFAC, ECORRs, EQUADs (following Equation~\ref{eq:TNEQUAD}) and DM
  and red noise models compatible with the \textsc{temponest} code.
\end{itemize}

\begin{figure*}
  \includegraphics[width=23cm,angle=-90.0]{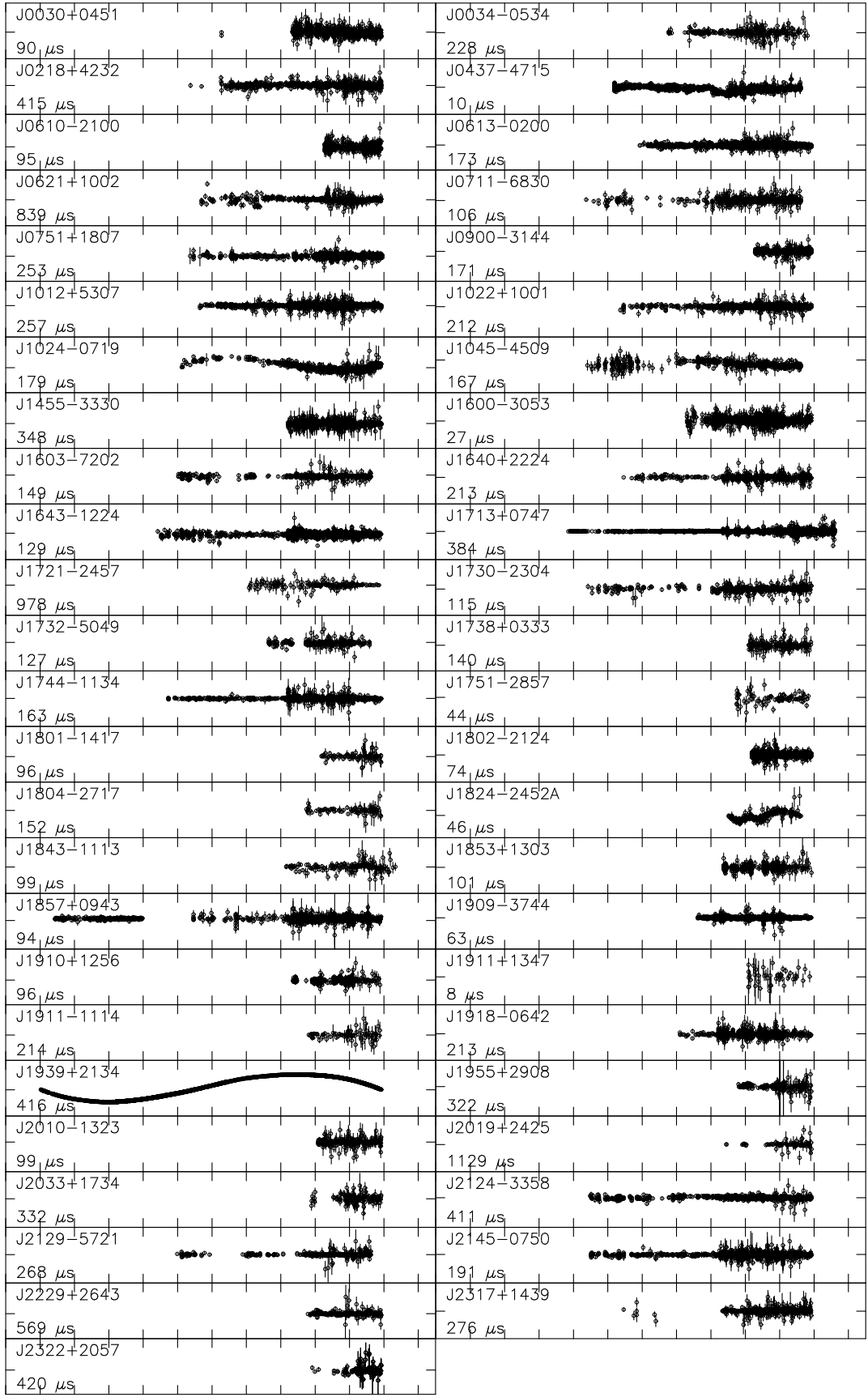}
  \caption{Plot of all timing residuals. Shown are the residuals
    (i.e.\ observed ToA minus model-predicted ToA) after subtraction
    of the DM model, but with inclusion of any modelled red noise. The
    X range is as in Figure~\ref{fig:Coverage}, covering the MJD range
    45000 to 57500 (31 January 1982 to 22 April 2016) with tick marks
    at 1000-day intervals; the Y range is
    different for each pulsar: the numbers in the plot indicate the
    full plotted Y range for each pulsar.}
  \label{fig:Res}
\end{figure*}

The post-fit timing residuals with the maximum likelihood DM-variation
signal subtracted are shown in Figure~\ref{fig:Res}. Red noise that
was inconsistent with DM variations was assumed to be intrinsic in
nature and was not subtracted.
Some fundamental characteristics describing these post-fit data are
summarised in Table~\ref{tab:Data}. A brief summary of the results
of our analysis along with some comments on the limitations and
specificities of this data set and analysis are given below.

\begin{table*}
  \begin{center}
    \caption{Summary table of the combined data set. The columns give respectively the pulsar name, the time span in years, the MJD range, the weighted RMS of the timing residuals (after
      subtraction of the timing model and DM variations), the number of ToAs, the average time between days on which the pulsar is observed and the number of telescopes from which data were
      included in the current data set. The final two columns show whether DM variations or timing noise (i.e.\ long-period noise intrinsic to the pulsar) were detected (`y') or not (`n') in
      the data set. In cases where long-period noise was detected but no distinction could be made between DM or intrinsic noise, the label ``Undetermined'' is used across both the ``DM
      Variations'' and ``Timing Noise'' columns. Five pulsars had linear or quadratic trends in DM that were formally significant at the $1-\sigma$ level, but which were strongly correlated
      to the pulsar's spin period or period derivative. Those five pulsars are identified with ``Undetermined'' only in the ``DM Variations'' column. For six pulsars system-dependent
      long-period noise was detected. These are marked as `s' in the ``Timing noise'' column, as this system-dependent noise has the characteristics of timing noise, but is likely
      instrumental.}
    \label{tab:Data}
    \begin{tabular}{lrcrrrccc}
      \hline
      \multicolumn{1}{c}{Pulsar} &\multicolumn{1}{c}{Time}   &MJD  &\multicolumn{1}{c}{Residual}&\multicolumn{1}{c}{Number}&\multicolumn{1}{c}{Average}&Number    &DM        &Timing\\
      \multicolumn{1}{c}{Name}   &\multicolumn{1}{c}{Span}   &Range&\multicolumn{1}{c}{RMS}     &\multicolumn{1}{c}{of}    &\multicolumn{1}{c}{Cadence}&of        &Variations&Noise \\
      \multicolumn{1}{c}{(J2000)}&\multicolumn{1}{c}{(years)}&     &\multicolumn{1}{c}{($\mu$s)}&\multicolumn{1}{c}{ToAs}  &\multicolumn{1}{c}{(days)} &Telescopes&          &      \\
      \hline                                                              
      J0030+0451   &   12.7  & 51275--55924 &        1.9 &   1250 &     6.6 &     3      & \multicolumn{2}{c}{Undetermined}\\
      J0034$-$0534 &   11.1  & 51770--55808 &        4.4 &    267 &    24.0 &     2      & y          & n      \\
      J0218+4232   &   15.2  & 50370--55924 &        6.7 &   1005 &     7.6 &     4      & y          & n      \\
      J0437$-$4715 &   14.9  & 50190--55619 &        0.3 &   5052 &     5.1 &     1      & y          & s      \\       
      J0610$-$2100 &    4.5  & 54270--55925 &        5.2 &    347 &    10.9 &     2      & n          & n      \\
      \\
      J0613$-$0200 &   13.7  & 50931--55926 &        1.2 &   2940 &     4.3 &     6      & y          & y      \\
      J0621+1002   &   14.3  & 50693--55921 &       11.5 &    637 &    10.6 &     4      & y          & y      \\
      J0711$-$6830 &   17.1  & 49373--55619 &        2.0 &    549 &    18.2 &     1      & y          & n      \\
      J0751+1807   &   15.3  & 50363--55948 &        3.5 &   1129 &    10.4 &     4      & \multicolumn{2}{c}{Undetermined}\\
      J0900$-$3144 &    4.5  & 54284--55922 &        3.4 &    575 &     3.1 &     2      & \multicolumn{2}{c}{Undetermined}\\
      \\
      J1012+5307   &   14.4  & 50647--55924 &        1.7 &   2910 &     6.3 &     5      & y          & y      \\
      J1022+1001   &   15.2  & 50361--55923 &        2.2 &   1375 &     6.5 &     5      & y          & s      \\
      J1024$-$0719 &   15.9  & 50117--55922 &        5.9 &    918 &     8.4 &     5      & y          & y      \\
      J1045$-$4509 &   17.0  & 49405--55619 &        3.3 &    635 &    16.9 &     1      & y          & n      \\
      J1455$-$3330 &    7.4  & 53217--55926 &        4.0 &   1495 &     5.9 &     3      & y          & s      \\
      \\
      J1600$-$3053 &    9.9  & 52301--55919 &        0.8 &   1697 &     5.1 &     4      & y          & s      \\
      J1603$-$7202 &   15.3  & 50026--55618 &        2.3 &    483 &    19.3 &     1      & y          & n      \\
      J1640+2224   &   15.0  & 50459--55924 &        2.0 &   1139 &    12.9 &     5      & y          & n      \\
      J1643$-$1224 &   17.8  & 49421--55919 &        2.7 &   2395 &     6.9 &     6      & y          & s      \\
      J1713+0747   &   21.2  & 48850--56598 &        0.3 &  19972 &     5.1 &     7      & y          & y      \\
      \\
      J1721$-$2457 &   10.3  & 52076--55853 &       25.5 &    152 &    24.9 &     2      & n          & n      \\
      J1730$-$2304 &   17.8  & 49421--55920 &        2.1 &    563 &    15.9 &     4      & y          & s      \\
      J1732$-$5049 &    8.0  & 52647--55582 &        2.5 &    242 &    18.8 &     1      & y          & n      \\
      J1738+0333   &    4.9  & 54103--55905 &        2.6 &    206 &    27.7 &     1      & n          & n      \\
      J1744$-$1134 &   17.0  & 49729--55925 &        1.1 &   2589 &     8.4 &     6      & \multicolumn{2}{c}{Undetermined}\\
      \\
      J1751$-$2857 &    5.7  & 53746--55836 &        2.4 &     78 &    26.8 &     1      & n          & n      \\
      J1801$-$1417 &    4.8  & 54184--55920 &        4.6 &     86 &    20.2 &     2      & \multicolumn{2}{c}{Undetermined}\\
      J1802$-$2124 &    4.7  & 54188--55916 &        4.3 &    433 &    24.8 &     2      & \multicolumn{2}{c}{Undetermined}\\
      J1804$-$2717 &    5.9  & 53747--55914 &        4.5 &     76 &    28.9 &     2      & Undetermined & n      \\
      J1824$-$2452A&    5.8  & 53518--55619 &        2.4 &    298 &    13.6 &     1      & y          & y      \\
      \\
      J1843$-$1113 &    8.7  & 53156--56331 &        1.7 &    186 &    17.5 &     3      & \multicolumn{2}{c}{Undetermined}\\
      J1853+1303   &    7.0  & 53370--55922 &        1.1 &    566 &    24.5 &     3      & n          & n      \\
      J1857+0943   &   26.0  & 46437--55916 &        1.3 &   1641 &    13.4 &     6      & y          & n      \\
      J1909$-$3744 &   10.8  & 53041--56980 &        0.2 &   2623 &     4.4 &     3      & y          & n      \\
      J1910+1256   &    6.9  & 53370--55886 &        3.0 &    597 &    25.2 &     3      & \multicolumn{2}{c}{Undetermined}\\
      \\
      J1911+1347   &    4.9  & 54092--55868 &        0.6 &     45 &    40.4 &     1      & Undetermined & n      \\
      J1911$-$1114 &    5.7  & 53815--55880 &        5.2 &     81 &    25.5 &     2      & n          & n      \\
      J1918$-$0642 &   10.5  & 52095--55914 &        1.5 &   1522 &    13.4 &     4      & y          & n      \\
      J1939+2134   &   27.1  & 46024--55924 &       70.0 &   3905 &     4.6 &     6      & y          & y      \\
      J1955+2908   &    5.8  & 53798--55918 &        5.0 &    319 &    16.6 &     3      & n          & n      \\
      \\
      J2010$-$1323 &    5.0  & 54086--55917 &        1.9 &    296 &     6.3 &     2      & y          & n      \\
      J2019+2425   &    6.8  & 53446--55920 &        8.8 &     80 &    31.7 &     2      & Undetermined & n      \\
      J2033+1734   &    5.5  & 53894--55917 &       13.3 &    130 &    15.6 &     2      & Undetermined & n      \\
      J2124$-$3358 &   17.6  & 49489--55924 &        3.0 &   1115 &     7.7 &     3      & y          & n      \\
      J2129$-$5721 &   15.4  & 49987--55618 &        1.2 &    447 &    19.2 &     1      & y          & n      \\
      \\
      J2145$-$0750 &   17.5  & 49517--55922 &        1.2 &   2347 &     7.0 &     6      & y          & y      \\
      J2229+2643   &    5.8  & 53790--55920 &        3.8 &    234 &     9.6 &     3      & y          & n      \\
      J2317+1439   &   14.9  & 50458--55917 &        1.6 &    867 &    13.5 &     5      & y          & n      \\
      J2322+2057   &    5.5  & 53916--55920 &        6.9 &    199 &    15.0 &     2      & Undetermined & n      \\
      \hline
    \end{tabular}
  \end{center}
\end{table*}

\paragraph*{ToA Selection}
In combining the IPTA data set, an attempt was made to limit the
analysis to a simple combination of the data, without further
selection. However, in a few cases ToAs that were included in existing
data sets have been removed or flagged for future
reference. Specifically, this includes the following three types of
ToAs:
\begin{description}
\item[Simultaneous:] ToAs that were observed at the same observatory
  with different instruments that operated at identical or (partially)
  overlapping frequency bands, have not been removed, but have been
  identified with ``-simul'' 
  flags on their ToA lines. This is particularly relevant for 64 ToAs
  from the PSR~J1713+0747 data set from \citet{zsd+15ltd}, where during the
  years 1998--2004 the ABPP and Mark 4 recorders were used
  simultaneously at Arecibo \citep[see][for more details]{sns+05}.
\item[Solar wind:] When the line of sight to a pulsar comes close to
  the Sun, the increased electron density of the solar wind causes
  additional dispersive delays. Therefore, ToAs that are taken along
  lines of sight that are within $5^{\circ}$ of the Sun have been
  commented out\footnote{These ToAs are undesirable for most
    experiments, but might be used to investigate solar-wind
    effects. Hence, they were not deleted from the data set, but
    inserted as comments in the data files, thereby excluding them
    from any standard analysis whilst keeping them available for
    potential solar-wind investigations.}. The $5^{\circ}$ threshold
  is somewhat arbitrary but is a conservative value based on the model
  predictions presented by \citet{ojs07}.
\item[Small groups:] Systems with fewer than five ToAs have been
  removed from the analysis (see Section~\ref{sssec:jumps}) as they
  increase the complexity by adding systemic offsets, but do not add
  sufficient information to reliably allow determination of their
  uncertainties (EFAC and EQUAD values). Such systems with few ToAs
  occur particularly in the PPTA data sets, which were originally
  analysed using a larger set of simultaneous ToAs that are no longer
  available. Since the IPTA data set improves the pulsar timing
  models, a renewed evaluation of systemic offsets and ToA
  uncertainties is in order, but cannot be performed on such limited
  systems without the inclusion of the simultaneous data (which were
  unavailable for the present work).
\end{description}

\paragraph*{Systemic Offsets.}
In principle, the large number of pulsars and long overlapping time
span of the data analysed should make it possible to identify
instrumental offsets more precisely by averaging the offsets measured
in different pulsars, as long as the differences in instrumental
delays are within a pulse period. There are both practical and
technical reasons why this does not work in the present data set.

Practically this can be done only if the reference phase used for
timing is identical for all pulsars. This can be accomplished by
phase-aligning the template profiles for the different systems through
cross-correlation. While this has been done to some degree for each
PTA separately, the phase-offsets between PTAs were not measured based
on the template profiles -- and in either case such information was
unavailable for the historic data (sub)sets from \citet{ktr94} and
\citet{vbc+09}. Technically this situation is complicated by the wide
variety of recording systems. Various (mostly older) systems
experience different time delays depending on the pulse period and DM
of the pulsar being observed. In particular, differences between older
systems where dedispersion may have been performed in hardware and
newer systems where this is done in software would produce variable
time offsets for different pulsars. This makes the measurement of
systemic offsets nearly intractable.

\paragraph*{Measurement Uncertainties.}
As introduced in Section~\ref{sssec:Unc}, underestimated uncertainties
on pulse ToAs are accounted for using uncertainty-multiplication
factors (EFACs, $F$), uncertainties added in quadrature (EQUADs, $Q$)
and additional correlated error factors (ECORRS). Physically the
primary source of EQUADs and ECORRs is expected to be pulse phase
jitter noise \citep{sod+14ltd}, while EFACs are most likely caused by
imperfections in the algorithm chosen for the uncertainty
determination (see Appendix~\ref{sec:Format}). As with systemic
offsets, the size and variety of the combined IPTA data should allow a
more detailed investigation of these factors. However, as with the
systemic offsets, such an exercise is complicated by the many
parameters that affect these values, as described below.

We find that for most pulsars the $F$ values derived for different
observing systems follow a Gaussian distribution centred near unity,
with a spread of order 0.3. The majority (57\%) of systems have $Q$
values below \unit[10]{ns}, indicating that little or no evidence
exists for additional white noise. For the significant $Q$
measurements, typical values were on the order of microseconds or
less, with maximum $Q$ values between 20 and $\unit[40]{\mu s}$ found
for a few fainter pulsars at observing bands with less
sensitivity. For 16 pulsars two ECORR values were used (one per
observing band) but for PSR~J1713+0747 14 ECORR values were needed,
given the large number of highly sensitive systems present in the
\citet{zsd+15ltd} data set. The ECORR values were detected in the vast
majority of these cases, with maxima around $\unit[3]{\mu s}$ and a
median of \unit[270]{ns}.

A few pulsars have wider distributions for their $F$ values, for two
possible reasons. Firstly, pulsars like PSRs J1713+0747 and J1939+2134
have extended data sets with early data from old observing systems
that are not present in the data sets from the other pulsars. Since
the technical specifications of observing systems have dramatically
improved over the past few decades, it should not be surprising that
systematic effects linked to limited resolution and sensitivity led to
lower-quality data in the past, thereby causing less reliable ToA
uncertainties. Therefore, data sets containing both recent and 20-year
old data are likely to have a wider spread for $F$. A second
contributing factor is the possible correlation between $F$ and
potentially unquantified white noise, $Q$. Specifically for observing
systems with only few ToAs and for weakly scintillating sources (i.e.\
if all ToAs have comparable measurement uncertainty), it is
mathematically impossible to disentangle $F$ from $Q$. In
these cases anomalously low values for $F$ (of order 0.1) are possible
in combination with comparably large values for $Q$ (of order
$\unit[10]{\mu s}$ or more).

Pulsars that show significant values for $Q$ mostly do so for only a
single or few observing systems (and typically not the most sensitive
systems), indicating that these significant values for $Q$ are
fundamentally artefacts of correlations in the analysis (e.g.\
correlations between $F$ and $Q$ as described above). A few of the
brightest pulsars, including PSR~J1909$-$3744, show significant
values for $Q$ for many observing systems. For PSR~J1909$-$3744 this
result stands in sharp contrast to the more advanced research of
\citet{sod+14ltd}, who found the pulse phase jitter noise in this
pulsar to be limited to \unit[10]{ns} or less (in hour-long
observations). This again indicates our poor understanding of the
systematics that cause ToA uncertainties to be underestimated and
requires further investigations, which go beyond the capabilities of
our data. 

In summary, a large majority of the pulsars observed did not require
significant EFAC, EQUAD or ECORR values. In the pulsars with the
longest timing baselines, a clear improvement has been observed with
lower $F$ and $Q$ values for more recent observing systems. Some
pulsars, however, require inexplicably high values (for $Q$ in
particular), well in excess of independently measured bounds on pulse
phase jitter. These pulsars warrant more detailed investigation as an
unknown noise source appears to be contributing to their timing.

\paragraph*{DM Variability and Timing Instabilities.}
As described earlier, DM variations typically have a long-term
character, similar to intrinsic instabilities in pulsar timing (known
as ``timing noise''). Given the poor multi-frequency sampling on many
of our sources (see Figure~\ref{fig:Coverage}), it is in many cases
impossible to distinguish these two types of variations; even when
multiple frequencies are present, the possible mismatch in the timing
precision at these frequencies can make measurements of DM variability
in these data imprecise and highly covariant with timing noise
estimates. Consequently, the analysis of these two sources of
long-period noise is closely intertwined and complex and will not be
discussed in detail here, but referred to a companion paper
\citep{lea+15}. However below, we briefly summarise and comment on the main
findings of this research.

As listed in Table~\ref{tab:Data}, 17 of the 49 pulsars in the IPTA
data set do not show evidence of excess low-frequency noise (i.e.\
show neither DM variations nor timing instabilities). This is
to be expected if the data set in question is relatively short, as is
the case for all but one of these pulsars, which have data lengths of
less than 15 years. The remaining source, PSR~J2124$-$3358, has a data
set of 17.6 years with a residual RMS of $\unit[3.8]{\mu s}$ and is
therefore highly sensitive to low-frequency noise, so its absence
indicates that this pulsar is inherently a very stable rotator.

Eight pulsars in our sample have evidence of both DM variability and
timing instabilities. Not surprisingly, this group contains the
pulsars with the longest time spans: PSRs~J1939+2134 (27.1\,years) and
J1713+0747 (21.2\,years). Another eight pulsars 
show evidence for low-frequency noise, but have no sufficiently
sensitive multi-frequency data; therefore, no distinction can be made
between intrinsic pulsar timing noise and DM variations. (Even though
the data sets on PSRs~J0030+0451, J0751+1807, J0900$-$3144 and
J1744$-$1134 contain ToAs at multiple frequencies, the measurement
precision and cadence turn out to be insufficient in these cases.)
Fifteen pulsars show
significant DM variations but no frequency-independent timing noise.

A particularly powerful aspect of the combined IPTA data set is that
the timing instabilities of different telescopes and observing systems
can be checked against each other, thereby clarifying whether the
observed timing noise is caused by hardware issues, or whether it is
truly intrinsic to the pulsar. Such a test was already performed on a
smaller scale by \citet{vlj+11ltd}, who found that for the few pulsars
and telescopes they compared, low-frequency noise models were
consistent. A similar analysis based on the IPTA data set presented
here, also mostly finds consistent models, except for six pulsars that
show system-dependent low-frequency noise in addition to DM
variations. In some cases this system-dependent noise is not simply
dependent on the observing hardware, but on the frequency band in
which the observations were taken, suggesting a possible interstellar
origin other than dispersion. For the full analysis, we refer to
\citet{lea+15}.

In summary, of the 26 pulsars with more than a decade of data, a vast
majority (25 pulsars) show (possible) DM variations and just over a
third (10 pulsars) show (possible) system-independent timing
noise. Only one of these 26 pulsars (PSR~J1721$-$2457) shows no
evidence for DM variations or red noise at all, but the timing of this
pulsar has been exclusively undertaken at a single frequency, so that
any long-term DM variations would most likely be absorbed in fits for
pulse period and period derivative.

\section{Limits on the GWB Amplitude}\label{sec:GWlimits}
As discussed in detail above, the present data set is a useful testbed
for general IPTA-like data combination efforts. While this combination
was ongoing, however, individual PTAs have been updating their data
sets more rapidly and have meanwhile improved their sensitivity,
particularly to the GWB, which is telescope-independent (unlike
instrumental effects) and highly sensitive to the length of the data
set. A full, detailed analysis of the present IPTA data set with
regards to obtaining a limit on the strength of the GWB is therefore
not worthwhile at present. Instead, we present a simplified analysis
on both the combined and the constituent data sets, to illustrate the
potential impact an IPTA combination can provide, as this is
analytically intractable. Based on the work presented elsewhere in
this paper, combination of IPTA data will in the future become more
straightforward, allowing a shorter timeline and therefore more
significant GWB limits to be derived using IPTA data.

To derive an indicative limit on the GWB amplitude, we used the
\textsc{Piccard} software
package\footnote{\url{https://github.com/vhaasteren/piccard}}. This
code has been cross-checked with \textsc{TempoNEST} \citep{lah+14} and
uses the same likelihood functions. The noise model used is as
described elsewhere in this paper, i.e.\ including EFAC, EQUAD and
ECORR values to properly quantify the white noise, but with a more
general red-noise model that allowed the power spectral density
amplitudes to vary per frequency bin and did not implicitly assume a
power-law shape. This deviation from the more extensive noise models
presented by \citet{lea+15} was made in order to avoid a full
re-analysis of the \citet{lea+15} work including GW limits. For the
scope of this paper, an indicative bound that could be compared
between the different data sets, was sought rather than an exhaustive
GW-limit analysis. A combined GW-limit analysis with full noise
modelling is beyond the scope of this paper and is deferred to a
future and more competitive IPTA data release. The sampling was done
with the Gibbs sampler introduced by \citet{vv14}. To reduce computing
time and avoid complications caused by some less precisely timed
pulsars, only the four pulsars with the highest sensitivity (as
quantified through the length of their data set and the precision and
number of their ToAs) were included in this analysis. These are
PSRs~J0437$-$4715, J1713+0747, J1744$-$1134 and
J1909$-$3744. Furthermore, correlations of the GWB signal between
pulsars have been neglected and no advanced noise-modelling \citep[as
in][]{lea+15} was performed, i.e.\ no system-specific red noise was
included. Even though the individual PTAs have previously published
limits based on the constituent data sets, we perform our analysis
again on the individual PTA data sets, because the published limits
were derived using slightly different noise models than ours, so the
limits cannot be directly and self-consistently compared to our
IPTA-based limit. Efforts to find a more appropriate noise model are
an ongoing effort within the IPTA \citep[see, e.g.][]{lea+15}.

\begin{table}
  \begin{center}
    \caption{Limits on the GWB from the combined IPTA data set and its
      PTA-specific subsets. Given are the data set for which the limit
      was determined, the limit on the GWB amplitude resulting from
      our basic analysis; and the limit published based on the same
      sub-sets (with any differences described in
      Section~\ref{ssec:sets}), along with their bibliographic
      reference. Note our limits are generally slightly worse because
      of the basic nature of our analysis, with the exception of the
      NANOGrav data set, as this one was significantly extended by
      including the PSR~J1713+0747 data from \protect\citet{zsd+15ltd}.}
    \label{tab:PTAlimits}
    \begin{tabular}{lccl}
      \hline
      PTA    & GWB   & Published & Reference \\
      Subset & Limit & Limit     &           \\
             & ($\times 10^{-15}$) & ($\times 10^{-15}$) &  \\
      \hline
      NANOGrav & $4.5$ & $1.5$ &\protect\citet{abb+15bltd}\\
      EPTA   & $3.3$ & $3.0$&\protect\citet{ltm+15ltd}\\
      PPTA   & $2.8$ & $1.0$&\protect\citet{srl+15ltd}\\
      IPTA   & $1.7$ & \\
      \hline
    \end{tabular}
  \end{center}
\end{table}

Our results are summarised in Table~\ref{tab:PTAlimits}. For the
individual PTAs our limits are consistent with or slightly worse than
those published by the individual PTAs, which is expected given the
fact that our analysis is more basic and less detailed than those
published elsewhere. In the case of NANOGrav, the limit we calculate
is better than the one published by \citet{dfg+13ltd} because of the
inclusion of the long data set of PSR~J1713+0747 by
\citet{zsd+15ltd}. For the PPTA data set, our limit is less constraining than
their most recent limit, but we use all data available on the pulsars
used, including lower-frequency ToAs that are affected by more severe
(and not well-modelled) low-frequency noise; furthermore, the recent
limit by \citet{srl+15ltd} extended the timing baseline with
high-quality data, further improving the overall timing precision. As
expected, the IPTA limit beats the lowest limit by an individual PTA,
by as much as a factor of 1.6. While this is a basic analysis that lacks the
rigour of a full investigation, it can be expected that future IPTA
work would also improve limits on the GWB amplitude by a similar
factor. More importantly, though, since the IPTA contains a larger
number of pulsars than any of the constituent PTAs (a logical
consequence of the complete sky coverage) and given the strong scaling
of PTA sensitivity with the number of pulsars
(equations~\ref{eq:PTASens} and \ref{eq:PTASens2}), it is clear that
the IPTA has a unique advantage when it comes to carrying out the
first actual detection of GWs with pulsar-timing data.

\section{The Future of the IPTA}\label{sec:Future}
The present IPTA data combination is a relatively ad-hoc combination
of (largely archival) timing data from a variety of observing
projects. It uses a large number of (mostly old) instruments and
focuses on a relatively poorly defined set of MSPs that has been
discovered in the course of the past few decades. In
Section~\ref{sec:intro} we have described how the pulsar timing
sensitivity depends on the telescope's sensitivity and how the
sensitivity of PTAs depends on the pulsars in their sample. All of
these aspects are about to go through a revolution and in a few years
time the updated IPTA data set will greatly differ from the present
one and will likely be sensitive to GW backgrounds with amplitudes far
below $1\times 10^{-15}$. In the following we briefly describe the
main progress that can be expected for the coming decade, above and
beyond the addition of more recent data. This includes some technical
advances to the pulsar timing methodology
(Section~\ref{ssec:Improvement}), which are being developed now and
should bear fruit soon, the potential expansion of the pulsar sample
(Section~\ref{ssec:survey}) and the impact significantly more
sensitive telescopes could make (Section~\ref{ssec:SKA}) over the
course of coming decades.

\subsection{Beating Systematics}\label{ssec:Improvement}
Several aspects of pulsar timing require further research and
development in order to improve data quality and long-term data
usefulness. Some straightforward practical measures have been laid out
in Appendix~\ref{sec:Format}, but several more fundamental questions
remain to be solved in the coming few years, in preparation for the
leap in sensitivity the SKA will bring. Specifically, we identify four
main aspects of ongoing study of key relevance to PTA research.

\paragraph*{DM-Correction Methods.} As described in
Section~\ref{sssec:DMvar}, correction methods for temporal DM
variations have essentially always been ad-hoc, based on whatever
(limited) data were available and without a thorough understanding of
the processes that underly these variations. The analysis presented
in this paper is no exception to this rule. 

Early work by \citet{fc90} found that in order to correct DM
variations in pulsar timing data, regular multi-frequency observations
with less-sensitive telescopes would be more efficient in mitigating
the variable IISM effects than less regular but more sensitive
observations. However, with increased telescope sensitivity and
bandwidths since then, new questions have arisen. Most importantly,
because of the different refraction angles at the different frequencies,
the IISM sampled by observations at different wavelengths might differ
slightly \citep{css15}. It is yet unknown whether the magnitude of
this effect is relevant for the observations included in the IPTA, but
for the new generation of low-frequency telescopes this question is
key to evaluating their usefulness for PTA-type work. Initial work on
a limited set of slow pulsars by \citet{hsh+12ltd} found that no such
``frequency-dependent DM'' could be identified, but this test needs to
be reproduced for the lines of sight to the MSPs in the IPTA sample.

A second unknown on this topic is whether a single, ultra-wide
observing bandwidth (including potential issues with RFI and system
temperature) would be preferred to a set of simultaneous observations
at various, widely separated observing frequencies; or whether a fully
independent observing campaign at ultra-low frequencies with high
cadence (e.g.\ as aperture arrays could provide through multi-beaming)
would be more sensitive and therefore more beneficial. This likely
depends on the RFI environment and on the spectral index of the pulsar
as well as its pulse-shape evolution with frequency and may therefore
require a sizeable study to achieve clarity.

Finally, DM correction methods either interpolate or smooth the
measured DM values \citep{yhc+07,kcs+13ltd}; or assume a model that is
fitted to them \citep{cbl+95,lah+14}.  However, these approaches
inherently assume the DM variations are time-stationary with the
exception of a limited number of top-hat-like "events", but this is
demonstrably \emph{not} the case \citep[see, e.g.][]{mlc03,cks+15}. As
our sensitivity improves with lower-frequency telescopes, wider
bandwidths and longer data sets, the characterisation of the IISM's
numerous effects should improve. This would increase our understanding
of the IISM and should allow more accurate DM correction methods.

Also, as bandwidths increase and future generations of telescopes
become more sensitive, direct in-band DM determination as part of the
timing model, without interpolation or model assumptions \citep[as
already proposed by][]{dfg+13ltd}, may become more widely applicable.

\paragraph*{Higher-Order IISM Effects.} In addition to changes in
dispersion, density variations in the IISM can cause temporal
variations in scattering and thereby change the pulse shape as a
function of time \citep{hs08}. While this effect is mostly
undetectable at observing frequencies of a GHz or higher with present
telescopes, its amplitude is mostly unknown and this may affect more
sensitive observations with upcoming telescopes like FAST or the
SKA. Detailed experiments with mitigation methods such as cyclic
spectroscopy \citep{dem11} \citep[as performed at lower frequencies
by][]{wdv13,akhs14} are therefore required on a larger sample of MSPs,
particularly because any newly discovered pulsars are likely to be
fainter and therefore more distant than the currently known
population, making scattering effects more likely to have a
significant impact.

\paragraph*{Absolute System Offsets.} In principle, systemic offsets
can be determined with high precision using interferometric
fringe-fitting on baseband data, at least for the most recent
generation of digital recorders. Such efforts are ongoing
\citep{bjk+15,dlt+14ltd}. An alternative method that has recently been
developed, is based on correlating the identical noise in the data
from two data recorders on the same telescope, as introduced by
\citet{abb+15ltd} in their Appendix~A. This last technique could also
be used on multi-telescope data (as a form of intensity
interferometry), but is likely to give less precise results than the
actual interferometric efforts mentioned before.

\paragraph*{Improved Calibration.} In cases where the observations are
correctly polarisation-calibrated, the timing precision of some
pulsars may be significantly enhanced by using the polarimetric
information in the pulse profile \citep{van06}. While this method is
promising and has been used with good results already \citep{van13},
its application is still non-standard and somewhat marginal in the
current IPTA data set. This is likely because of the difficulty in
reliably modelling any impurities in the receiver system; and
time-variations thereof \citep[see, e.g.][]{van13}. Proper
characterisation and monitoring of receiver properties could therefore
provide further enhancements to pulsar timing precision. Especially at
lower observing frequencies and with highly sensitive, future
telescopes, frequency and time-dependent changes in the polarimetric
position angle of the pulsar radiation, as most significantly
introduced by time-variable Faraday rotation in the ionosphere
\citep{ssh+13ltd}, may also need to be corrected for, which is not
typically the case presently. (Note that ionospheric RM variation
measurements may be a side-product of advanced calibration techniques,
as shown in Figure~8 of \citet{ovdb13}.)

\paragraph*{Advanced White-Noise Modelling.} As discussed in
Appendix~\ref{sec:Format}, EFAC, EQUAD and ECORR determination methods
should be more extensive, to take into account certain expected
scaling relations (e.g. $Q \propto T^{-1/2}$). However, all of the
effects listed above also add impurities to the timing residuals,
which are not necessarily reflected in the ToA uncertainties. It is
therefore safer to measure the effect of phenomena like pulse-phase
jitter on the ToA uncertainty directly \citep[as recently done by][for
the PPTA pulsars]{sod+14ltd}, rather than to implement EQUAD
measurement methods that assume jitter as the key contributor. Such a
bottom-up approach also ensures the correct interpretation of ToA
uncertainty underestimation and thereby removes any possible but
unphysical correlations that might exist. A first step in that
direction is the determination of ECORR values, which by design
quantify the EQUAD part that correlates between simultaneous ToAs and
as such already move towards a more physical understanding of these
ad-hoc parameters.

\paragraph*{SWIMS Mitigation}
As described in Section~\ref{ssec:PT}, two noise sources affect pulsar
timing data: radiometer noise and pulse-phase jitter or SWIMS
\citep{ovh+11}. The former of these can only be reduced through
hardware upgrades, the impact of the latter can be reduced through
generalised least-squares template-matching techniques, like those
proposed by \citet{ovh+11}. Such techniques not have been fully
developed yet, but in the coming era of highly sensitive radio
telescopes this may well become a fundamental tool of radio pulsar
timing. For practical applications \citet{ovh+11} and \citet{ovdb13}
did propose a mitigation method that can presently be applied to
pulsar timing work.

\subsection{Pulsar Surveys}\label{ssec:survey}
Pulsar surveys are long-term undertakings as both observing and
processing requirements are extremely large. As a list of the most
prominent on-going pulsar surveys shows (Table~\ref{tab:surveys}),
many of the world's major radio telescopes are currently -- and have
been for multiple years -- involved in surveys for pulsars. This
concerted effort has led to an MSP discovery rate that is
unprecedented (see Figure~\ref{fig:Disc}) and even though none of
these recently discovered MSPs have made it into the first IPTA data
release, the monitoring and evaluation of these sources for IPTA use
is ongoing and is already lengthening the source lists of individual
PTAs \citep[see, e.g.][]{abb+15ltd}. This is particularly important
given the strong scaling of PTA sensitivity with the number of pulsars
(see equations~\ref{eq:PTASens} and \ref{eq:PTASens2}).

For the IPTA, there are three prime reasons to support ongoing pulsar
surveys. Firstly, the larger the number of pulsars in the IPTA, the
more sensitivity the IPTA has to any correlated signal. While this is
technically true (see the equations in Section~\ref{ssec:PTASens}), it
depends strongly on the \emph{timeability} of the pulsars in question,
i.e.\ mostly on their flux density and pulse width (or the integrated
derivative of the pulse profile, to be precise), as shown in
Equation~\ref{eq:radiom}. So while fainter, slower MSPs can still be
useful for the IPTA, they will be useful only if the observing time
dedicated to them is proportionally increased \citep{lbj+12}. This
means that the required observing time may become prohibitively
large. A second advantage, however, is that existing pulsars in the
array may be replaced by new discoveries. This is particularly
relevant since the strength of timing noise differs greatly from
pulsar to pulsar (see Figure~\ref{fig:Res}), so that for long-term
projects the stability of the pulsar will become more important than
its instantaneous timing precision. A third and final benefit of
ongoing pulsar surveys is their use for PTA experiments with the next
generation of radio telescopes (see Section~\ref{ssec:SKA}). As
telescope sensitivity increases, the radiometer noise will decrease
and a far larger set of pulsars will become useful \citep{lvk+11}.

\begin{table}
  \begin{center}
    \caption{List of major ongoing pulsar surveys. Given are the survey
      acronym, telescope used, centre frequency, starting year and
      literature reference.} 
    \label{tab:surveys}
    \begin{tabular}{llrrl}
      \hline
      Survey  & Telescope & Frequency & Start & Reference \\
      Acronym &  Used     & (MHz)     & Year  & \\
      \hline
      PALFA  &  AO       & 1400      & 2004 & \protect\citet{cfl+06ltd} \\ 
      GBNCC  & GBT       &  350      & 2009 & \protect\citet{slr+14ltd} \\
      Fermi PSC & various & various  & 2009 & \protect\citet{rap+12ltd} \\
      AO327  &  AO       &  327      & 2010 & \protect\citet{dsm+13}    \\
      HTRU-S & PKS       & 1352      & 2008 & \protect\citet{kjv+10}    \\
      HTRU-N & EFF       & 1360      & 2010 & \protect\citet{bck+13}    \\
      LOTAAS & LOFAR     &  135      & 2014 & \protect\citet{cvh+14ltd} \\
      \hline
    \end{tabular}
  \end{center}
\end{table}

\subsection{SKA and Pathfinder Telescopes}\label{ssec:SKA}
In the coming decade, the construction and use of the Square Kilometre
Array will commence and, with its order-of-magnitude increase in
sensitivity, it will revolutionise all aspects of the science
discussed in this paper. Specifically, pulsar surveys with the SKA
\citep{kbk+15ltd} will multiply the number of pulsars available for
PTA research; and PTA sensitivity based on both newly discovered and
already known pulsars will not merely enable GW detection, but likely
commence the field of low-frequency GW astronomy \citep{jhm+15}. In
anticipation of these events, a host of ``pathfinder'' telescopes are
currently being constructed, commissioned and used, paving the way
towards the SKA revolution in a wide range of aspects.

\paragraph*{Low-Frequency Pathfinders.} Three low-frequency SKA
pathfinders are currently operational for pulsar research. These are
the European LOw-Frequency ARray \citep[LOFAR,
][]{sha+11ltd,vwg+13ltd}, the Long-Wavelength Array (LWA) in New
Mexico \citep{drt+13} and the Murchison Widefield Array \citep[MWA,
][]{bot+14ltd} in Western Australia. Since the Galactic synchrotron
background emission has a steeper spectral index than the typical
pulsar \citep{blv13}, these low-frequency arrays are not optimal for
highly sensitive timing efforts, but given the strong frequency
dependence of interstellar effects \citep[see Equation~\ref{eq:DM} and
further effects in][]{lk05}, these pathfinders could prove to be
highly useful tools for monitoring and correcting variability in the
IISM \citep{kvh+15ltd}.

\paragraph*{FAST and LEAP.} The Five-hundred-metre Aperture Spherical
radio Telescope (FAST) is an Arecibo-type spherical telescope
currently being constructed in China; and will be the world's largest
and most sensitive single-dish radio telescope upon completion. Its
receiver platform is also moveable so that a substantial part of the
sky can be observed \citep{nlj+11}. Another sensitive project is the
Large European Array for Pulsars \citep[LEAP,][]{bjk+15}, which
coherently combines the data from the five major centimetre-wavelength
radio telescopes in Europe, thereby synthesising an Arecibo-sized
telescope that is able to point in any direction of the Northern
sky. With its unrivalled instantaneous sensitivity, FAST should be
able to make a major contribution to pulsar surveys \citep{yln13},
particularly if equipped with a multi-beam receiver of phased-array
feed, since the limited beamwidth will either necessitate vast amounts of
observing time to complete a survey of any part of the sky; or require
the survey to be undertaken at lower frequencies. More importantly,
the increased sensitivity of these telescopes will allow improved
timing precision which will enhance the sensitivity of PTAs to GWs
(and other signals) to levels beyond the reach of current technology
\citep{zzyz13}.

\paragraph*{MeerKAT.} The MeerKAT telescope \citep{bj12} is the
South-African SKA pathfinder, located in the Karoo desert where the
core of the mid and high-frequency parts of the SKA will be
located. MeerKAT will be more sensitive than the 100-m-class
telescopes of the Northern hemisphere and up to five times more
sensitive than Parkes, making it the most sensitive fully steerable
telescope in the world, placed in the Southern hemisphere, where many
of the most precisely timed MSPs reside (Table~\ref{tab:PSRs}). This
will make it an important addition to PTA efforts in the lead up to
the SKA.

\section{Conclusions}\label{sec:Conc}
In this paper, we present the creation of the first IPTA data release
by combining the data from the three constituent PTAs and illustrate
the importance of this for limits on GW backgrounds by comparing
straightforward results from the subsets and the combined set. This
indicates an IPTA combined limit on the GW background should be close
to twice as sensitive as any of the constituent data sets. Further
analyses of these data, particularly relating to the timing stability
of MSPs \citep{lea+15}, Solar-System ephemeris and clock errors, will
be published separately in due course. Beyond these specific projects,
though, this work can be seen as a primer, identifying pitfalls and
challenges with the formats and practices common in pulsar timing
today. Through this first analysis, we hope the quality and ease of
use of pulsar timing data can be vastly improved upon, so that
subsequent IPTA analyses will be performed in a more rigorous manner,
thereby preparing the field for both the advent of GW astronomy and
the SKA era.

\section*{Acknowledgments}

The National Radio Astronomy Observatory is a facility of the NSF
operated under cooperative agreement by Associated Universities,
Inc. The Arecibo Observatory is operated by SRI International under a
cooperative agreement with the NSF (AST-1100968), and in alliance with
Ana G. M\'endez-Universidad Metropolitana, and the Universities Space
Research Association.
The Parkes telescope is part of the Australia Telescope which is
funded by the Commonwealth Government for operation as a National
Facility managed by CSIRO.
Part of this work is based on observations with the 100-m telescope of
the Max-Planck-Institut f\"ur Radioastronomie (MPIfR) at
Effelsberg. Access to the Lovell Telescope and pulsar research at the
Jodrell Bank Centre for Astrophysics is supported through an
STFC consolidated grant.
The Nan\c{c}ay radio telescope is operated by the Paris Observatory,
associated with the Centre National de la Recherche Scientifique
(CNRS) and acknowledges financial support from the ``Programme
National de Cosmologie et Galaxies (PNCG)'' and ``Gravitation,
R\'ef\'erences, Astronomie, M\'etrologie (GRAM)'' programmes of
CNRS/INSU, France. We gratefully acknowledge the financial support
provided by the R\'egion Centre.
The Westerbork Synthesis Radio Telescope is operated by the
Netherlands Foundation for Research in Astronomy (ASTRON) with support
from the NWO.
Some of the work reported in this paper was supported by the ERC
Advanced Grant ``LEAP'', Grant Agreement Number 227947 (PI M.~Kramer).
This work was partially supported through the National Science
Foundation (NSF) PIRE program award number 0968296 and the NSF Physics
Frontier Center award number 1430284.
Part of this research was carried out at the Jet Propulsion
Laboratory, California Institute of Technology, under a contract with
the National Aeronautics and Space Administration.
Several plots in this paper were prepared based on data gathered from
the ATNF pulsar catalogue, available online at:
\url{http://www.atnf.csiro.au/research/pulsar/psrcat/}.
The authors acknowledge careful reading of and useful comments on the
draft by Bill Coles and an anonymous referee. 
RvH is supported by NASA Einstein Fellowship grant
PF3-140116. Portions of this research were carried out at the Jet
Propulsion Laboratory, California Institute of Technology, under a
contract with the National Aeronautics and Space Administration.
J-BW is supported by NSFC project No.\ 11403086 and the West Light
Foundation CAS XBBS201322.
RNC acknowledges the support of the International Max Planck Research
School Bonn/Cologne and the Bonn-Cologne Graduate School. 
NDRB is supported by a Curtin Research Fellowship.
TD was partially supported through the National Science Foundation
(NSF) PIRE program award number 0968296.
JAE acknowledges support by NASA through Einstein Fellowship grant
PF4-150120. 
JRG's work is supported by the Royal Society.
MEG was partly funded by an NSERC PDF award.
JWTH and SAS acknowledge funding from an NWO Vidi fellowship.
JWTH and CGB acknowledge funding from ERC Starting Grant ``DRAGNET''
(337062; PI Jason Hessels).
PDL and PR are supported by the Australian Research Council Discovery
Project DP140102578.
PL acknowledges the support of IMPRS Bonn/Cologne.
KJL gratefully acknowledges support from the National Basic Research
Program of China, 973 Program, 2015CB857101 and NSFC 11373011.
KL acknowledges the financial support by the European Research Council
for the ERC Synergy Grant BlackHoleCam under contract no. 610058.
CMFM was supported by a Marie-Curie International Outgoing Fellowship
within the European Union Seventh Framework Programme.
SO is supported by the Alexander von Humboldt Foundation.
AS is supported by a University Research Fellowship of the Royal Society.
JS was partly supported through the Wisconsin Space Grant Consortium.
Pulsar research at UBC is supported by an NSERC Discovery Grant and
Discovery Accelerator Supplement and by the Canadian Institute for
Advanced Research.
SRT is supported by an appointment to the NASA Postdoctoral Program at
the Jet Propulsion Laboratory, administered by the Oak Ridge
Associated Universities through a contract with NASA.
MV acknowledges support from the JPL RTD program. 
YW was supported by the National Science Foundation of China (NSFC)
award number 11503007.
LW and XJZ acknowledge funding support from the Australian Research
Council and computing support from the Pawsey Supercomputing Centre at
WA. 
XPY acknowledges support by NNSF of China (U1231120) and FRFCU
(XDJK2015B012). 
\bibliographystyle{mn2e}
\bibliography{journals,psrrefs,modrefs,crossrefs}

\begin{thebibliography}{}

\bibitem[\protect\citeauthoryear{{Abdo}, {Ackermann}, {Ajello}, {Allafort},
  {Baldini}, {Ballet}, {Barbiellini}, {Bastieri}, {Bechtol}, {Bellazzini},
  {Berenji}, {Blandford}, {Bloom} \& {Bonamente}}{{Abdo}
  et~al.}{2010}]{aaa+10bltd}
{Abdo} A.~A.,  {Ackermann} M.,  {Ajello} M.,  {Allafort} A.,  {Baldini} L.,
  {Ballet} J.,  {Barbiellini} G.,  {Bastieri} D.,  {Bechtol} K.,  {Bellazzini}
  R.,  {Berenji} B.,  {Blandford} R.~D.,  {Bloom} E.~D.,    {Bonamente} E.,
  2010, ApJ, 712, 957

\bibitem[\protect\citeauthoryear{{Abdo}, {Ackermann}, {Atwood}, {Axelsson},
  {Baldini}, {Ballet}, {Barbiellini}, {Bastieri}, {Battelino}, {Baughman},
  {Bechtol}, {Bellazzini} \& {Berenji}}{{Abdo} et~al.}{2009}]{aaa+09eltd}
{Abdo} A.~A.,  {Ackermann} M.,  {Atwood} W.~B.,  {Axelsson} M.,  {Baldini} L.,
  {Ballet} J.,  {Barbiellini} G.,  {Bastieri} D.,  {Battelino} M.,  {Baughman}
  B.~M.,  {Bechtol} K.,  {Bellazzini} R.,    {Berenji} B.,  2009, ApJ, 699,
  1171

\bibitem[\protect\citeauthoryear{{Archibald}, {Kondratiev}, {Hessels} \&
  {Stinebring}}{{Archibald} et~al.}{2014}]{akhs14}
{Archibald} A.~M.,  {Kondratiev} V.~I.,  {Hessels} J.~W.~T.,    {Stinebring}
  D.~R.,  2014, ApJ, 790, L22

\bibitem[\protect\citeauthoryear{Armstrong, Rickett \& Spangler}{Armstrong
  et~al.}{1995}]{ars95}
Armstrong J.~W.,  Rickett B.~J.,    Spangler S.~R.,  1995, ApJ, 443, 209

\bibitem[\protect\citeauthoryear{{Arzoumanian}, {Brazier}, {Burke-Spolaor},
  {Chamberlin}, {Chatterjee}, {Christy}, {Cordes}, {Cornish}, {Demorest},
  {Deng}, {Dolch}, {Ellis}, {Ferdman}, {Fonseca}, {Garver-Daniels} \& et
  al.}{{Arzoumanian} et~al.}{2015}]{abb+15bltd}
{Arzoumanian} Z.,  {Brazier} A.,  {Burke-Spolaor} S.,  {Chamberlin} S.,
  {Chatterjee} S.,  {Christy} B.,  {Cordes} J.,  {Cornish} N.,  {Demorest} P.,
  {Deng} X.,  {Dolch} T.,  {Ellis} J.,  {Ferdman} R.,  {Fonseca} E.,
  {Garver-Daniels} N.,    et al. 2015, ApJ submitted

\bibitem[\protect\citeauthoryear{{Arzoumanian}, {Brazier}, {Burke-Spolaor},
  {Chamberlin}, {Chatterjee}, {Cordes}, {Demorest} \& {Deng}}{{Arzoumanian}
  et~al.}{2014}]{abb+14ltd}
{Arzoumanian} Z.,  {Brazier} A.,  {Burke-Spolaor} S.,  {Chamberlin} S.~J.,
  {Chatterjee} S.,  {Cordes} J.~M.,  {Demorest} P.~B.,    {Deng} X.,  2014,
  ApJ, 794, 141

\bibitem[\protect\citeauthoryear{{Babak}, {Petiteau}, {Sesana}, {Brem},
  {Rosado}, {Taylor}, {Lassus}, {Hessels}, {Bassa} \& et al.}{{Babak}
  et~al.}{2015}]{bps+15ltd}
{Babak} S.,  {Petiteau} A.,  {Sesana} A.,  {Brem} P.,  {Rosado} P.~A.,
  {Taylor} S.~R.,  {Lassus} A.,  {Hessels} J.~W.~T.,  {Bassa} C.~G.,    et al.
  2015, MNRAS, 455, 1665

\bibitem[\protect\citeauthoryear{Backer, Kulkarni, Heiles, Davis \&
  Goss}{Backer et~al.}{1982}]{bkh+82}
Backer D.~C.,  Kulkarni S.~R.,  Heiles C.,  Davis M.~M.,    Goss W.~M.,  1982,
  Nature, 300, 615

\bibitem[\protect\citeauthoryear{Bailes, Harrison, Lorimer, Johnston, Lyne,
  Manchester, D'Amico, Nicastro, Tauris \& Robinson}{Bailes
  et~al.}{1994}]{bhl+94}
Bailes M.,  Harrison P.~A.,  Lorimer D.~R.,  Johnston S.,  Lyne A.~G.,
  Manchester R.~N.,  D'Amico N.,  Nicastro L.,  Tauris T.~M.,    Robinson C.,
  1994, ApJ, 425, L41

\bibitem[\protect\citeauthoryear{Bailes, Johnston, Bell, Lorimer, Stappers,
  Manchester, Lyne, D'Amico \& Gaensler}{Bailes et~al.}{1997}]{bjb+97}
Bailes M.,  Johnston S.,  Bell J.~F.,  Lorimer D.~R.,  Stappers B.~W.,
  Manchester R.~N.,  Lyne A.~G.,  D'Amico N.,    Gaensler B.~M.,  1997, ApJ,
  481, 386

\bibitem[\protect\citeauthoryear{{Barr}, {Champion}, {Kramer}, {Eatough},
  {Freire}, {Karuppusamy}, {Lee}, {Verbiest}, {Bassa}, {Lyne}, {Stappers},
  {Lorimer} \& {Klein}}{{Barr} et~al.}{2013}]{bck+13}
{Barr} E.~D.,  {Champion} D.~J.,  {Kramer} M.,  {Eatough} R.~P.,  {Freire}
  P.~C.~C.,  {Karuppusamy} R.,  {Lee} K.~J.,  {Verbiest} J.~P.~W.,  {Bassa}
  C.~G.,  {Lyne} A.~G.,  {Stappers} B.,  {Lorimer} D.~R.,    {Klein} B.,  2013,
  MNRAS, 435, 2234

\bibitem[\protect\citeauthoryear{{Bassa}, {Janssen}, Karuppusamy, Kramer, Lee,
  Liu, McKee, Perrodin, Purver, Sanidas, Smits \& Stappers}{{Bassa}
  et~al.}{2015}]{bjk+15}
{Bassa} C.~G.,  {Janssen} G.~H.,  Karuppusamy R.,  Kramer M.,  Lee K.~J.,  Liu
  K.,  McKee J.,  Perrodin D.,  Purver M.,  Sanidas S.,  Smits R.,    Stappers
  B.~W.,  2015, MNRAS

\bibitem[\protect\citeauthoryear{{Bates}, {Lorimer} \& {Verbiest}}{{Bates}
  et~al.}{2013}]{blv13}
{Bates} S.~D.,  {Lorimer} D.~R.,    {Verbiest} J.~P.~W.,  2013, MNRAS, 431,
  1352

\bibitem[\protect\citeauthoryear{{Bhat}, {Ord}, {Tremblay} \& et al.}{{Bhat}
  et~al.}{2014}]{bot+14ltd}
{Bhat} N.~D.~R.,  {Ord} S.~M.,  {Tremblay} S.~E.,    et al. 2014, ApJ, 791, L32

\bibitem[\protect\citeauthoryear{{Booth} \& {Jonas}}{{Booth} \&
  {Jonas}}{2012}]{bj12}
{Booth} R.~S.,  {Jonas} J.~L.,  2012, African Skies, 16, 101

\bibitem[\protect\citeauthoryear{Boriakoff, Buccheri \& Fauci}{Boriakoff
  et~al.}{1983}]{bbf83}
Boriakoff V.,  Buccheri R.,    Fauci F.,  1983, Nature, 304, 417

\bibitem[\protect\citeauthoryear{{Boyle} \& {Buonanno}}{{Boyle} \&
  {Buonanno}}{2008}]{bb08}
{Boyle} L.~A.,  {Buonanno} A.,  2008, Phys. Rev. D, 78, 043531

\bibitem[\protect\citeauthoryear{Boynton, Groth, Hutchinson, Nanos, Partridge
  \& Wilkinson}{Boynton et~al.}{1972}]{bgh+72}
Boynton P.~E.,  Groth E.~J.,  Hutchinson D.~P.,  Nanos G.~P.,  Partridge R.~B.,
     Wilkinson D.~T.,  1972, ApJ, 175, 217

\bibitem[\protect\citeauthoryear{Britton}{Britton}{2000}]{bri00}
Britton M.~C.,  2000, ApJ, 532, 1240

\bibitem[\protect\citeauthoryear{Burgay, Joshi, {D'Amico}, Possenti, Lyne,
  Manchester, McLaughlin, Kramer, Camilo \& Freire}{Burgay
  et~al.}{2006}]{bjd+06}
Burgay M.,  Joshi B.~C.,  {D'Amico} N.,  Possenti A.,  Lyne A.~G.,  Manchester
  R.~N.,  McLaughlin M.~A.,  Kramer M.,  Camilo F.,    Freire P.~C.~C.,  2006,
  MNRAS, 368, 283

\bibitem[\protect\citeauthoryear{Camilo}{Camilo}{1995}]{cam95a}
Camilo F.,  1995, PhD thesis, Princeton University

\bibitem[\protect\citeauthoryear{Camilo, Nice, Shrauner \& Taylor}{Camilo
  et~al.}{1996}]{cnst96}
Camilo F.,  Nice D.~J.,  Shrauner J.~A.,    Taylor J.~H.,  1996, ApJ, 469, 819

\bibitem[\protect\citeauthoryear{Camilo, Nice \& Taylor}{Camilo
  et~al.}{1993}]{cnt93}
Camilo F.,  Nice D.~J.,    Taylor J.~H.,  1993, ApJ, 412, L37

\bibitem[\protect\citeauthoryear{Camilo, Nice \& Taylor}{Camilo
  et~al.}{1996}]{cnt96}
Camilo F.,  Nice D.~J.,    Taylor J.~H.,  1996, ApJ, 461, 812

\bibitem[\protect\citeauthoryear{{Champion}, {Hobbs}, {Manchester}, {Edwards},
  {Backer}, {Bailes}, {Bhat}, {Burke-Spolaor} \& {Coles}}{{Champion}
  et~al.}{2010}]{chm+10Ltd}
{Champion} D.~J.,  {Hobbs} G.~B.,  {Manchester} R.~N.,  {Edwards} R.~T.,
  {Backer} D.~C.,  {Bailes} M.,  {Bhat} N.~D.~R.,  {Burke-Spolaor} S.,
  {Coles} W.,  2010, ApJ, 720, L201

\bibitem[\protect\citeauthoryear{{Coenen}, {van Leeuwen}, {Hessels},
  {Stappers}, {Kondratiev}, {Alexov}, {Breton}, {Bilous}, {Cooper}, {Falcke},
  {Fallows}, {Gajjar} \& et al.}{{Coenen} et~al.}{2014}]{cvh+14ltd}
{Coenen} T.,  {van Leeuwen} J.,  {Hessels} J.~W.~T.,  {Stappers} B.~W.,
  {Kondratiev} V.~I.,  {Alexov} A.,  {Breton} R.~P.,  {Bilous} A.,  {Cooper}
  S.,  {Falcke} H.,  {Fallows} R.~A.,  {Gajjar} V.,    et al. 2014, A\&A, 570,
  A60

\bibitem[\protect\citeauthoryear{Cognard, Bourgois, Lestrade, Biraud, Aubry,
  Darchy \& Drouhin}{Cognard et~al.}{1995}]{cbl+95}
Cognard I.,  Bourgois G.,  Lestrade J.~F.,  Biraud F.,  Aubry D.,  Darchy B.,
   Drouhin J.~P.,  1995, A\&A, 296, 169

\bibitem[\protect\citeauthoryear{{Coles}, {Hobbs}, {Champion}, {Manchester} \&
  {Verbiest}}{{Coles} et~al.}{2011}]{chc+11}
{Coles} W.,  {Hobbs} G.,  {Champion} D.~J.,  {Manchester} R.~N.,    {Verbiest}
  J.~P.~W.,  2011, MNRAS, 418, 561

\bibitem[\protect\citeauthoryear{{Coles}, {Kerr}, {Shannon}, {Hobbs},
  {Manchester}, {You}, {Bailes}, {Bhat}, {Burke-Spolaor}, {Dai}, {Keith},
  {Levin}, {Os{\l}owski}, {Ravi}, {Reardon}, {Toomey}, {van Straten}, {Wang},
  {Wen} \& {Zhu}}{{Coles} et~al.}{2015}]{cks+15}
{Coles} W.~A.,  {Kerr} M.,  {Shannon} R.~M.,  {Hobbs} G.~B.,  {Manchester}
  R.~N.,  {You} X.-P.,  {Bailes} M.,  {Bhat} N.~D.~R.,  {Burke-Spolaor} S.,
  {Dai} S.,  {Keith} M.~J.,  {Levin} Y.,  {Os{\l}owski} S.,  {Ravi} V.,
  {Reardon} D.,  {Toomey} L.,  {van Straten} W.,  {Wang} J.~B.,  {Wen} L.,
  {Zhu} X.~J.,  2015, ApJ, 808, 113

\bibitem[\protect\citeauthoryear{{Cordes}, {Freire}, {Lorimer}, {Camilo},
  {Champion}, {Nice}, {Ramachandran}, {Hessels}, {Vlemmings}, {van Leeuwen},
  {Ransom}, {Bhat} \& et al.}{{Cordes} et~al.}{2006}]{cfl+06ltd}
{Cordes} J.~M.,  {Freire} P.~C.~C.,  {Lorimer} D.~R.,  {Camilo} F.,  {Champion}
  D.~J.,  {Nice} D.~J.,  {Ramachandran} R.,  {Hessels} J.~W.~T.,  {Vlemmings}
  W.,  {van Leeuwen} J.,  {Ransom} S.~M.,  {Bhat} N.~D.~R.,    et al. 2006,
  ApJ, 637, 446

\bibitem[\protect\citeauthoryear{{Cordes} \& {Lazio}}{{Cordes} \&
  {Lazio}}{2002}]{cl02}
{Cordes} J.~M.,  {Lazio} T.~J.~W.,  2002, {ArXiv:astro-ph/0207156}

\bibitem[\protect\citeauthoryear{{Cordes}, {Shannon} \& {Stinebring}}{{Cordes}
  et~al.}{2015}]{css15}
{Cordes} J.~M.,  {Shannon} R.~M.,    {Stinebring} D.~R.,  2015, ArXiv e-prints

\bibitem[\protect\citeauthoryear{{Damour} \& {Vilenkin}}{{Damour} \&
  {Vilenkin}}{2000}]{dv00}
{Damour} T.,  {Vilenkin} A.,  2000, Physical Review Letters, 85, 3761

\bibitem[\protect\citeauthoryear{Deller, Verbiest, Tingay \& Bailes}{Deller
  et~al.}{2008}]{dvtb08}
Deller A.~T.,  Verbiest J.~P.~W.,  Tingay S.~J.,    Bailes M.,  2008, ApJ, 685,
  L67

\bibitem[\protect\citeauthoryear{{Demorest}}{{Demorest}}{2011}]{dem11}
{Demorest} P.~B.,  2011, MNRAS, 416, 2821

\bibitem[\protect\citeauthoryear{{Demorest}, {Ferdman}, {Gonzalez} \& et.
  al.}{{Demorest} et~al.}{2013}]{dfg+13ltd}
{Demorest} P.~B.,  {Ferdman} R.~D.,  {Gonzalez} M.~E.,    et. al. 2013, ApJ,
  762, 94

\bibitem[\protect\citeauthoryear{{Deneva}, {Stovall}, {McLaughlin}, {Bates},
  {Freire}, {Martinez}, {Jenet} \& {Bagchi}}{{Deneva} et~al.}{2013}]{dsm+13}
{Deneva} J.~S.,  {Stovall} K.,  {McLaughlin} M.~A.,  {Bates} S.~D.,  {Freire}
  P.~C.~C.,  {Martinez} J.~G.,  {Jenet} F.,    {Bagchi} M.,  2013, ApJ, 775, 51

\bibitem[\protect\citeauthoryear{{Desvignes}, Caballero, Lentati, Verbiest,
  Stappers, Champion, Janssen, Lazarus \& et al.}{{Desvignes}
  et~al.}{2016}]{dcl+16ltd}
{Desvignes} G.,  Caballero R.~N.,  Lentati L.,  Verbiest J.~P.~W.,  Stappers
  B.~W.,  Champion D.~J.,  Janssen G.~H.,  Lazarus P.,    et al. 2016, MNRAS,
  submitted

\bibitem[\protect\citeauthoryear{{Dolch}, {Lam}, {Cordes}, {Chatterjee},
  {Bassa}, {Bhattacharyya}, {Champion}, {Cognard} \& {Crowter}}{{Dolch}
  et~al.}{2014}]{dlt+14ltd}
{Dolch} T.,  {Lam} M.~T.,  {Cordes} J.,  {Chatterjee} S.,  {Bassa} C.,
  {Bhattacharyya} B.,  {Champion} D.~J.,  {Cognard} I.,    {Crowter} K.,  2014,
  ApJ, 794, 21

\bibitem[\protect\citeauthoryear{{Dowell}, {Ray}, {Taylor}, {Blythe}, {Clarke},
  {Craig}, {Ellingson}, {Helmboldt}, {Henning}, {Lazio}, {Schinzel}, {Stovall}
  \& {Wolfe}}{{Dowell} et~al.}{2013}]{drt+13}
{Dowell} J.,  {Ray} P.~S.,  {Taylor} G.~B.,  {Blythe} J.~N.,  {Clarke} T.,
  {Craig} J.,  {Ellingson} S.~W.,  {Helmboldt} J.~F.,  {Henning} P.~A.,
  {Lazio} T.~J.~W.,  {Schinzel} F.,  {Stovall} K.,    {Wolfe} C.~N.,  2013,
  ApJ, 775, L28

\bibitem[\protect\citeauthoryear{{Du}, {Yang}, {Campbell}, {Janssen},
  {Stappers} \& {Chen}}{{Du} et~al.}{2014}]{dyc+14}
{Du} Y.,  {Yang} J.,  {Campbell} R.~M.,  {Janssen} G.,  {Stappers} B.,
  {Chen} D.,  2014, ApJ, 782, L38

\bibitem[\protect\citeauthoryear{Edwards \& Bailes}{Edwards \&
  Bailes}{2001}]{eb01b}
Edwards R.~T.,  Bailes M.,  2001, ApJ, 553, 801

\bibitem[\protect\citeauthoryear{Edwards, Hobbs \& Manchester}{Edwards
  et~al.}{2006}]{ehm06}
Edwards R.~T.,  Hobbs G.~B.,    Manchester R.~N.,  2006, MNRAS, 372, 1549

\bibitem[\protect\citeauthoryear{{Enoki} \& {Nagashima}}{{Enoki} \&
  {Nagashima}}{2007}]{en07}
{Enoki} M.,  {Nagashima} M.,  2007, Progress of Theoretical Physics, 117, 241

\bibitem[\protect\citeauthoryear{{Faulkner}, {Stairs}, {Kramer}, {Lyne},
  {Hobbs}, {Possenti}, {Lorimer}, {Manchester}, {McLaughlin}, {D'Amico},
  {Camilo} \& {Burgay}}{{Faulkner} et~al.}{2004}]{fsk+04}
{Faulkner} A.~J.,  {Stairs} I.~H.,  {Kramer} M.,  {Lyne} A.~G.,  {Hobbs} G.,
  {Possenti} A.,  {Lorimer} D.~R.,  {Manchester} R.~N.,  {McLaughlin} M.~A.,
  {D'Amico} N.,  {Camilo} F.,    {Burgay} M.,  2004, MNRAS, 355, 147

\bibitem[\protect\citeauthoryear{{Favata}}{{Favata}}{2009}]{fav09}
{Favata} M.,  2009, ApJ, 696, L159

\bibitem[\protect\citeauthoryear{{Ferdman}, {Stairs}, {Kramer}, {McLaughlin},
  {Lorimer}, {Nice}, {Manchester}, {Hobbs}, {Lyne}, {Camilo}, {Possenti},
  {Demorest}, {Cognard}, {Desvignes}, {Theureau}, {Faulkner} \&
  {Backer}}{{Ferdman} et~al.}{2010}]{fsk+10}
{Ferdman} R.~D.,  {Stairs} I.~H.,  {Kramer} M.,  {McLaughlin} M.~A.,  {Lorimer}
  D.~R.,  {Nice} D.~J.,  {Manchester} R.~N.,  {Hobbs} G.,  {Lyne} A.~G.,
  {Camilo} F.,  {Possenti} A.,  {Demorest} P.~B.,  {Cognard} I.,  {Desvignes}
  G.,  {Theureau} G.,  {Faulkner} A.,    {Backer} D.~C.,  2010, ApJ, 711, 764

\bibitem[\protect\citeauthoryear{{Finn} \& {Lommen}}{{Finn} \&
  {Lommen}}{2010}]{fl10}
{Finn} L.~S.,  {Lommen} A.~N.,  2010, ApJ, 718, 1400

\bibitem[\protect\citeauthoryear{{Folkner}, {Williams} \& {Boggs}}{{Folkner}
  et~al.}{2009}]{fwb09}
{Folkner} W.~M.,  {Williams} J.~G.,    {Boggs} D.~H.,  2009, IPN Progress
  Report, 42-178

\bibitem[\protect\citeauthoryear{Foster \& Backer}{Foster \&
  Backer}{1990}]{fb90}
Foster R.~S.,  Backer D.~C.,  1990, ApJ, 361, 300

\bibitem[\protect\citeauthoryear{Foster \& Cordes}{Foster \&
  Cordes}{1990}]{fc90}
Foster R.~S.,  Cordes J.~M.,  1990, ApJ, 364, 123

\bibitem[\protect\citeauthoryear{Foster, Wolszczan \& Camilo}{Foster
  et~al.}{1993}]{fwc93}
Foster R.~S.,  Wolszczan A.,    Camilo F.,  1993, ApJ, 410, L91

\bibitem[\protect\citeauthoryear{Frail \& Weisberg}{Frail \&
  Weisberg}{1990}]{fw90}
Frail D.~A.,  Weisberg J.~M.,  1990, AJ, 100, 743

\bibitem[\protect\citeauthoryear{{Freire}, {Wex}, {Esposito-Far{\`e}se},
  {Verbiest}, {Bailes}, {Jacoby}, {Kramer}, {Stairs}, {Antoniadis} \&
  {Janssen}}{{Freire} et~al.}{2012}]{fwe+12}
{Freire} P.~C.~C.,  {Wex} N.,  {Esposito-Far{\`e}se} G.,  {Verbiest} J.~P.~W.,
  {Bailes} M.,  {Jacoby} B.~A.,  {Kramer} M.,  {Stairs} I.~H.,  {Antoniadis}
  J.,    {Janssen} G.~H.,  2012, MNRAS, 423, 3328

\bibitem[\protect\citeauthoryear{{Gonzalez}, {Stairs}, {Ferdman}, {Freire},
  {Nice}, {Demorest}, {Ransom}, {Kramer}, {Camilo}, {Hobbs}, {Manchester} \&
  {Lyne}}{{Gonzalez} et~al.}{2011}]{gsf+11}
{Gonzalez} M.~E.,  {Stairs} I.~H.,  {Ferdman} R.~D.,  {Freire} P.~C.~C.,
  {Nice} D.~J.,  {Demorest} P.~B.,  {Ransom} S.~M.,  {Kramer} M.,  {Camilo} F.,
   {Hobbs} G.,  {Manchester} R.~N.,    {Lyne} A.~G.,  2011, ApJ, 743, 102

\bibitem[\protect\citeauthoryear{{Grishchuk}}{{Grishchuk}}{2005}]{gri05}
{Grishchuk} L.~P.,  2005, Phys. Uspekhi, pp 1235--1247

\bibitem[\protect\citeauthoryear{{Hassall}, {Stappers}, {Hessels} \& et
  al.}{{Hassall} et~al.}{2012}]{hsh+12ltd}
{Hassall} T.~E.,  {Stappers} B.~W.,  {Hessels} J.~W.~T.,    et al. 2012, A\&A,
  543, A66

\bibitem[\protect\citeauthoryear{Hellings \& Downs}{Hellings \&
  Downs}{1983}]{hd83}
Hellings R.~W.,  Downs G.~S.,  1983, ApJ, 265, L39

\bibitem[\protect\citeauthoryear{Hemberger \& Stinebring}{Hemberger \&
  Stinebring}{2008}]{hs08}
Hemberger D.~A.,  Stinebring D.~R.,  2008, ApJ, 674, L37

\bibitem[\protect\citeauthoryear{{Hobbs}, Archibald, Arzoumanian, Backer,
  Bailes, Bhat, Burgay, Burke-Spolaor, Champion, Cognard \& Coles}{{Hobbs}
  et~al.}{2010}]{haa+10ltd}
{Hobbs} G.,  Archibald A.,  Arzoumanian Z.,  Backer D.,  Bailes M.,  Bhat
  N.~D.~R.,  Burgay M.,  Burke-Spolaor S.,  Champion D.,  Cognard I.,    Coles
  W.,  2010, Class. Quant Grav., 27, 084013

\bibitem[\protect\citeauthoryear{{Hobbs}, {Coles}, {Manchester}, {Keith},
  {Shannon}, {Chen}, {Bailes}, {Bhat} \& {Burke-Spolaor}}{{Hobbs}
  et~al.}{2012}]{hcm+12Ltd}
{Hobbs} G.,  {Coles} W.,  {Manchester} R.~N.,  {Keith} M.~J.,  {Shannon} R.~M.,
   {Chen} D.,  {Bailes} M.,  {Bhat} N.~D.~R.,    {Burke-Spolaor} S.,  2012,
  MNRAS, 427, 2780

\bibitem[\protect\citeauthoryear{{Hobbs}, {Faulkner}, {Stairs}, {Camilo},
  {Manchester}, {Lyne}, {Kramer}, {D'Amico}, {Kaspi}, {Possenti}, {McLaughlin},
  {Lorimer}, {Burgay}, {Joshi} \& {Crawford}}{{Hobbs} et~al.}{2004}]{hfs+04}
{Hobbs} G.,  {Faulkner} A.,  {Stairs} I.~H.,  {Camilo} F.,  {Manchester} R.~N.,
   {Lyne} A.~G.,  {Kramer} M.,  {D'Amico} N.,  {Kaspi} V.~M.,  {Possenti} A.,
  {McLaughlin} M.~A.,  {Lorimer} D.~R.,  {Burgay} M.,  {Joshi} B.~C.,
  {Crawford} F.,  2004, MNRAS, 352, 1439

\bibitem[\protect\citeauthoryear{{Hobbs}, {Lyne} \& {Kramer}}{{Hobbs}
  et~al.}{2010}]{hlk10}
{Hobbs} G.,  {Lyne} A.~G.,    {Kramer} M.,  2010, MNRAS, 402, 1027

\bibitem[\protect\citeauthoryear{Hobbs, Lyne, Kramer, Martin \& Jordan}{Hobbs
  et~al.}{2004}]{hlk+04}
Hobbs G.,  Lyne A.~G.,  Kramer M.,  Martin C.~E.,    Jordan C.,  2004, MNRAS,
  353, 1311

\bibitem[\protect\citeauthoryear{{Hobbs}, {Edwards} \& {Manchester}}{{Hobbs}
  et~al.}{2006}]{hem06}
{Hobbs} G.~B.,  {Edwards} R.~T.,    {Manchester} R.~N.,  2006, MNRAS, 369, 655

\bibitem[\protect\citeauthoryear{Hotan, Bailes \& Ord}{Hotan
  et~al.}{2006}]{hbo06}
Hotan A.~W.,  Bailes M.,    Ord S.~M.,  2006, MNRAS, 369, 1502

\bibitem[\protect\citeauthoryear{{Hotan}, {van Straten} \&
  {Manchester}}{{Hotan} et~al.}{2004}]{hvm04}
{Hotan} A.~W.,  {van Straten} W.,    {Manchester} R.~N.,  2004, PASA, 21, 302

\bibitem[\protect\citeauthoryear{Jacoby}{Jacoby}{2004}]{jac04a}
Jacoby B.~A.,  2004, PhD thesis, California Institute of Technology

\bibitem[\protect\citeauthoryear{Jacoby, Bailes, Ord, Knight \& Hotan}{Jacoby
  et~al.}{2007}]{jbo+07}
Jacoby B.~A.,  Bailes M.,  Ord S.~M.,  Knight H.~S.,    Hotan A.~W.,  2007,
  ApJ, 656, 408

\bibitem[\protect\citeauthoryear{{Jacoby}, {Bailes}, {van Kerkwijk}, {Ord},
  {Hotan}, {Kulkarni} \& {Anderson}}{{Jacoby} et~al.}{2003}]{jbv+03}
{Jacoby} B.~A.,  {Bailes} M.,  {van Kerkwijk} M.~H.,  {Ord} S.,  {Hotan} A.,
  {Kulkarni} S.~R.,    {Anderson} S.~B.,  2003, ApJ, 599, L99

\bibitem[\protect\citeauthoryear{{Janssen}, {Hobbs}, {McLaughlin}, {Bassa},
  {Deller}, {Kramer}, {Lee}, {Mingarelli}, {Rosado}, {Sanidas}, {Sesana},
  {Shao}, {Stairs}, {Stappers} \& {Verbiest}}{{Janssen} et~al.}{2015}]{jhm+15}
{Janssen} G.,  {Hobbs} G.,  {McLaughlin} M.,  {Bassa} C.,  {Deller} A.,
  {Kramer} M.,  {Lee} K.,  {Mingarelli} C.,  {Rosado} P.,  {Sanidas} S.,
  {Sesana} A.,  {Shao} L.,  {Stairs} I.,  {Stappers} B.,    {Verbiest}
  J.~P.~W.,  2015, Advancing Astrophysics with the Square Kilometre Array
  (AASKA14), p.~37

\bibitem[\protect\citeauthoryear{{Janssen}, {Stappers}, {Bassa}, {Cognard},
  {Kramer} \& {Theureau}}{{Janssen} et~al.}{2010}]{jsb+10}
{Janssen} G.~H.,  {Stappers} B.~W.,  {Bassa} C.~G.,  {Cognard} I.,  {Kramer}
  M.,    {Theureau} G.,  2010, A\&A, 514, A74+

\bibitem[\protect\citeauthoryear{{Jenet}, {Hobbs}, {Lee} \&
  {Manchester}}{{Jenet} et~al.}{2005}]{jhlm05}
{Jenet} F.~A.,  {Hobbs} G.~B.,  {Lee} K.~J.,    {Manchester} R.~N.,  2005, ApJ,
  625, L123

\bibitem[\protect\citeauthoryear{Johnston, Lorimer, Harrison, Bailes, Lyne,
  Bell, Kaspi, Manchester, D'Amico, Nicastro \& Jin}{Johnston
  et~al.}{1993}]{jlh+93}
Johnston S.,  Lorimer D.~R.,  Harrison P.~A.,  Bailes M.,  Lyne A.~G.,  Bell
  J.~F.,  Kaspi V.~M.,  Manchester R.~N.,  D'Amico N.,  Nicastro L.,    Jin S.,
   1993, Nature, 361, 613

\bibitem[\protect\citeauthoryear{Kaspi, Taylor \& Ryba}{Kaspi
  et~al.}{1994}]{ktr94}
Kaspi V.~M.,  Taylor J.~H.,    Ryba M.,  1994, ApJ, 428, 713

\bibitem[\protect\citeauthoryear{{Keane}, {Bhattacharyya}, {Kramer},
  {Stappers}, {Bates} \& et al.}{{Keane} et~al.}{2015}]{kbk+15ltd}
{Keane} E.~F.,  {Bhattacharyya} B.,  {Kramer} M.,  {Stappers} B.~W.,  {Bates}
  S.~D.,    et al. 2015, Advancing Astrophysics with the Square Kilometre Array
  (AASKA14), p.~40

\bibitem[\protect\citeauthoryear{{Keith}, {Coles}, {Shannon}, {Hobbs},
  {Manchester} \& et al.}{{Keith} et~al.}{2013}]{kcs+13ltd}
{Keith} M.~J.,  {Coles} W.,  {Shannon} R.~M.,  {Hobbs} G.~B.,  {Manchester}
  R.~N.,    et al. 2013, MNRAS, 429, 2161

\bibitem[\protect\citeauthoryear{{Keith}, {Jameson}, {van Straten}, {Bailes},
  {Johnston}, {Kramer}, {Possenti}, {Bates}, {Bhat}, {Burgay}, {Burke-Spolaor},
  {D'Amico}, {Levin}, {McMahon}, {Milia} \& {Stappers}}{{Keith}
  et~al.}{2010}]{kjv+10}
{Keith} M.~J.,  {Jameson} A.,  {van Straten} W.,  {Bailes} M.,  {Johnston} S.,
  {Kramer} M.,  {Possenti} A.,  {Bates} S.~D.,  {Bhat} N.~D.~R.,  {Burgay} M.,
  {Burke-Spolaor} S.,  {D'Amico} N.,  {Levin} L.,  {McMahon} P.~L.,  {Milia}
  S.,    {Stappers} B.~W.,  2010, MNRAS, 409, 619

\bibitem[\protect\citeauthoryear{{Khmelnitsky} \& {Rubakov}}{{Khmelnitsky} \&
  {Rubakov}}{2014}]{kr14}
{Khmelnitsky} A.,  {Rubakov} V.,  2014, Journal of Cosmology and Astroparticle
  Physics, 2, 19

\bibitem[\protect\citeauthoryear{{Kondratiev}, {Verbiest}, {Hessels}, {Bilous},
  {Stappers}, {Kramer}, {Keane}, {Noutsos}, {Os{\l}owski}, {Breton}, {Hassall},
  {Alexov} \& et al.}{{Kondratiev} et~al.}{2015}]{kvh+15ltd}
{Kondratiev} V.~I.,  {Verbiest} J.~P.~W.,  {Hessels} J.~W.~T.,  {Bilous} A.~V.,
   {Stappers} B.~W.,  {Kramer} M.,  {Keane} E.~F.,  {Noutsos} A.,
  {Os{\l}owski} S.,  {Breton} R.~P.,  {Hassall} T.~E.,  {Alexov} A.,    et al.
  2015, A\&A submitted; arXiv:1508.02948

\bibitem[\protect\citeauthoryear{Kopeikin}{Kopeikin}{1995}]{kop95}
Kopeikin S.~M.,  1995, ApJ, 439, L5

\bibitem[\protect\citeauthoryear{Kopeikin}{Kopeikin}{1996}]{kop96}
Kopeikin S.~M.,  1996, ApJ, 467, L93

\bibitem[\protect\citeauthoryear{Kramer, Xilouris, Lorimer, Doroshenko,
  Jessner, Wielebinski, Wolszczan \& Camilo}{Kramer et~al.}{1998}]{kxl+98}
Kramer M.,  Xilouris K.~M.,  Lorimer D.~R.,  Doroshenko O.,  Jessner A.,
  Wielebinski R.,  Wolszczan A.,    Camilo F.,  1998, ApJ, 501, 270

\bibitem[\protect\citeauthoryear{{L{\" o}hmer}, {Lewandowski}, {Wolszczan} \&
  {Wielebinski}}{{L{\" o}hmer} et~al.}{2005}]{llww05}
{L{\" o}hmer} O.,  {Lewandowski} W.,  {Wolszczan} A.,    {Wielebinski} R.,
  2005, ApJ, 621, 388

\bibitem[\protect\citeauthoryear{Lazaridis, Wex, Jessner, Kramer, Stappers,
  Janssen, Desvignes, Purver, Cognard, Theureau, Lyne, Jordan \&
  Zensus}{Lazaridis et~al.}{2009}]{lwj+09}
Lazaridis K.,  Wex N.,  Jessner A.,  Kramer M.,  Stappers B.~W.,  Janssen
  G.~H.,  Desvignes G.,  Purver M.~B.,  Cognard I.,  Theureau G.,  Lyne A.~G.,
  Jordan C.~A.,    Zensus J.~A.,  2009, MNRAS, 400, 805

\bibitem[\protect\citeauthoryear{{Lee}, {Bassa}, {Janssen}, {Karuppusamy},
  {Kramer}, {Smits} \& {Stappers}}{{Lee} et~al.}{2012}]{lbj+12}
{Lee} K.~J.,  {Bassa} C.~G.,  {Janssen} G.~H.,  {Karuppusamy} R.,  {Kramer} M.,
   {Smits} R.,    {Stappers} B.~W.,  2012, MNRAS, 423, 2642

\bibitem[\protect\citeauthoryear{{Lee}, {Wex}, {Kramer}, {Stappers}, {Bassa},
  {Janssen}, {Karuppusamy} \& {Smits}}{{Lee} et~al.}{2011}]{lwk+11}
{Lee} K.~J.,  {Wex} N.,  {Kramer} M.,  {Stappers} B.~W.,  {Bassa} C.~G.,
  {Janssen} G.~H.,  {Karuppusamy} R.,    {Smits} R.,  2011, MNRAS, 414, 3251

\bibitem[\protect\citeauthoryear{{Lentati}, {Alexander}, {Hobson}, {Feroz},
  {van Haasteren}, {Lee} \& {Shannon}}{{Lentati} et~al.}{2014}]{lah+14}
{Lentati} L.,  {Alexander} P.,  {Hobson} M.~P.,  {Feroz} F.,  {van Haasteren}
  R.,  {Lee} K.~J.,    {Shannon} R.~M.,  2014, MNRAS, 437, 3004

\bibitem[\protect\citeauthoryear{{Lentati}, Shannon, Coles, Verbiest \&
  }{{Lentati} et~al.}{2015}]{lea+15}
{Lentati} L.,  Shannon R.~M.,  Coles W.~A.,  Verbiest J.~P.~W.,      e.~a.,
  2015, In preparation, 1000, 100

\bibitem[\protect\citeauthoryear{{Lentati}, {Taylor}, {Mingarelli}, {Sesana},
  {Sanidas} \& et al.}{{Lentati} et~al.}{2015}]{ltm+15ltd}
{Lentati} L.,  {Taylor} S.~R.,  {Mingarelli} C.~M.~F.,  {Sesana} A.,  {Sanidas}
  S.~A.,    et al. 2015, MNRAS, 453, 2576

\bibitem[\protect\citeauthoryear{{Liu}, {Desvignes}, {Cognard}, {Stappers},
  {Verbiest}, {Lee}, {Champion}, {Kramer}, {Freire} \& {Karuppusamy}}{{Liu}
  et~al.}{2014}]{ldc+14}
{Liu} K.,  {Desvignes} G.,  {Cognard} I.,  {Stappers} B.~W.,  {Verbiest}
  J.~P.~W.,  {Lee} K.~J.,  {Champion} D.~J.,  {Kramer} M.,  {Freire} P.~C.~C.,
    {Karuppusamy} R.,  2014, MNRAS, 443, 3752

\bibitem[\protect\citeauthoryear{{Liu}, {Keane}, {Lee}, {Kramer}, {Cordes} \&
  {Purver}}{{Liu} et~al.}{2012}]{lkl+12}
{Liu} K.,  {Keane} E.~F.,  {Lee} K.~J.,  {Kramer} M.,  {Cordes} J.~M.,
  {Purver} M.~B.,  2012, MNRAS, 420, 361

\bibitem[\protect\citeauthoryear{{Liu}, {Verbiest}, {Kramer}, {Stappers}, {van
  Straten} \& {Cordes}}{{Liu} et~al.}{2011}]{lvk+11}
{Liu} K.,  {Verbiest} J.~P.~W.,  {Kramer} M.,  {Stappers} B.~W.,  {van Straten}
  W.,    {Cordes} J.~M.,  2011, MNRAS, 417, 2916

\bibitem[\protect\citeauthoryear{{Lommen}, {Kipphorn}, {Nice}, {Splaver},
  {Stairs} \& {Backer}}{{Lommen} et~al.}{2006}]{lkn+06}
{Lommen} A.~N.,  {Kipphorn} R.~A.,  {Nice} D.~J.,  {Splaver} E.~M.,  {Stairs}
  I.~H.,    {Backer} D.~C.,  2006, ApJ, 642, 1012

\bibitem[\protect\citeauthoryear{{Lommen}, {Zepka}, {Backer}, {McLaughlin},
  {Cordes}, {Arzoumanian} \& {Xilouris}}{{Lommen} et~al.}{2000}]{lzb+00}
{Lommen} A.~N.,  {Zepka} A.,  {Backer} D.~C.,  {McLaughlin} M.,  {Cordes}
  J.~M.,  {Arzoumanian} Z.,    {Xilouris} K.,  2000, ApJ, 545, 1007

\bibitem[\protect\citeauthoryear{Lorimer, Faulkner, Lyne, Manchester, Kramer,
  McLaughlin, Hobbs, Possenti, Stairs, Camilo, Burgay, {D'Amico}, Corongiu \&
  Crawford}{Lorimer et~al.}{2006}]{lfl+06}
Lorimer D.~R.,  Faulkner A.~J.,  Lyne A.~G.,  Manchester R.~N.,  Kramer M.,
  McLaughlin M.~A.,  Hobbs G.,  Possenti A.,  Stairs I.~H.,  Camilo F.,  Burgay
  M.,  {D'Amico} N.,  Corongiu A.,    Crawford F.,  2006, MNRAS, 372, 777

\bibitem[\protect\citeauthoryear{Lorimer \& Kramer}{Lorimer \&
  Kramer}{2005}]{lk05}
Lorimer D.~R.,  Kramer M.,  2005, {Handbook of Pulsar Astronomy}.
Cambridge University Press

\bibitem[\protect\citeauthoryear{Lorimer, Lyne, Bailes, Manchester, D'Amico,
  Stappers, Johnston \& Camilo}{Lorimer et~al.}{1996}]{llb+96}
Lorimer D.~R.,  Lyne A.~G.,  Bailes M.,  Manchester R.~N.,  D'Amico N.,
  Stappers B.~W.,  Johnston S.,    Camilo F.,  1996, MNRAS, 283, 1383

\bibitem[\protect\citeauthoryear{Lorimer, Nicastro, Lyne, Bailes, Manchester,
  Johnston, Bell, D'Amico \& Harrison}{Lorimer et~al.}{1995}]{lnl+95}
Lorimer D.~R.,  Nicastro L.,  Lyne A.~G.,  Bailes M.,  Manchester R.~N.,
  Johnston S.,  Bell J.~F.,  D'Amico N.,    Harrison P.~A.,  1995, ApJ, 439,
  933

\bibitem[\protect\citeauthoryear{Lundgren, Zepka \& Cordes}{Lundgren
  et~al.}{1995}]{lzc95}
Lundgren S.~C.,  Zepka A.~F.,    Cordes J.~M.,  1995, ApJ, 453, 419

\bibitem[\protect\citeauthoryear{Lyne, Brinklow, Middleditch, Kulkarni, Backer
  \& Clifton}{Lyne et~al.}{1987}]{lbm+87}
Lyne A.~G.,  Brinklow A.,  Middleditch J.,  Kulkarni S.~R.,  Backer D.~C.,
  Clifton T.~R.,  1987, Nature, 328, 399

\bibitem[\protect\citeauthoryear{{Maitia}, {Lestrade} \& {Cognard}}{{Maitia}
  et~al.}{2003}]{mlc03}
{Maitia} V.,  {Lestrade} J.-F.,    {Cognard} I.,  2003, ApJ, 582, 972

\bibitem[\protect\citeauthoryear{{Manchester}, {Hobbs}, {Bailes}, {Coles}, {van
  Straten} \& {Keith}}{{Manchester} et~al.}{2013}]{mhb+13ltd}
{Manchester} R.~N.,  {Hobbs} G.,  {Bailes} M.,  {Coles} W.~A.,  {van Straten}
  W.,    {Keith} M.~J.,  2013, PASA, 30, 17

\bibitem[\protect\citeauthoryear{{Manchester}, {Hobbs}, {Teoh} \&
  {Hobbs}}{{Manchester} et~al.}{2005}]{mhth05}
{Manchester} R.~N.,  {Hobbs} G.~B.,  {Teoh} A.,    {Hobbs} M.,  2005, AJ, 129,
  1993

\bibitem[\protect\citeauthoryear{{Manchester} \& {IPTA}}{{Manchester} \&
  {IPTA}}{2013}]{man13}
{Manchester} R.~N.,  {IPTA} 2013, Class. Quant Grav., 30, 224010

\bibitem[\protect\citeauthoryear{{Nan}, {Li}, {Jin}, {Wang}, {Zhu}, {Zhu},
  {Zhang}, {Yue} \& {Qian}}{{Nan} et~al.}{2011}]{nlj+11}
{Nan} R.,  {Li} D.,  {Jin} C.,  {Wang} Q.,  {Zhu} L.,  {Zhu} W.,  {Zhang} H.,
  {Yue} Y.,    {Qian} L.,  2011, International Journal of Modern Physics D, 20,
  989

\bibitem[\protect\citeauthoryear{Navarro, de Bruyn, Frail, Kulkarni \&
  Lyne}{Navarro et~al.}{1995}]{nbf+95}
Navarro J.,  de Bruyn G.,  Frail D.,  Kulkarni S.~R.,    Lyne A.~G.,  1995,
  ApJ, 455, L55

\bibitem[\protect\citeauthoryear{Nicastro, Lyne, Lorimer, Harrison, Bailes \&
  Skidmore}{Nicastro et~al.}{1995}]{nll+95}
Nicastro L.,  Lyne A.~G.,  Lorimer D.~R.,  Harrison P.~A.,  Bailes M.,
  Skidmore B.~D.,  1995, MNRAS, 273, L68

\bibitem[\protect\citeauthoryear{Nice, Splaver \& Stairs}{Nice
  et~al.}{2001}]{nss01}
Nice D.~J.,  Splaver E.~M.,    Stairs I.~H.,  2001, ApJ, 549, 516

\bibitem[\protect\citeauthoryear{{Nice}, {Splaver}, {Stairs}, {L{\"o}hmer},
  {Jessner}, {Kramer} \& {Cordes}}{{Nice} et~al.}{2005}]{nss+05}
{Nice} D.~J.,  {Splaver} E.~M.,  {Stairs} I.~H.,  {L{\"o}hmer} O.,  {Jessner}
  A.,  {Kramer} M.,    {Cordes} J.~M.,  2005, ApJ, 634, 1242

\bibitem[\protect\citeauthoryear{Nice \& Taylor}{Nice \& Taylor}{1995}]{nt95}
Nice D.~J.,  Taylor J.~H.,  1995, ApJ, 441, 429

\bibitem[\protect\citeauthoryear{Nice, Taylor \& Fruchter}{Nice
  et~al.}{1993}]{ntf93}
Nice D.~J.,  Taylor J.~H.,    Fruchter A.~S.,  1993, ApJ, 402, L49

\bibitem[\protect\citeauthoryear{{Ord}, {Johnston} \& {Sarkissian}}{{Ord}
  et~al.}{2007}]{ojs07}
{Ord} S.~M.,  {Johnston} S.,    {Sarkissian} J.,  2007, Solar Phys., 245, 109

\bibitem[\protect\citeauthoryear{{Os{\l}owski}, {van Straten}, {Demorest} \&
  {Bailes}}{{Os{\l}owski} et~al.}{2013}]{ovdb13}
{Os{\l}owski} S.,  {van Straten} W.,  {Demorest} P.,    {Bailes} M.,  2013,
  MNRAS, 430, 416

\bibitem[\protect\citeauthoryear{{Os{\l}owski}, {van Straten}, {Hobbs},
  {Bailes} \& {Demorest}}{{Os{\l}owski} et~al.}{2011}]{ovh+11}
{Os{\l}owski} S.,  {van Straten} W.,  {Hobbs} G.~B.,  {Bailes} M.,
  {Demorest} P.,  2011, MNRAS, 418, 1258

\bibitem[\protect\citeauthoryear{{Pennucci}, {Demorest} \& {Ransom}}{{Pennucci}
  et~al.}{2014}]{pdr+14}
{Pennucci} T.~T.,  {Demorest} P.~B.,    {Ransom} S.~M.,  2014, ApJ, 790, 93

\bibitem[\protect\citeauthoryear{{Porayko} \& {Postnov}}{{Porayko} \&
  {Postnov}}{2014}]{pp14}
{Porayko} N.~K.,  {Postnov} K.~A.,  2014, Phys. Rev. D, 90, 062008

\bibitem[\protect\citeauthoryear{{Pshirkov}, {Baskaran} \&
  {Postnov}}{{Pshirkov} et~al.}{2010}]{pbp10}
{Pshirkov} M.~S.,  {Baskaran} D.,    {Postnov} K.~A.,  2010, MNRAS, 402, 417

\bibitem[\protect\citeauthoryear{{Rajagopal} \& {Romani}}{{Rajagopal} \&
  {Romani}}{1995}]{rr95a}
{Rajagopal} M.,  {Romani} R.~W.,  1995, ApJ, 446, 543

\bibitem[\protect\citeauthoryear{{Ray}, {Abdo}, {Parent}, {Bhattacharya},
  {Bhattacharyya}, {Camilo}, {Cognard}, {Theureau}, {Ferrara} \& et al.}{{Ray}
  et~al.}{2012}]{rap+12ltd}
{Ray} P.~S.,  {Abdo} A.~A.,  {Parent} D.,  {Bhattacharya} D.,  {Bhattacharyya}
  B.,  {Camilo} F.,  {Cognard} I.,  {Theureau} G.,  {Ferrara} E.~C.,    et al.
  2012, ArXiv e-prints

\bibitem[\protect\citeauthoryear{Ray, Thorsett, Jenet, van Kerkwijk, Kulkarni,
  Prince, Sandhu \& Nice}{Ray et~al.}{1996}]{rtj+96}
Ray P.~S.,  Thorsett S.~E.,  Jenet F.~A.,  van Kerkwijk M.~H.,  Kulkarni S.~R.,
   Prince T.~A.,  Sandhu J.~S.,    Nice D.~J.,  1996, ApJ, 470, 1103

\bibitem[\protect\citeauthoryear{Rickett}{Rickett}{1990}]{ric90}
Rickett B.~J.,  1990, Ann. Rev. Astr. Ap., 28, 561

\bibitem[\protect\citeauthoryear{Romani}{Romani}{1989}]{rom89}
Romani R.~W.,  1989, in {\"O}gelman H.,  van~den Heuvel E.~P.~J.,  eds, Timing
  Neutron Stars Timing a millisecond pulsar array.
Kluwer, Dordrecht, p.~113

\bibitem[\protect\citeauthoryear{{Rosado}, {Sesana} \& {Gair}}{{Rosado}
  et~al.}{2015}]{rsg15}
{Rosado} P.~A.,  {Sesana} A.,    {Gair} J.,  2015, MNRAS, 451, 2417

\bibitem[\protect\citeauthoryear{Sandhu, Bailes, Manchester, Navarro, Kulkarni
  \& Anderson}{Sandhu et~al.}{1997}]{sbm+97}
Sandhu J.~S.,  Bailes M.,  Manchester R.~N.,  Navarro J.,  Kulkarni S.~R.,
  Anderson S.~B.,  1997, ApJ, 478, L95

\bibitem[\protect\citeauthoryear{{Sanidas}, {Battye} \& {Stappers}}{{Sanidas}
  et~al.}{2012}]{sbs12}
{Sanidas} S.~A.,  {Battye} R.~A.,    {Stappers} B.~W.,  2012, Phys. Rev. D, 85,
  122003

\bibitem[\protect\citeauthoryear{Segelstein, Rawley, Stinebring, Fruchter \&
  Taylor}{Segelstein et~al.}{1986}]{srs+86}
Segelstein D.~J.,  Rawley L.~A.,  Stinebring D.~R.,  Fruchter A.~S.,    Taylor
  J.~H.,  1986, Nature, 322, 714

\bibitem[\protect\citeauthoryear{{Sesana}}{{Sesana}}{2013}]{ses13}
{Sesana} A.,  2013, MNRAS, 433, L1

\bibitem[\protect\citeauthoryear{{Sesana}, {Haardt}, {Madau} \&
  {Volonteri}}{{Sesana} et~al.}{2004}]{shmv04}
{Sesana} A.,  {Haardt} F.,  {Madau} P.,    {Volonteri} M.,  2004, ApJ, 611, 623

\bibitem[\protect\citeauthoryear{{Sesana}, {Vecchio} \& {Volonteri}}{{Sesana}
  et~al.}{2009}]{svv09}
{Sesana} A.,  {Vecchio} A.,    {Volonteri} M.,  2009, MNRAS, 394, 2255

\bibitem[\protect\citeauthoryear{{Seto}}{{Seto}}{2009}]{set09}
{Seto} N.,  2009, MNRAS, 400, L38

\bibitem[\protect\citeauthoryear{{Shannon} \& {Cordes}}{{Shannon} \&
  {Cordes}}{2010}]{sc10}
{Shannon} R.~M.,  {Cordes} J.~M.,  2010, ApJ, 725, 1607

\bibitem[\protect\citeauthoryear{{Shannon}, {Os{\l}owski}, {Dai}, {Bailes},
  {Hobbs}, {Manchester} \& et al.}{{Shannon} et~al.}{2014}]{sod+14ltd}
{Shannon} R.~M.,  {Os{\l}owski} S.,  {Dai} S.,  {Bailes} M.,  {Hobbs} G.,
  {Manchester} R.~N.,    et al. 2014, MNRAS, 443, 1463

\bibitem[\protect\citeauthoryear{{Shannon}, {Ravi}, {Lentati}, {Lasky},
  {Hobbs}, {Kerr}, {Manchester}, {Coles}, {Levin}, {Bailes}, {Bhat} \& et
  al.}{{Shannon} et~al.}{2015}]{srl+15ltd}
{Shannon} R.~M.,  {Ravi} V.,  {Lentati} L.~T.,  {Lasky} P.~D.,  {Hobbs} G.,
  {Kerr} M.,  {Manchester} R.~N.,  {Coles} W.~A.,  {Levin} Y.,  {Bailes} M.,
  {Bhat} N.~D.~R.,    et al. 2015, Science, 349, 1522

\bibitem[\protect\citeauthoryear{{Siemens}, {Ellis}, {Jenet} \&
  {Romano}}{{Siemens} et~al.}{2013}]{sejr13}
{Siemens} X.,  {Ellis} J.,  {Jenet} F.,    {Romano} J.~D.,  2013, Classical and
  Quantum Gravity, 30, 224015

\bibitem[\protect\citeauthoryear{{Sotomayor-Beltran}, {Sobey}, {Hessels}, {de
  Bruyn}, {Noutsos}, {Alexov}, {Anderson}, {Asgekar}, {Avruch}, {Beck}, {Bell},
  {Bell} \& et al.}{{Sotomayor-Beltran} et~al.}{2013}]{ssh+13ltd}
{Sotomayor-Beltran} C.,  {Sobey} C.,  {Hessels} J.~W.~T.,  {de Bruyn} G.,
  {Noutsos} A.,  {Alexov} A.,  {Anderson} J.,  {Asgekar} A.,  {Avruch} I.~M.,
  {Beck} R.,  {Bell} M.~E.,  {Bell} M.~R.,    et al. 2013, A\&A, 552, A58

\bibitem[\protect\citeauthoryear{Splaver}{Splaver}{2004}]{spl04}
Splaver E.~M.,  2004, PhD thesis, Princeton University, Princeton, N. J.,
  U.S.A.

\bibitem[\protect\citeauthoryear{Splaver, Nice, Arzoumanian, Camilo, Lyne \&
  Stairs}{Splaver et~al.}{2002}]{sna+02}
Splaver E.~M.,  Nice D.~J.,  Arzoumanian Z.,  Camilo F.,  Lyne A.~G.,    Stairs
  I.~H.,  2002, ApJ, 581, 509

\bibitem[\protect\citeauthoryear{Splaver, Nice, Stairs, Lommen \&
  Backer}{Splaver et~al.}{2005}]{sns+05}
Splaver E.~M.,  Nice D.~J.,  Stairs I.~H.,  Lommen A.~N.,    Backer D.~C.,
  2005, ApJ, 620, 405

\bibitem[\protect\citeauthoryear{Stairs, Faulkner, Lyne, Kramer, Lorimer,
  McLaughlin, Manchester, Hobbs, Camilo, Possenti, Burgay, D'Amico, Freire,
  Gregory \& Wex}{Stairs et~al.}{2005}]{sfl+05}
Stairs I.~H.,  Faulkner A.~J.,  Lyne A.~G.,  Kramer M.,  Lorimer D.~R.,
  McLaughlin M.~A.,  Manchester R.~N.,  Hobbs G.~B.,  Camilo F.,  Possenti A.,
  Burgay M.,  D'Amico N.,  Freire P.~C.~C.,  Gregory P.~C.,    Wex N.,  2005,
  ApJ, 632, 1060

\bibitem[\protect\citeauthoryear{{Stappers}, {Hessels}, {Alexov}, {Anderson} \&
  {Coenen}}{{Stappers} et~al.}{2011}]{sha+11ltd}
{Stappers} B.~W.,  {Hessels} J.~W.~T.,  {Alexov} A.,  {Anderson} K.,
  {Coenen} T.,  2011, A\&A, 530, A80

\bibitem[\protect\citeauthoryear{{Stinebring}}{{Stinebring}}{2013}]{sti13}
{Stinebring} D.,  2013, Class. Quant Grav., 30, 224006

\bibitem[\protect\citeauthoryear{{Stovall}, {Lynch}, {Ransom}, {Archibald},
  {Banaszak}, {Biwer}, {Boyles}, {Dartez}, {Day} \& {Ford}}{{Stovall}
  et~al.}{2014}]{slr+14ltd}
{Stovall} K.,  {Lynch} R.~S.,  {Ransom} S.~M.,  {Archibald} A.~M.,  {Banaszak}
  S.,  {Biwer} C.~M.,  {Boyles} J.,  {Dartez} L.~P.,  {Day} D.,    {Ford}
  A.~J.,  2014, ApJ, 791, 67

\bibitem[\protect\citeauthoryear{Taylor}{Taylor}{1992}]{tay92}
Taylor J.~H.,  1992, Philos. Trans. Roy. Soc. London A, 341, 117

\bibitem[\protect\citeauthoryear{{Taylor}, {Mingarelli}, {Gair}, {Sesana},
  {Theureau}, {Babak}, {Bassa}, {Brem}, {Burgay}, {Caballero}, {Champion},
  {Cognard}, {Desvignes} \& et al.}{{Taylor} et~al.}{2015}]{tmg+15ltd}
{Taylor} S.~R.,  {Mingarelli} C.~M.~F.,  {Gair} J.~R.,  {Sesana} A.,
  {Theureau} G.,  {Babak} S.,  {Bassa} C.~G.,  {Brem} P.,  {Burgay} M.,
  {Caballero} R.~N.,  {Champion} D.~J.,  {Cognard} I.,  {Desvignes} G.,    et
  al. 2015, Physical Review Letters, 115, 041101

\bibitem[\protect\citeauthoryear{{Arzoumanian}, {Brazier},
    {Burke-Spolaor}, {Chamberlin} \& et al.}{{Arzoumanian}
    et~al.}{2015}]{abb+15ltd} {Arzoumanian} Z., {Brazier} A.,
  {Burke-Spolaor} S., {Chamberlin} S., et al. 2015, ApJ, 813, 65

\bibitem[\protect\citeauthoryear{{Tiburzi}, {Hobbs}, {Kerr}, {Coles}, {Dai},
  {Manchester}, {Possenti}, {Shannon} \& {You}}{{Tiburzi}
  et~al.}{2015}]{thk+15}
{Tiburzi} C.,  {Hobbs} G.,  {Kerr} M.,  {Coles} W.,  {Dai} S.,  {Manchester}
  R.,  {Possenti} A.,  {Shannon} R.,    {You} X.,  2015, MNRAS accepted, ArXiv
  e-prints

\bibitem[\protect\citeauthoryear{Toscano, Bailes, Manchester \& Sandhu}{Toscano
  et~al.}{1998}]{tbms98}
Toscano M.,  Bailes M.,  Manchester R.,    Sandhu J.,  1998, ApJ, 506, 863

\bibitem[\protect\citeauthoryear{{Toscano}, {Sandhu}, {Bailes}, {Manchester},
  {Britton}, {Kulkarni}, {Anderson} \& {Stappers}}{{Toscano}
  et~al.}{1999}]{tsb+99}
{Toscano} M.,  {Sandhu} J.~S.,  {Bailes} M.,  {Manchester} R.~N.,  {Britton}
  M.~C.,  {Kulkarni} S.~R.,  {Anderson} S.~B.,    {Stappers} B.~W.,  1999,
  MNRAS, 307, 925

\bibitem[\protect\citeauthoryear{{van Haarlem}, {Wise}, {Gunst}, {Heald},
  {McKean}, {Hessels}, {de Bruyn}, {Nijboer}, {Swinbank}, {Fallows},
  {Brentjens}, {Nelles}, {Beck}, {Falcke}, {Fender}, {H{\"o}randel}, {Koopmans}
  \& et al.}{{van Haarlem} et~al.}{2013}]{vwg+13ltd}
{van Haarlem} M.~P.,  {Wise} M.~W.,  {Gunst} A.~W.,  {Heald} G.,  {McKean}
  J.~P.,  {Hessels} J.~W.~T.,  {de Bruyn} A.~G.,  {Nijboer} R.,  {Swinbank} J.,
   {Fallows} R.,  {Brentjens} M.,  {Nelles} A.,  {Beck} R.,  {Falcke} H.,
  {Fender} R.,  {H{\"o}randel} J.,  {Koopmans} L.~V.~E.,    et al. 2013, A\&A,
  556, A2

\bibitem[\protect\citeauthoryear{{van Haasteren} \& {Levin}}{{van Haasteren} \&
  {Levin}}{2010}]{vl10}
{van Haasteren} R.,  {Levin} Y.,  2010, MNRAS, 401, 2372

\bibitem[\protect\citeauthoryear{{van Haasteren}, {Levin}, {Janssen},
  {Lazaridis} \& {Kramer}}{{van Haasteren} et~al.}{2011}]{vlj+11ltd}
{van Haasteren} R.,  {Levin} Y.,  {Janssen} G.~H.,  {Lazaridis} K.,    {Kramer}
  M.,  2011, MNRAS, 414, 3117

\bibitem[\protect\citeauthoryear{{van Haasteren} \& {Vallisneri}}{{van
  Haasteren} \& {Vallisneri}}{2014}]{vv14}
{van Haasteren} R.,  {Vallisneri} M.,  2014, Phys. Rev. D, 90, 104012

\bibitem[\protect\citeauthoryear{van Straten}{van Straten}{2006}]{van06}
van Straten W.,  2006, ApJ, 642, 1004

\bibitem[\protect\citeauthoryear{{van Straten}}{{van Straten}}{2013}]{van13}
{van Straten} W.,  2013, ApJS, 204, 13

\bibitem[\protect\citeauthoryear{{Verbiest}, {Bailes}, {Bhat}, {Burke-Spolaor}
  \& {Champion}}{{Verbiest} et~al.}{2010}]{vbb+10ltd}
{Verbiest} J.~P.~W.,  {Bailes} M.,  {Bhat} N.~D.~R.,  {Burke-Spolaor} S.,
  {Champion} D.~J.,  2010, Classical and Quantum Gravity, 27, 084015

\bibitem[\protect\citeauthoryear{Verbiest, Bailes, Coles, Hobbs, {van Straten},
  Champion, Jenet, Manchester, Bhat, Sarkissian, Yardley, {Burke-Spolaor},
  Hotan \& You}{Verbiest et~al.}{2009}]{vbc+09}
Verbiest J.~P.~W.,  Bailes M.,  Coles W.~A.,  Hobbs G.~B.,  {van Straten} W.,
  Champion D.~J.,  Jenet F.~A.,  Manchester R.~N.,  Bhat N.~D.~R.,  Sarkissian
  J.~M.,  Yardley D.,  {Burke-Spolaor} S.,  Hotan A.~W.,    You X.~P.,  2009,
  MNRAS, 400, 951

\bibitem[\protect\citeauthoryear{{Verbiest}, {Bailes}, {van Straten}, {Hobbs},
  {Edwards}, {Manchester}, {Bhat}, {Sarkissian}, {Jacoby} \&
  {Kulkarni}}{{Verbiest} et~al.}{2008}]{vbv+08}
{Verbiest} J.~P.~W.,  {Bailes} M.,  {van Straten} W.,  {Hobbs} G.~B.,
  {Edwards} R.~T.,  {Manchester} R.~N.,  {Bhat} N.~D.~R.,  {Sarkissian} J.~M.,
  {Jacoby} B.~A.,    {Kulkarni} S.~R.,  2008, ApJ, 679, 675

\bibitem[\protect\citeauthoryear{{Verbiest} \& {Lorimer}}{{Verbiest} \&
  {Lorimer}}{2014}]{vl14}
{Verbiest} J.~P.~W.,  {Lorimer} D.~R.,  2014, MNRAS, 444, 1859

\bibitem[\protect\citeauthoryear{{Verbiest}, {Weisberg}, {Chael}, {Lee} \&
  {Lorimer}}{{Verbiest} et~al.}{2012}]{vwc+12}
{Verbiest} J.~P.~W.,  {Weisberg} J.~M.,  {Chael} A.~A.,  {Lee} K.~J.,
  {Lorimer} D.~R.,  2012, ApJ, 755, 39

\bibitem[\protect\citeauthoryear{{Walker}, {Demorest} \& {van
  Straten}}{{Walker} et~al.}{2013}]{wdv13}
{Walker} M.~A.,  {Demorest} P.~B.,    {van Straten} W.,  2013, ApJ, 779, 99

\bibitem[\protect\citeauthoryear{{Wang}, {Hobbs}, {Coles}, {Shannon}, {Zhu},
  {Madison}, {Kerr}, {Ravi}, {Keith}, {Manchester}, {Levin}, {Bailes}, {Bhat},
  {Burke-Spolaor}, {Dai}, {Os{\l}owski}, {van Straten}, {Toomey}, {Wang} \&
  {Wen}}{{Wang} et~al.}{2015}]{whc+15}
{Wang} J.~B.,  {Hobbs} G.,  {Coles} W.,  {Shannon} R.~M.,  {Zhu} X.~J.,
  {Madison} D.~R.,  {Kerr} M.,  {Ravi} V.,  {Keith} M.~J.,  {Manchester} R.~N.,
   {Levin} Y.,  {Bailes} M.,  {Bhat} N.~D.~R.,  {Burke-Spolaor} S.,  {Dai} S.,
  {Os{\l}owski} S.,  {van Straten} W.,  {Toomey} L.,  {Wang} N.,    {Wen} L.,
  2015, MNRAS, 446, 1657

\bibitem[\protect\citeauthoryear{Wolszczan, Doroshenko, Konacki, Kramer,
  Jessner, Wielebinski, Camilo, Nice \& Taylor}{Wolszczan
  et~al.}{2000}]{wdk+00}
Wolszczan A.,  Doroshenko O.,  Konacki M.,  Kramer M.,  Jessner A.,
  Wielebinski R.,  Camilo F.,  Nice D.~J.,    Taylor J.~H.,  2000, ApJ, 528,
  907

\bibitem[\protect\citeauthoryear{You, Hobbs, Coles, Manchester, Edwards,
  Bailes, Sarkissian, Verbiest, {van Straten}, Hotan, Ord, Jenet, Bhat \&
  Teoh}{You et~al.}{2007}]{yhc+07}
You X.-P.,  Hobbs G.,  Coles W.,  Manchester R.~N.,  Edwards R.,  Bailes M.,
  Sarkissian J.,  Verbiest J.~P.~W.,  {van Straten} W.,  Hotan A.,  Ord S.,
  Jenet F.,  Bhat N.~D.~R.,    Teoh A.,  2007, MNRAS, 378, 493

\bibitem[\protect\citeauthoryear{You, Hobbs, Coles, Manchester \& Han}{You
  et~al.}{2007}]{yhc+07b}
You X.~P.,  Hobbs G.~B.,  Coles W.~A.,  Manchester R.~N.,    Han J.~L.,  2007,
  ApJ, 671, 907

\bibitem[\protect\citeauthoryear{{Yue}, {Li} \& {Nan}}{{Yue}
  et~al.}{2013}]{yln13}
{Yue} Y.,  {Li} D.,    {Nan} R.,  2013, in {van Leeuwen} J.,  ed., IAU
  Symposium Vol.~291 of IAU Symposium, {FAST low frequency pulsar survey}.
pp 577--579

\bibitem[\protect\citeauthoryear{{Zhao}, {Zhang}, {You} \& {Zhu}}{{Zhao}
  et~al.}{2013}]{zzyz13}
{Zhao} W.,  {Zhang} Y.,  {You} X.-P.,    {Zhu} Z.-H.,  2013, Phys. Rev. D, 87,
  124012

\bibitem[\protect\citeauthoryear{{Zhu}, {Stairs}, {Demorest}, {Nice}, {Ellis}
  \& et al.}{{Zhu} et~al.}{2015}]{zsd+15ltd}
{Zhu} W.~W.,  {Stairs} I.~H.,  {Demorest} P.~B.,  {Nice} D.~J.,  {Ellis} J.~A.,
     et al. 2015, ApJ, 809, 41

\bibitem[\protect\citeauthoryear{{Zhu}, {Hobbs}, {Wen}, {Coles}, {Wang},
  {Shannon}, {Manchester}, {Bailes}, {Bhat}, {Burke-Spolaor}, {Dai}, {Keith},
  {Kerr}, {Levin}, {Madison}, {Os{\l}owski}, {Ravi}, {Toomey} \& {van
  Straten}}{{Zhu} et~al.}{2014}]{zhw+14}
{Zhu} X.-J.,  {Hobbs} G.,  {Wen} L.,  {Coles} W.~A.,  {Wang} J.-B.,  {Shannon}
  R.~M.,  {Manchester} R.~N.,  {Bailes} M.,  {Bhat} N.~D.~R.,  {Burke-Spolaor}
  S.,  {Dai} S.,  {Keith} M.~J.,  {Kerr} M.,  {Levin} Y.,  {Madison} D.~R.,
  {Os{\l}owski} S.,  {Ravi} V.,  {Toomey} L.,    {van Straten} W.,  2014,
  MNRAS, 444, 3709

\end{thebibliography}

\appendix

\section{The IPTA Timing Format}\label{sec:Format}
Because of the unprecedented size, diversity and precision of the IPTA
data set, the shortcomings of present pulsar timing practices have
come out very clearly during this combination. We therefore present a
series of guidelines detailing ``good pulsar timing practice'', aimed
at streamlining and optimising future pulsar timing efforts, below.

\paragraph*{Systemic Offsets.}
As described earlier, the determination of systemic offsets between
different telescopes and recording systems, is difficult. However,
with an increased homogeneity in data recording systems (presently
most pulsar data are created in software-based coherent-dedispersion
systems), systemic offsets may become far more tractable. In order to
ensure more accurate determination of these offsets in the future as
well as to ensure the usefulness of present data sets for future use,
we propose the following pulsar timing standard practices:
\begin{itemize}
\item When combining multiple systems, the \emph{reference system
    should be chosen as that system with the lowest value for
    $\sigma/\sqrt{N}$}, where $\sigma$ is the median ToA uncertainty
  for the system and $N$ is the number of ToAs for this system in the
  data set. Since any offsets are measured with respect to the
  reference system, choosing a system with worse precision or fewer
  ToAs will increase the uncertainty of all measured systemic offsets.
\item Systemic offsets are part of the timing solution and are
  therefore stored as part of the pulsar timing model. To ease
  combination and for increased clarity and convenience, we recommend
  that \emph{for the reference system an unfitted offset with zero
    value is also included} in the timing model.
\item Since absolute alignment of data sets can be assured only by
  cross-correlation of the standard templates used for creating these
  data sets, any \emph{ToAs should be accompanied by the template
    profile} used to create them.
\item In cases where simultaneous ToAs are used to determine systemic
  offsets, \emph{all simultaneous ToAs should be contained in the
    released data}. To properly weight these correlated ToAs, ideally
  information on their simultaneity would be included in a covariance
  matrix, though this is not effectively implemented as yet in the
  \textsc{tempo2} software.
\item \emph{Offsets between systems should never be absorbed in the
    ToAs}. This is to avoid corruptions of the most basic measurement
  data (i.e.\ the ToAs) and to provide transparency and clarity.
\end{itemize}

\paragraph*{Measurement Uncertainties.}
The causes behind underestimation of ToA uncertainties are not fully
clear yet, but a few aspects are understood and should be
accounted for in pulsar timing investigations. In particular we
therefore suggest the following:
\begin{itemize}
\item For observations of scintillating pulsars across a bandwidth
  that is large enough to encompass significant frequency-dependent
  variations in the profile shape, biases to the ToAs would be
  introduced by variations in the brightness distribution across the
  observing band. This problem can be averted by reducing the
  frequency range per ToA (as done for the NANOGrav data), or by using
  frequency-dependent template profiles. This latter method has been
  simultaneously and independently implemented by two groups:
  \citet{pdr+14} and \citet{ldc+14}.
\item The Fourier phase gradient method for ToA determination as
  proposed by \citet{tay92} measures the ToA of an observation by
  performing a phase-gradient fit to the Fourier transform of the
  cross-correlation of the observation and template profile. This
  traditional approach derives the ToA uncertainty from the second
  derivative of the $\chi^2$ at the best-fit point, like any standard
  $\chi^2$ optimisation routine. An alternative approach is to derive
  the uncertainty from a simple one-dimensional Markov-Chain
  Monte-Carlo based on the likelihood as a function of phase-shift. In
  the \textsc{psrchive} software package \citep{hvm04}, this method is
  implemented as \emph{``-A FDM'' and should be the default ToA
    determination method}. While the differences are negligible for
  high-S/N data, the standard $\chi^2$ fit tends to underestimate ToA
  uncertainties for low-S/N data (see \citet{lvk+11} and \citet[App.\
  B]{abb+15ltd}) so in these cases the FDM method is clearly
  preferred.
\item While often ignored, the additive white noise caused by pulse
  phase jitter scales with the square-root of the number of pulses
  averaged. In order to create more reliable measurement uncertainties
  (and lower any EQUAD values), this jitter noise should be included
  in the timing analysis, which requires \emph{the integration time}
  related to the ToAs.
  This information, along with other descriptors of individual ToAs
  (such as bandwidth, number of time bins and number of frequency
  channels, all of which can affect the sensitivity of the system and
  therefore the uncertainties) \emph{needs to be stored as part of the
    raw data}. To this end, the \textsc{psrchive} package has recently
  implemented the so-called ``IPTA'' ToA format, which extends the
  ToAs with such meta-data.
\item To quantitatively assess outlier ToAs, \emph{a goodness-of-fit
    value} (describing the template-to-observation fit) \emph{could be
    added to the meta data} provided along with the ToAs. In
  \textsc{psrchive}, this option already exists in combination with
  the FDM method described above (and other methods such as MTM) and
  can be invoked through ``-c gof''.
\end{itemize}

\paragraph*{Dispersion Measurements.}
As pulsar timing data sets become longer and more precise, they become
ever more sensitive to long-term DM variations. This means that
multi-frequency observing is crucial in the long-term, even for
pulsars in which DM variations are yet to be observed. The optimal way
of measuring and correcting variable DMs is still unclear, so the
principal aim in pulsar timing should be to \emph{provide the basic
  multi-frequency ToAs; and not derived DM values or models}.

\paragraph*{Intrinsic Pulsar Timing Instabilities.}
Long-period variations seen in some MSPs are mostly consistent between
telescopes \citep{lea+15}, indicating they are true astrophysical
signals. However, some observing systems have been shown to be
unreliable, producing signals which mimic intrinsic pulsar timing
instabilities. Such unreliability can only be identified and remedied
when comparable data sets from other telescopes exist.

\section*{Affiliations}
{\footnotesize
  $^{1}$Fakult\"at f\"ur Physik, Universit\"at Bielefeld, Postfach
  100131, 33501 Bielefeld, Germany\\
  $^{2}$Max-Planck-Institut f\"ur Radioastronomie, Auf dem H\"ugel 69,
  53121 Bonn, Germany\\
  $^{3}$Astrophysics Group, Cavendish Laboratory, JJ Thomson Avenue,
  Cambridge CB3 0HE, UK\\
  $^{4}$CSIRO Astronomy and Space Science, Australia Telescope
  National Facility, PO Box 76, Epping NSW 1710, Australia\\ 
  $^{5}$Jet Propulsion Laboratory, California Institute of Technology,
  4800 Oak Grove Dr., M/S 67-201, Pasadena, CA 91109, USA\\ 
  $^{6}$National Radio Astronomy Observatory, P.O. Box O, Socorro, NM
  87801, USA\\ 
  $^{7}$ASTRON, the Netherlands Institute for Radio Astronomy, Postbus
  2, 7990 AA, Dwingeloo, the Netherlands\\ 
  $^{8}$Xinjiang Astronomical Observatory, Chinese Academy of
  Sciences, 150 Science 1-Street, Urumqi, Xinjiang 830011, China\\ 
  $^{9}$Jodrell Bank Centre for Astrophysics, School of Physics and
  Astronomy, The University of Manchester, Manchester M13 9PL, UK\\ 
  $^{10}$Center for Research and Exploration in Space Science and
  Technology/USRA and X-Ray Astrophysics Laboratory, NASA Goddard
  Space Flight Center, Code 662, Greenbelt, MD 20771, USA\\
  $^{11}$MPI for Gravitational Physics (Albert Einstein Institute),
  Golm-Potsdam 14476, Germany\\ 
  $^{12}$International Centre for Radio Astronomy Research, Curtin
  University, Bentley, WA 6102, Australia\\
  $^{13}$Cornell Center for Advanced Computing, Cornell University,
  Ithaca, NY 14853, USA\\
  $^{14}$Cornell Center for Astrophysics and Planetary Science,
  Cornell University, Ithaca, NY 14853, USA\\
  $^{15}$INAF-Osservatorio Astronomico di Cagliari, via della Scienza
  5, 09047, Selargius (CA), Italy\\ 
  $^{16}$Department of Physics, The Pennsylvania State University,
  University Park, PA 16802, USA\\
  $^{17}$Astronomy Department, Cornell University, Ithaca, NY 14853,
  USA\\
  $^{18}$Notre Dame of Maryland University, 4701 N.\ Charles Street,
  Baltimore, MD 21210, USA\\
  $^{19}$Laboratoire de Physique et Chimie de l'Environnement et de
  l'Espace LPC2E CNRS-Universit\'e d'Orl\'eans, F-45071 Orl\'eans,
  France\\
  $^{20}$Station de radioastronomie de Nan{\c c}ay, Observatoire de
  Paris, CNRS/INSU F-18330 Nan{\c c}ay, France\\ 
  $^{21}$Department of Astronomy, School of Physics, Peking University,
  Beijing, 100871, China\\
  $^{22}$Department of Physics, Hillsdale College, 33 E. College
  Street, Hillsdale, Michigan 49242, USA\\
  $^{23}$McGill University, Department of Physics, Rutherford Physics
  Building, 3600 University Street, Montreal, QC, H3A 2T8, Canada\\ 
  $^{24}$Dept. of Physics and Astronomy, University of British
  Columbia, 6224 Agricultural Road, Vancouver, BC V6T 1Z1, Canada\\
  $^{25}$School of Mathematics, University of Edinburgh, King's
  Buildings, Edinburgh EH9 3JZ, UK\\
  $^{26}$Department of Physics and Astronomy, West Virginia
  University, Morgantown, WV 26506, USA\\
  $^{27}$Vancouver Coastal Health, Department of Nuclear Medicine, 899
  W 12th Ae, Vancouver, BC, V5Z 1M9, Canada\\
  $^{28}$Anton Pannekoek Institute for Astronomy, University of
  Amsterdam, Science Park 904, 1098 XH Amsterdam, the Netherlands\\
  $^{29}$Department of Physics, Columbia University, New York, NY
  10027, USA\\
  $^{30}$Monash Centre for Astrophysics (MoCA), School of Physics and
  Astronomy, Monash University, VIC 3800, Australia\\
  $^{31}$Kavli institute for astronomy and astrophysics, Peking
  University, Beijing 100871, P.~R.~China\\
  $^{32}$National Radio Astronomy Observatory, P.O.~Box 2, Green Bank,
  WV, 24944, USA\\
  $^{33}$National Radio Astronomy Observatory, 520 Edgemont Rd.,
  Charlottesville, VA 22903, USA\\
  $^{34}$TAPIR, California Institute of Technology, MC 350-17,
  Pasadena, CA 91125, USA\\
  $^{35}$Physics Department, Lafayette College, Easton, PA 18042,
  USA\\ 
  $^{36}$Texas Tech University, Physics Department, Box 41051,
  Lubbock, TX 79409\\ 
  $^{37}$University of Virginia, Department of Astronomy, P.O.~Box
  400325 Charlottesville, VA 22904-4325, USA\\
  $^{38}$Universit\'e Paris-Diderot-Paris7 APC - UFR de Physique,
  Batiment Condorcet, 10 rue Alice Domont et L\'eonie Duquet, 75205
  Paris Cedex 13, France\\
  $^{39}$Centre for Astrophysics and Supercomputing, Swinburne
  University of Technology, PO Box 218, Hawthorn VIC 3122, Australia\\
  $^{40}$School of Physics and Astronomy, University of Birmingham,
  Edgbaston, Birmingham, B15 2TT, UK\\
  $^{41}$Center for Gravitation, Cosmology and Astrophysics,
  Department of Physics, University of Wisconsin-Milwaukee, P.O.~Box
  413, Milwaukee, WI 53201, USA\\
  $^{42}$Physics and Astronomy Dept., Oberlin College, Oberlin OH
  44074, USA\\
  $^{43}$Department of Physics and Astronomy, University of New
  Mexico, Albuquerque, NM 87131, USA\\
  $^{44}$Laboratoire Univers et Th\'eories LUTh, Observatoire de
  Paris, CNRS/INSU, Universit{\'e} Paris Diderot, 5 place Jules
  Janssen,\\ 92190 Meudon, France\\ 
  $^{45}$School of Physics, Huazhong University of Science and
  Technology, Wuhan, Hubei Province 430074, China\\
  $^{46}$School of Physics, University of Western Australia, 35
  Stirling Hwy, Crawley WA 6009, Australia\\
  $^{47}$School of Physical Science and Technology, Southwest
  University, Chongqing 400715, China\\
}

\label{lastpage}

\end{document}